\documentclass[a4paper]{article}

\usepackage[english]{babel}
\usepackage[utf8]{inputenc}
\usepackage{amsmath}
\usepackage{url}
\usepackage{graphicx}
\usepackage[colorinlistoftodos]{todonotes}
\usepackage{longtable}
\usepackage{hyperref}
\usepackage{authblk}

\usepackage[margin=0.7in]{geometry}

\title{Inferring Mechanisms for Global Constitutional Progress}

\author[1]{Alex Rutherford}
\author[2]{Yonatan Lupu}
\author[3]{Manuel Cebrian}
\author[4]{Iyad Rahwan}
\author[5]{Brad LeVeck}
\author[1]{Manuel Garcia-Herranz}

\affil[1]{UNICEF Office of Innovation, New York}
\affil[2]{George Washington University, Washngton DC}
\affil[3]{NICTA, Melbourne}
\affil[4]{Massachusetts Institute of Technology, Cambridge}
\affil[5]{University of California, Merced}

\date{}
\begin{document}

  \maketitle


\begin{abstract}
Constitutions help define domestic political orders, but are known to be influenced by two international mechanisms: one that reflects global temporal trends in legal development, and another that reflects international network dynamics such as shared colonial history. We introduce the provision space; the growing set of all legal provisions existing in the world's constitutions over time. Through this we uncover a third mechanism influencing constitutional change: hierarchical dependencies between legal provisions, under which the adoption of essential, fundamental provisions precedes more advanced provisions. This third mechanism appears to play an especially important role in the emergence of new political rights, and may therefore provide a useful roadmap for advocates of those rights. We further characterise each legal provision in terms of the strength of these mechanisms.

\end{abstract}

\section{Introduction}

Constitutions detail literally what constitutes an entity. In the case of nation-states, formal constitutions describe the fundamental principles by which the state will be governed, the political and legal state institutions, the powers, procedures, and duties of those institutions, and the rights and responsibilities of individuals.  The roles of the constitution are several; as a `rule book` for how the country should operate, as a reflection of the values and character of that country as observed by the rest of the world as well as a means for accountability through international law~\cite{practicalguide}.\\ \newline

There is great diversity in the processes used to create and amend these documents~\cite{elster}. In some cases, the same document is slowly refined over many decades or even centuries.  In others, constitutions are dissolved and replaced with new documents as part of a broader political transition. This may coincide with the installation of a new political regime, particularly when the constitution itself declares amendment to be unconstitutional. The processes for constitutional creation and evolution take many forms, including executive action, constituent assembly, referendum, or some combination of these~\cite{process}. The constitution writing process generally involves some degree of compromise and coordination between elected representatives, institutions, and/or citizens, such as the case of Iceland's 2008 constitutional reform~\cite{iceland}. This diversity in origin persists in present day constitutional documents; in a small number of countries the constitution is partially unwritten, and legal rules and principles are derived in part from judicial precedent.\\ \newline

In addition to the influence of these domestic processes, the present-day versions of formal constitutions also represent the result of long-term and complex international interdependencies.  Affiliation or potential affiliation with a formal bloc, union, international organization, or informal community of states --  such as the European Union or United Nations -- can influence constitution writing.  Likewise, wartime turbulence and post-war occupation and reconstruction can result in dramatic changes to formal constitutions~\cite{japan}.  Ongoing international events, trends, and processes can influence the content and style of constitutions being drafted or modified at a given time.  Changing social and legal norms, cultural trends, and new empirical evidence bases likely yield profound and enduring influences on constitution drafting, whereas coordinated political movements -- such as decolonisation -- may give rise to a less sustained influence.  Seven distinct “waves” of constitutional activity have been identified~\cite{elster}, including, during the 20th century, in the aftermath of the First World War, the Second World War, decolonisation beginning with the Indian subcontinent and culminating in the 1960s, the fall of Southern European dictatorships in the 1970s, and finally the fall of the Soviet Union. Throughout this period and into the future, globalization is also likely to provide pressure on countries to adopt the legal norms of influential and powerful neighbors.\\ \newline

Ongoing pressures for constitutions to evolve are nonetheless tempered by historical legacy and the often cumbersome processes for constitutional amendment. Many constitutions belonging to former colonies were based on or were influenced by the laws and constitutions of former Imperial powers~\cite{go2002,billrights}. The colonial relationship may also continue to exert influence after independence~\cite{transplants}. Extant constitutions also influence new constitutions by forming a baseline relative to which changes are made or new provisions added.  It is an open question as to whether concerted global policy efforts by international bodies such as the UN e.g. the Convention on the Rights of the Child~\cite{rightschild}, can affect these processes in a durable way. This combination of interacting influences between a large set of actors with a series of abrupt perturbations and trends acting at different time-scales are typical of a complex system~\cite{complexsystems}.\\ \newline

Comparative law is the study of differing national legal systems.  This vast literature has yielded several classifications of legal systems such as Arminjon et al~\cite{arminjon}, La Porta et al.~\cite{laporta} and Zweigert et al~\cite{zweigert}, and adjudicating between those categorizations is beyond the scope of this work. These classifications are based largely on qualitative assessments focusing on religion, national history, and differing applications of common law and civil law.  The application of computational techniques to constitutional documents and the historical records of their changes provide an attractive alternative toolset with which to systematically analyze and measure the dynamics of global constitutionalism.\\ \newline

Our work uses a range of quantitative techniques from the field of computational social science to analyse the content and structure of national constitutions. Our focus is on public law, which governs relations between persons and the state, and not on private law (such as contract law), which governs relations between persons.  Computational social science is now a well-developed field in which computational methods (and particularly those of network science~\cite{barabasi13}) are applied to analyze social systems with numerable and diverse applications~\cite{css}.\\ \newline

The contributions of this work are as follows. Firstly we introduce a new mechanism by which provisions are introduced into a growing provision space of legal rights that are included in constitutional documents. We identify hierarchical dependencies between provisions, under which related provisions are adopted in a natural sequential order. Secondly, we empirically quantify the well understood phenomena of horizontal influence between countries (particularly along former Imperial relations) as well as temporal trends in constitutional development. We are able to confirm, using computational techniques, theoretical results from comparative law concerning families of legal systems and also a novel classification of provisions in terms of the relative strengths of competing these effects.\\ \newline

Although the use of computational tools is relatively new in this literature, it is not without precedent.  In~\cite{constitutionalism}, the authors analysed the presence or absence of social rights provisions over a 60-year period. They used this to classify countries on an ideological scale with a libertarian, common law nature at one extreme and a more statist nature at the other.  Later work~\cite{transnational} found evidence that the adoption of social rights is influenced by former colonizers, other countries with the same legal system and the same dominant foreign aid donor. Melton et al.~\cite{melton} investigated the interpretability of national constitutions and found that textual features such as sentence length and the Flesch index measure of complexity were more important than contextual factors such as the geographical region and applicability of common law.  More recent work  has used automatic content analysis and topic modeling~\cite{preambles,archetypes} to classify and assess formal constitutions, although this work has examined only the preambles (or introductions) to these documents rather than the full texts. Rockmore et al analysed the diffusion of legal concepts between constitutions over time using a biological framework in which some countries' constitutions inherit from one another~\cite{evolution}. It was found that several distinct epochs of cultural evolution exist and while most constitutions are only influential for a short period, several have a sustained influence.\\ \newline

Henceforth, we consider two different data sets: the current text of the constitutions of 194 nation-states and expert labeled provisions within them (\textbf{dataset I}); the historical timeline of amendments and new writing events to such texts along with the presence or absence of 234 constitutional provisions that address specific issues or rights in these historical documents (\textbf{dataset II}).

\section{Methods}
\subsection{Temporal Dynamics}

Methods:
Temporal Dynamics
We seek to characterise the temporal dynamics of constitutional development with a view to uncovering the influence of time on which provisions are included in a constitution. We begin by introducing the concept of a provision space in analogy with the product space~\cite{productspace}, by which a country may only produce complex goods by first producing other related goods of incrementally increasing complexity. We define the provision space as the set of provisions included across all national constitutions in a given year. Formally the provision space at time t, $\rho(t)$ depends on the provisions of each of the $n^{c}(t)$ countries in existence in year t, each of which is given by $n^{c}(t)$ as

\begin{equation}
\rho(t)=\{\rho(t)^{1} \cup \rho(t)^{2} \dots \cup \rho(t)^{n^{c}(t)}\}
\end{equation}

Henceforth we consider only 234 provisions from the survey undertaken in~\cite{comparative} with a binary yes/no answer (see SI Table 10 for a full list) (dataset II). In figure (1) we plot the growth of the provision space over time as new provisions, representing legal concepts and rights, are introduced and first adopted. We see a clear increasing, although not strictly monotonic, trend. We also note the trend for increasing numbers of new countries over time. Periods of particularly rapid growth arise in the provision space, often coinciding with significant events of constitutional history such as World War I/II . However we also note the opposite situation: rapid growth in the number of constitutions during periods of geo-political change but without any growth in the provision space e.g. African decolonisation in the period 1960-70.

\begin{figure*}[h]
\centerline{\includegraphics[width=1.0\linewidth]{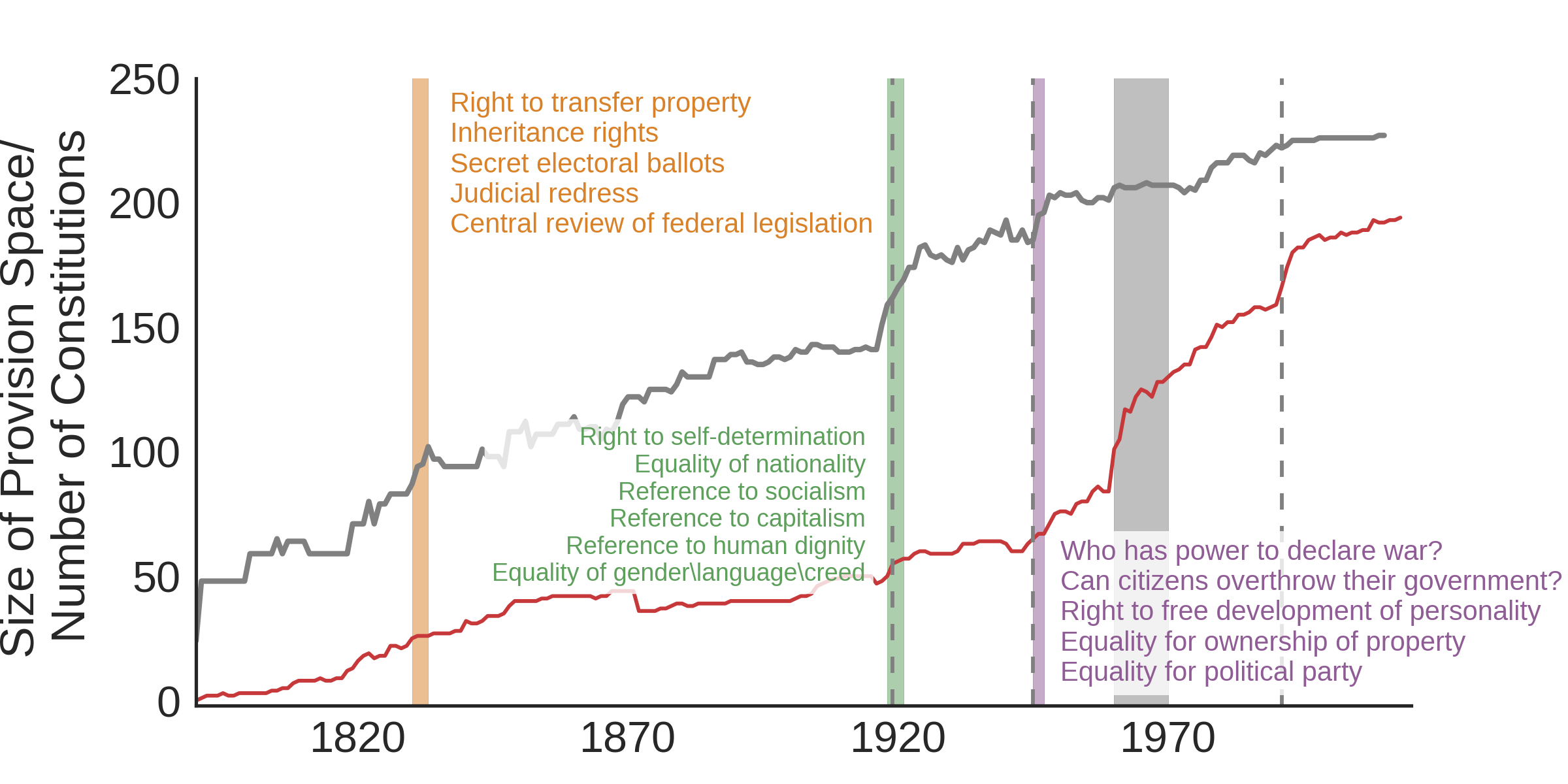}}
\caption{The size of the provision space over time (red line) compared to the growth in the number of countries (grey line). The introduction of specific provisions are highlighted.}
\end{figure*}

\subsection{Provision Co-Occurrence and Hierarchy}

Next we consider dependencies between provisions to see if the adoption of one co-occurs or is more likely to be adopted with respect to another. We begin by focusing on constitutional provisions relating to young people and children. We identify the 9 provisions most clearly affecting children and young people; Access to higher education, Guarantee of rights of children, Limits on child employment, Free education, General mention of education, Freedom from parentage discrimination, Freedom from age discrimination, Financial support for children/orphans and Special status to juveniles in court.
We consider the provision that is most frequently co-occurring with each of these child specific provisions within the set of current constitutions (Table 1). Most of these relations are intuitive; for example limiting the employment of children is associated with the right to join a trade union. Intuitively, it follows that to consider the protection of children from work, adults should also be protected from exploitative employment.\\\newline

\begin{table}
\small
\centering
\label{child_provisions}
\begin{tabular}{|l|l|}

Child Provision                      & Top Co-occurring Provision    \\\hline
Access to higher education                & Right to join trade unions \\
Privileges for juveniles in criminal process  & Protection from ex post facto laws \\
Compulsory education                 & Free education                        \\
Free education                       & Protection of environment             \\
Limits in the employment of children & Right to join trade unions           \\
Equality regardless of parentage     & General guarantee of equality  \\
Right to found a family              & General guarantee of equality        \\
State support for children           & State support for the disabled        \\
Rights of children guaranteed        & General guarantee of equality        \\\hline

\end{tabular}
\end{table}

The relationship between employment protection laws for adults and children hints at the presence of hierarchical dependencies within the provisions. Such a hierarchy is intuitively appealing; a fledgling state will set out rules for the basic functioning of the state through structural provisions before addressing other issues as the state matures. In addition, legal activists often use the strategy of “norm grafting’’ to advocate for the adopting of new rights by associating them with existing rights~\cite{norm}.\\\newline

We explore possible hierarchical structure by constructing a directed network between provisions, such that an edge from i to j is weighted by the number of countries for which i has been adopted before j. From this we find the Minimal Violation Ranking; the ranking most consistent with this network as in~\cite{hiring}. We find strong evidence that the provisions represent a hierarchical structure with 24\% of edges being violated (50\% represents `no better than random`, see SI). This is consistent with a `rights-creep`~\cite{constitutionalism} under which rights are increasingly likely to be adopted, and in the case of younger states, increasingly likely to be adopted at their inception. This is supported by the increasing mutual similarity over time in provisions adopted (see S27 in SI and the fact that provisions are nearly 3 times more likely to be adopted when a country's constitution is first written than to be inserted by amendment. The imitation evident in the text of Imperial powers and colonies along with the affect of global policy efforts support this (see next section).
\\\newline

\begin{figure*}[h]
\centerline{\includegraphics[width=1.0\linewidth]{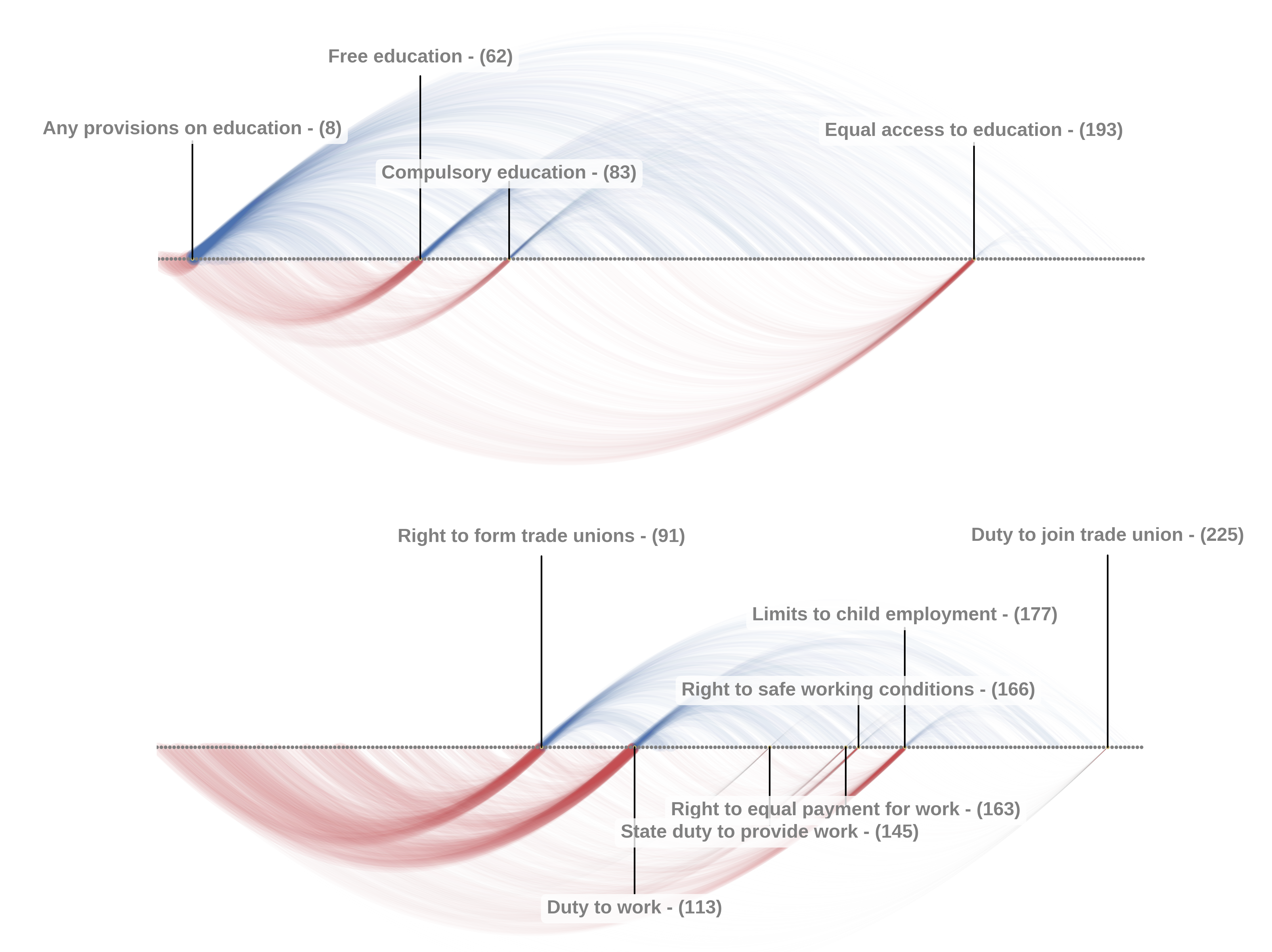}}
\caption{A representation of the hierarchical dependencies of provisions based on the sequential adoption by all countries over time. Provisions on the left are adopted before provisions to the right. Provisions related to (a) education and  (b) work are labeled with their numerical position in the hierarchy (out of 234) and their connections to provisions below them (blue) and above them (red) in the hierarchy. Provisions at the top of the hierarchy will have more blue edges than red and vica versa for those at the bottom of the hierarchy.}
\end{figure*}

The provision space is, by definition, a fluid set of rights and concepts that grows as the result of many societal and global level processes. These include new evidence bases, technological advances and continued advocacy on behalf of marginalised groups which can account for provisions such as protection of the environment, the right to privacy and protection of children respectively. Therefore, by definition the provision space contains both very mature provisions that are accepted as building blocks of a constitution as well as very young and potentially controversial concepts. Thus the position of a provision in the hierarchy is somewhat determined by the year in which the provision first entered the provision space (Spearman (r,p)=(0.69,$<10^{-32}$)).\\\newline

Specific examples of provisions related to work and education are visualised in Fig 2. We see that while a general reference to education is fundamental to constitutions (appearing at 8th position in the ranking), explicit mention of equal access to education is adopted much later (193rd). Likewise, the right to form trade unions is a foundational provision that precedes more specific labor rights such as equal pay for equal work and the right to a safe working environment. This latent structure suggests that efforts to impose emerging rights on young states will be more fruitful if an incremental approach is used.

\subsection{Constitutional Similarity Network}

It is well understood that constitutions of newly independent states typically inherit from extant constitutions and particularly from those of former colonial powers. In order to illuminate the effect of this mechanism we construct a document similarity network in which countries are nodes that are linked by edges determined by the similarity of their text as measured by the Jaccard similarity. Here we consider the entire constitutional text, including text which may not be attributable to a specific subject e.g. a preamble.
We find strong text similarity values between the constitutions of countries with shared Imperial histories despite considerable diversity in terms of word length, time of writing and other characteristics. There are several possible explanations for this similarity, it could be attributed to wholesale inheritance of text at the moment of independence or an influence that persists after independence due to academic and immigrant links between countries and their former Imperial powers. More simply, it may be a byproduct of the decision to include the same provisions, with the exact wording being largely a function of this choice. However, we find low similarity values between the language dedicated to the same provision in different constitutions. Further, we find only an intermediate correlation between the similarity matrices based on full constitutional text on one hand and the concatenation of the all provisions on the other (Pearson r=0.39, p$<10^{-12}$). This suggests that the word choice similarity is driven by both similarity within non-provisional content and similar wording between different provisions.
Upon the application of network community detection, the network partitions into clusters of mutually similar countries. 4 clusters emerge, the membership of which is listed in full in the SI.

\begin{figure*}[h]
\centerline{\includegraphics[width=1.0\linewidth]{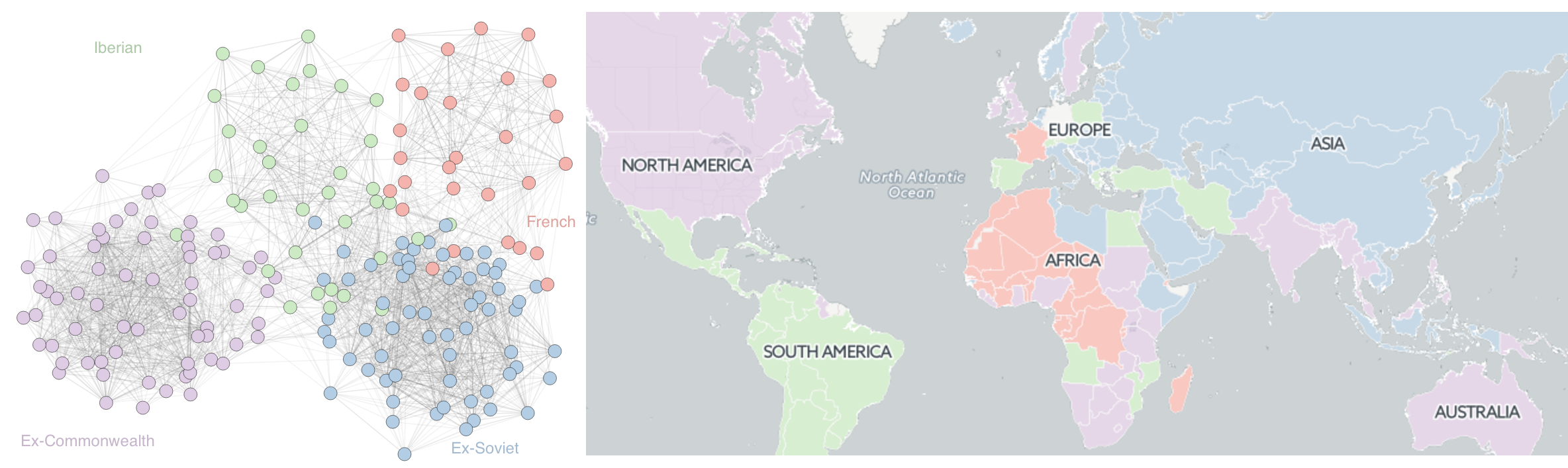}}
\caption{Clusters derived from constitutional similarity network (a) as a geographical map (b) as a network}
\end{figure*}

We note that the clusters reveal consistency with historical trends and some qualitative characterizations of legal system type.  One cluster contains the majority of former French colonial countries along with France itself. The former British and Spanish empires along with the UK and Spain themselves separate into distinct clusters. The final cluster does not reveal the same clear historical relation; however, we note that these countries include the majority of former members of the Soviet Union along with Russia itself. Another clear component of this cluster is Middle Eastern, especially Persian Gulf countries. The final component are Scandinavian countries. We note that this measure produces some cluster memberships that are not immediately intuitive e.g. Greece and Germany associated in the Iberian cluster, however there is remarkable explanatory power for a simple text similarity measure. We emphasise that the purpose is to uncover macroscopic structure in a robust manner, rather than derive a fully consistent set of individual dyadic relations between countries.\\\newline
We also find consistencies in the timeline of constitutional amendment activity broken down by cluster using historical data from dataset II based on clusters described above (Fig 4). Some common features are shared between each cluster; such as a spike in activity following World Wars and the collapse of the Soviet Union. Remarkably, significant coordination is observed between clusters despite limited geographic proximity or obvious political similarity. A simple pairwise correlation over time confirms this; with statistically significant correlation values in the range (0.45-0.68), more details in the SI.\\\newline

\begin{figure*}[h]
\centerline{\includegraphics[width=1.0\linewidth]{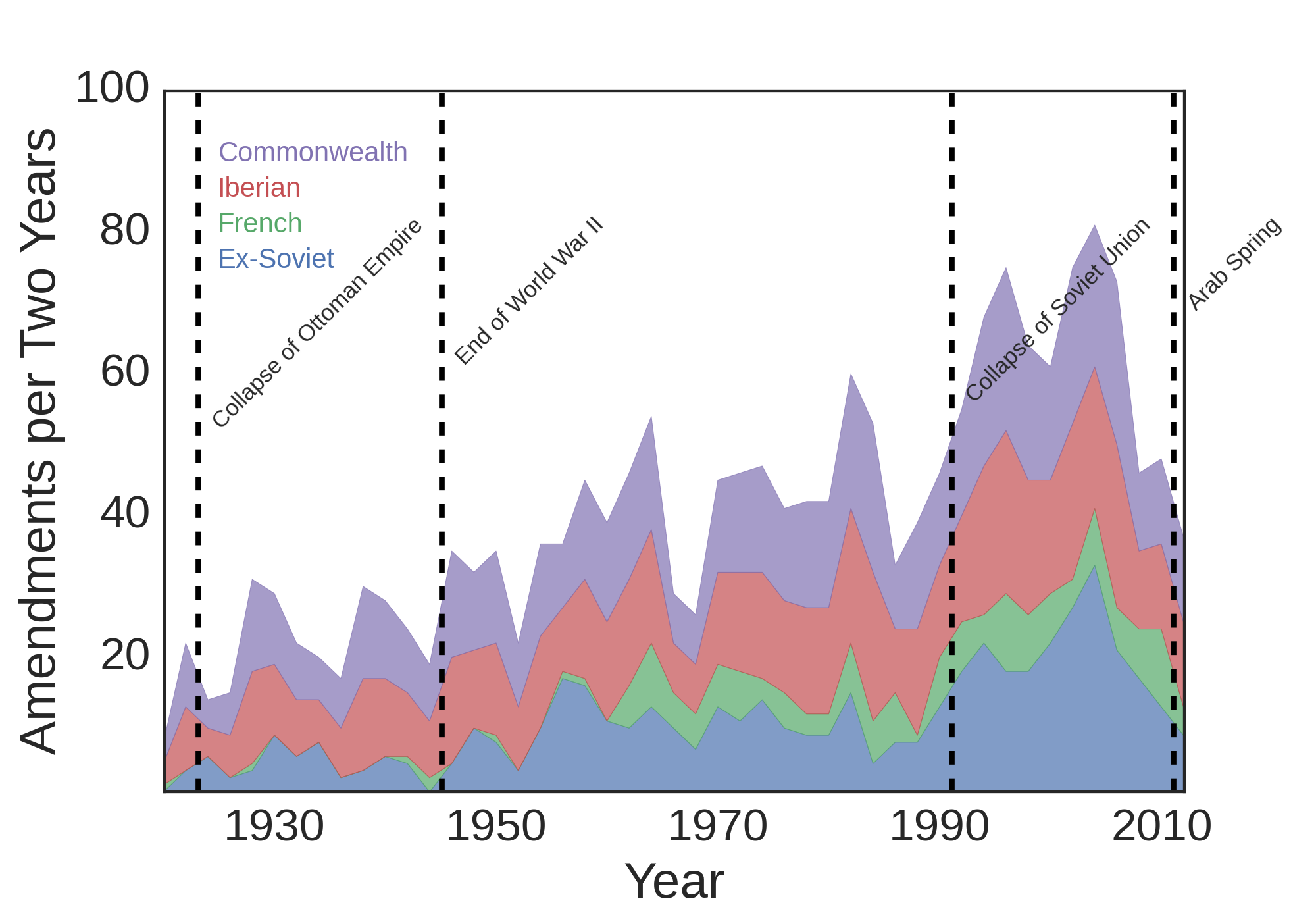}}
\caption{Timeline of constitutional amendment activity over time broken down by language based cluster}
\end{figure*}

In order to quantify the correspondence between this empirical structure and theoretical classifications of legal systems, we analyze the clusters using a series of multinomial logit models in which the cluster ID is the dependent variable. The clusters appear to correlate well with the legal system types identified by~\cite{laporta}. They divide legal system into 5 categories: UK (which is correlated with our Commonwealth cluster), French (our Francophone cluster), Spanish (our Iberian cluster), former Socialist (our Soviet Cluster), and Scandinavian (our Soviet Cluster).  A model that includes only the La Porta classifications yields an expected percent correctly predicted (ePCP) of 58.48\%, so these data incorrectly predict many of our results.  Likewise, a model that includes only indicators of former colonial powers, using data from~\cite{chilton}, has an ePCP of 65.01\%.  A model that includes both sets of variables yields an ePCP of 76.24\%, which suggests that colonial history and legal system type correctly predict the word choice of constitutional texts, but nonetheless also suggests that other factors may play important roles in influencing such texts.\\\newline

We now address the question of whether this inheritance of constitutional language generally implies inheritance of the same individual provisions as are present in a parent constitution. We compare text similarity and provisional similarity as follows; for each constitution we define a constitutional provisional fingerprint, a binary valued vector of length nprovisions indicating the presence or absence of a provision based on human coding. The provisional similarity between two countries is computed as the cosine distance between their provisional fingerprints, the text similarity is defined as before. For each country we measure the textual and provisional similarity of the remaining $n_{countries}$-1=193. That is for each country i we have a set of textual similarities and provisional similarities. From these two measures of similarity, we construct two country rankings; from most similar to least similar. Using the Spearman rank correlation coefficient we derive a value in the range [-1,1] that describes how these rankings compare. Values close to one suggest that the constitutions that have the most similar text content also have similar provisions (and vica versa) as would be expected if constitutions were perfectly inherited with no new provisions inserted or deleted. Surprisingly, we find low correlations between these two rankings; $<\rho>$=0.227 ($\sigma$=0.146). Correlating the raw numerical values rather than ranks using a Pearson correlation yields similar results with $<\rho>$=0.257 ($\sigma$=0.149). A final examination of provision clusters finds little correspondence between similar provisions and country clusters (see SI S14-16).\\\newline

These findings suggest that constitutions do not inherit perfectly; preserving both wording and substantive content. While the similarities in wording are consistent in terms of colonial history and geography, these findings suggest changes to the individual provisions despite imitation of wording. In addition we note that when countries first create a constitution there is a strong tendency to add provisions across all colonial based clusters (see SI S20). This suggests that other contextual factors, exogenous to the precedent or influence of other countries with shared history or legal system, encourage the adoption of provisions in new constitutions.

\subsection{Disentangling Global and Network Effects}

The results above have shown that strong coordination occurs between countries with shared colonial history and legal systems, which we refer to as `network effects`, which manifest in word choice and provisional topics. On the other hand we see strong temporal trends driven by significant historical events and policy efforts leading to correlated global behaviour independent from network structure.\\\newline

Having found coordination between clusters, we investigate correlated temporal behaviours within provisions (Fig 5b). Our dataset of provisional adoption of countries suffers from a changing denominator; new constitutions enter when a country achieves independence and may leave for several reasons; the country may simply no longer exist due to annexation or unification or the constitution may be suspended for example during a temporary period of military rule. With this variability in mind, henceforth we consider the proportion of constitutions in existence which include each provision over time. We perform a clustering of provisions based on a dimensionality reduction on the yearly time series of proportional adoption of each provision (see S21-26 in SI for more details).\\\newline

Although these provisions are not arranged in topics or a hierarchy, we can extract some themes by inspection of the provision labels (Fig 5b). Cluster 1 mostly describes social rights of citizens and increases steadily over time. Cluster 2 also describes social rights as well as religion and privacy and is distinguished from the other clusters by a relatively high and constant adoption from the early 20th century. Cluster 3 describes legal obligations of the state and enjoys a peak around 1945. All clusters are characterized by an increase and convergence over time with the exception of cluster 4 which contains a few obscure and sparsely adopted provisions including the right to bear arms and same-sex marriage (the full list is found in SI. Each component is distinguished by behaviour in specific periods corresponding to World War II and the collapse of the Soviet Union. The trend for convergence and increasing adoption is observed not only across the components found above, but across the set of constitutions as a whole (see SI S17).\\\newline

Having demonstrated that both network and temporal effects are significant, we next quantify the effects of each on the adoption of individual provisions. We consider each provision in dataset II and compare the probability of two countries co-adopting each provision, under the condition that they belong to the same cluster based on word choice identified above (manifesting network effects) and the probability of coadoption in the same year (manifesting global temporal effects).\\\newline

If the adoption of a provision is largely determined by colonial history and legal system, the probability of coadoption conditioned on cluster would be relatively high. Conversely, if the adoption of a provision is influenced by time-varying trends and social norms, then the probability of coadoption by year would be higher. These two effects are intuitively far from independent; countries sharing colonial history may well undergo constitutional amendment at the same time e.g. a coordinated period of decolonisation. This dependence is confirmed by the scatter plot of these quantities per provision (Fig 5a) and a linear fit ((r,p)=(0.936,<$10^{-12}$) showing that the probability of temporal coadoption and network coadoption increase together.\\\newline

\begin{figure*}[h]
\centerline{\includegraphics[width=1.0\linewidth]{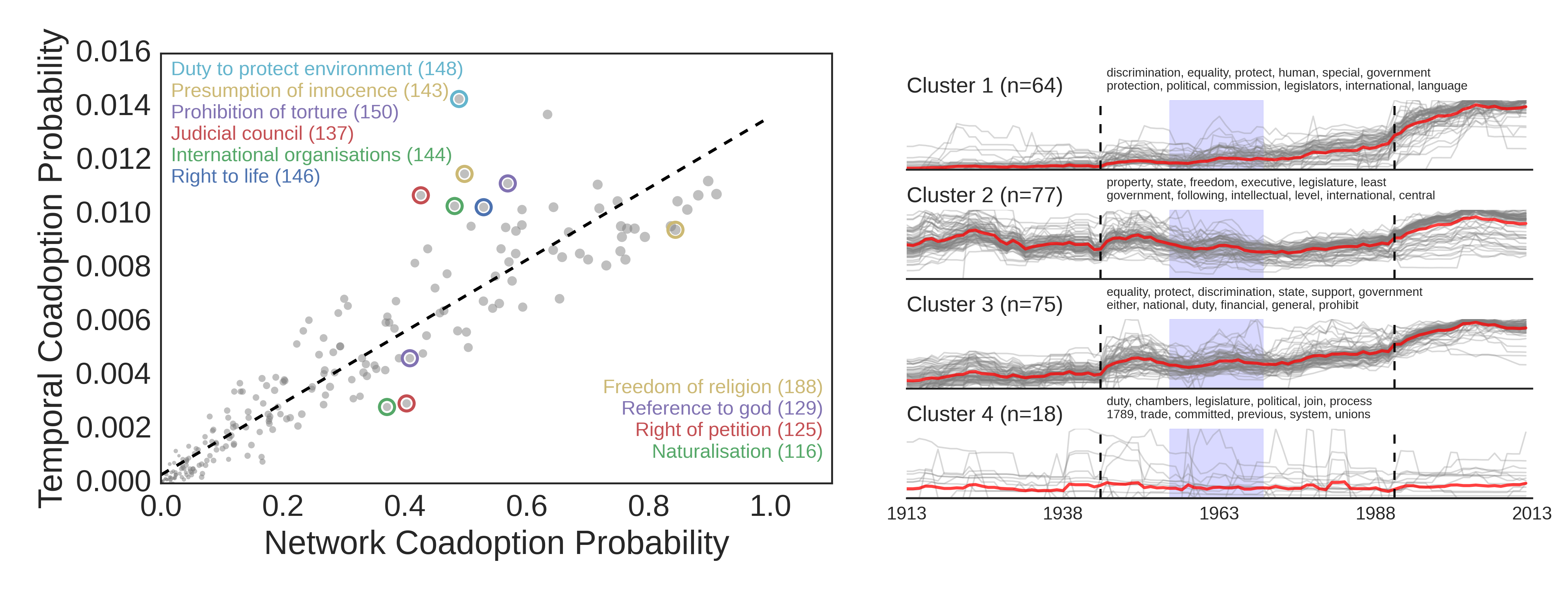}}
\caption{Scatter plot of conditional probability of coadoption of a provision given a common year of writing and a common network cluster. Dashed line indicates a linear regression; the adoption of provisions above this line is driven by time, and those below by network effects. The size of the points indicates the number of countries adopting in the final year of the dataset (also indicated by the number in brackets) (b) Clustering of time series of adoption of provisions. The number of provisions in each cluster is labeled along with the most common terms in the labels of the cluster members. The mean of each cluster is marked in red}
\end{figure*}

However we note several outlying provisions, which may now be classified by the residual difference from the regression line and associated z-score. For example, reference to god (z=\,-\,0.89) is more strongly coadopted within clusters compared to the average since the role of religion in the state varies significantly with differing legal systems. Whereas coadoption of a duty to protect the environment (z=6.00), a provision intuitively linked to changing norms over time, is favourably coadopted in the same year.\\\newline

We isolate provisions relevant to children in the context of the 1991 Convention on the Rights of the Child  (the underlying adoption of these 9 provisions is shown in S27 in (22)). We note that a general guarantee of the rights of the child (z=1.1) and financial support for orphans (z=0.69) is coadopted signficantly more in time than within clusters suggesting a strong temporal dependence. However, a general mention of education (z=\,-\,1.73) and special status for juveniles in court (z=\,-\,0.45) are dominated by network effects determined by legal system and history. A full list of provisions and associated z-scores is provided in SI.

\section{Discussion}

In this work we have demonstrated that computational techniques can help to understand and quantify mechanisms of constitutional change. To understand the emergence of new legal provisions, we introduce the provision space as the set of provisions that are included among all constitutions at a given time. The provision space grows almost monotonically over two centuries as new constitutional legal rights and concepts are introduced, with three main temporal patterns of adoption emerging.\\\newline
We uncover strong hierarchical dependencies among provisions such that adoption of one provision is statistically more likely to be adopted following the adoption of other related and foundational rights. For example we find that limits on employment of children is preceded by the right to safe working conditions which is in turn preceded by the right to form trade unions. This allows us to articulate a roadmap to advocating for social rights; introduction of new concepts into provision space that are incremental advances on previously adopted provisions.\\\newline

We have used a computational linguistic measure to quantify the similarity between pairs of constitutions from which clusters of mutually similar constitutions emerge. While these findings largely align with accepted and theoretically derived legal families, we also show that this similarity is not only due to perfect replication of the provisions of the parent constitution. Rather there is a tendency for different wording of existing provisions and insertion of new provisions at independence that are not present in parent constitutions.
While both network based and temporal trends determine the nature of constitutions, the relative importance of these mechanisms varies between provisions. We identify provisions such as environmental protection that are strongly time dependent and conversely others such as reference to god that are strongly network dependent.\\\newline

We acknowledge that this study is constrained by the nature of constitutional law. Laws which are conferred by precedent or legislature are not included here. Compelling future work would consider the texts of other appropriate legal documents. We are further limited by the lack of availability of historical versions of constitutions. This is on account of copyright issues. Such an historical corpus containing each version of a constitution would provide time varying snapshot of the structure of constitutional similarities allowing for a rich dynamical study of contagion and influence. Since constitutional and other legal documents tend to use more formal and strict language, these texts will likely be amenable to more complex semantic level NLP approaches. This will allow us to measure similarity between the content and meaning of different countries` specific provisions, beyond noting only that a pair of countries both refer to the same topic.\\\newline

A natural extension to the question of which constitutional provisions are adopted and how, is to the question of their efficacy. Evidence exists~/cite{chilton} that rights that are supported with organisations i.e. the right to join a political party or the right to join a trade union, do lead to increased protection of social rights when compared to intrinsically individual rights e.g. freedom of movement or freedom of expression. Further evidence exists of a `rights creep` by which signature of the UN International Convention on Human Rights becomes de rigeur for UN member states, yet does not necessarily lead to statistically significant improvements in treatment of human rights~\cite{authoritarianism}. The validity of such a `top-down` approach to development is part of a larger discussion on the efficacy and causality of international development processes and metrics of evaluation~\cite{fail}.\\\newline

Nevertheless, this work sets forth a precedent that inter-disciplinary quantitative techniques can provide rich insight into constitution making in young states and global policy making procedures. The findings presented here can be particularly informative in light of discussions of self-determination through partition of extant states into new states.

\section{Data Availability}

All data is freely available at \url{https://www.constituteproject.org/} and \url{http://comparativeconstitutionsproject.org/}

\section{Acknowledgements}

The authors are grateful to Natalia Adler for helpful conversations and to Aaron Clauset for MATLAB code to derive the Minimal Violation Ranking.



\begin{thebibliography}{20}

  \bibitem{practicalguide}
  Hedling N (2011)
  A Practical Guide to Constitution Building: Principles and Cross-Cutting Themes
  {\it International Institute for Democracy and Electoral Assistance, Stockholm}
  Available at \url{http://www.constitutionnet.org/files/cb-handbook-chapter-2-low_0.pdf}

  \bibitem{elster}
      Elster J (1995)
      Forces and Mechanisms in the Constitution Making Process, {\it Duke Law Journal 45 \bf
      (2)} p 364-396

      \bibitem{process}
      Ginsburgh T, Elkins Z and Blount J(2009)
      Does the Process of Constitution-Making Matter?, {\it Annual Review of Social Science and Law \bf
      X}, 5:5.1-5.23.

      \bibitem{iceland}
      \url{http://rendezvous.blogs.nytimes.com/2012/10/24/crowdsourcing-icelands-constitution/?\_r=0}

      \bibitem{japan}
      \url{http://www.nytimes.com/2013/01/02/world/asia/beate-gordon-feminist-heroine-in-japan-dies-at-89.html?pagewanted=all\&\_r=1}


      \bibitem{go2002}
      Go J (2002)
      Modeling the state: post-colonial consitutions in Asia and Africa {\it Southeast Asian Studies \bf 39 (4)}

      \bibitem{billrights}
      Parkinson C (2008) Bill of Rights and Decolonization, Oxford University Press

      \bibitem{transplants}
      Spamman H (2009) Contemporary Legal Transplants: Legal Families and the Diffusion of (Corporate) Law, Brigham Young University Law Review (1813)

      \bibitem{rightschild}
      The United Nations
      Convention on the rights of the child (1989)
      {\it Treaty Series (3)}
      Available at \url{http://www.cirp.org/library/ethics/UN-convention/} (Retrieved 10th July 2017)

      \bibitem{complexsystems}
      Bar-Yam Y
      Dynamics of Complex Systems (1997) {\it in Studies in Non-linearity, Westview Press }

      \bibitem{arminjon}
      Arminjon (1950)
      Traité de droit comparé


      \bibitem{laporta}
      La Porta E, Lopez-de-Silanes F, Shleifer A, Vishny R (1998) Law and Finance Journal of Political Economy 106 p111355.

      \bibitem{zweigert}
      Zweigert K and Kotz H
      An Introduction to Comparative Law, translation from the German original: Einführung in die Rechtsvergleichung auf dem Gebiete des Privatrechts {\it Oxford press 1998}

      \bibitem{barabasi13}
      Barabasi L The network takeover. Nature Physics 8, 14-16 (2013) p36 14.

      \bibitem{css}
      Lazer, David and Pentland, Alex and Adamic, Lada and Aral, Sinan and Barabási, Albert-László and Brewer, Devon and Christakis, Nicholas and Contractor, Noshir and Fowler, James and Gutmann, Myron and Jebara, Tony and King, Gary and Macy, Michael and Roy, Deb and Van Alstyne, Marshall (2009)
      Computational Social Science, {\it Science \bf 323} p721-723

      \bibitem{constitutionalism}
      David Law and Meera Versteeg (2011)
      The Evolution and Ideology of Global Constitutionalism, {\it California Law Review \bf 99} p1163

      \bibitem{transnational}
      Goderis Benedikt and Verseteeg Mila (2014)
      The Diffusion of Constitutional Rights {\it International Review Law and Economics 39} 1-19

      \bibitem{melton}
      Ginsburg T (2013)
      On the Interpretability of Law: Lessons from the Decoding of National Constitutions, {\it INSTITUTE FOR LAW AND ECONOMICS WORKING PAPER\bf 624}

      \bibitem{preambles}
      Ginsburg T, Foti N, Rockmore D (2014) We the Peoples: The Global Origins of Constitutional Preambles, The George Washington International Law Review 46 p305.

      \bibitem{archetypes}
      Law D (2016) Constitutional Archetypes Texas Law Review 9

      \bibitem{evolution}
      Rockmore D, Fang C, Foti N, Ginsburg T, Krakauer D (2016) The Cultural Evolution of National Constitutions.

      \bibitem{productspace}
      C. Hidalgo, B. Klinger, A Barabasi, R. Hausmann (2007) The Product Space Conditions the Development of Nations.

      \bibitem{comparative}
      Elkins Z, Ginsburg T and Melton J(2014) Characteristics of National Constitutions Version 2, \url{comparativeconstitutionsproject.org}

      \bibitem{norm}
      Price R (1998) Reversing the Gun Sights: Transnational Civil Society Targets Land Mines, International Organization 52 (3) p 613-644.

      \bibitem{hiring}
      Clauset A and Arbesman S and Larremore D (2015)
      Systematic inequality and hierarchy in faculty hiring networks, {\it Science Advances \bf
      1:e1400005}

      \bibitem{chilton}
      Chilton A and Versteeg M (2015)
      Do constitutional rights make a difference?, {\it American Journal of Political Science \bf}

      \bibitem{authoritarianism}
      Law D and Versteeg M (2013)
      Constitutional variation among strains of authoritarianism {\it in Constitutions in Authoritarian Regimes eds. Ginsburg T and Simpser A}

      \bibitem{fail}
      Why Nations Fail, Robinson J and Acemoglu D (2012) Crown Publishing Group.



\end{thebibliography}
\end{document}


\maketitle
\tableofcontents
\section{Description of Datasets}

\begin{itemize}
  \item \textbf{Current Constitutions}
 	We make use of the English translations of 194 national constitutions~\cite{1} along with the year of writing and last amendment. In the case where a single constitutional document does not exist and are instead distributed among many statutes and laws, satisfaction of at least one of the following three requirements must be met
  \textit{(a)The document is identified explicitly as the Constitution, Fundamental Law, or Basic Law of a country.
(b) The document contains explicit provisions that establish it as the highest law, either through entrenchment or limits on future law.
(c) The document contains provisions which define the basic pattern of authority, either by establishing or suspending an executive, legislative or judicial branch of government, or by protecting the rights and freedoms of individuals.}

  \item \textbf{Current Constitutional Provisions}
Content in these constitutional documents pertaining to 330 specific topics have been labeled by domain experts in a redundant manner~\cite{1}. Thus we have both a binary label for each of 330 provisions for each of 194 countries as well as the text content for each provision.

  \item \textbf{Historical Constitutional Provisions}
  A similar coding procedure has been repeated for historical constitutional texts~\cite{ccp}. The provisions labeled in this case are not identical to those in the previous dataset and include a number of non-binary indicators e.g. \textit{What is the minimum age for becoming ahead of state?} In this analysis only binary valued indicators of the form \textit{Does the constitution contain a provisions for X?}. The dataset considers all historical versions of constitutions of independent nation states; therefore not all of the countries in the first dataset are present throughout. In fact the number of countries changes dramatically. The basic unit of analysis is the `constitution-year' describing the characteristics of each nation's constitution in each year. However, owing to copyright restrictions, the full text of these historical documents is not publicly available.
  The states included conform to a reconciled list of independent states that satisfy consistent criteria~\cite{gleditsch}. On account of apparent missing values in the more recent years of the data, we restrict our analysis to years before 2010.

\end{itemize}

\section{Summary Statistics}
\subsection{Document Length}
The UK has the longest constitution (over 145,000 words) with the shortest belonging to Libya with just over 2,000 words.\\

\begin{figure}[ht]
\label{wordcount}
\centerline{\includegraphics[width=0.5\linewidth]{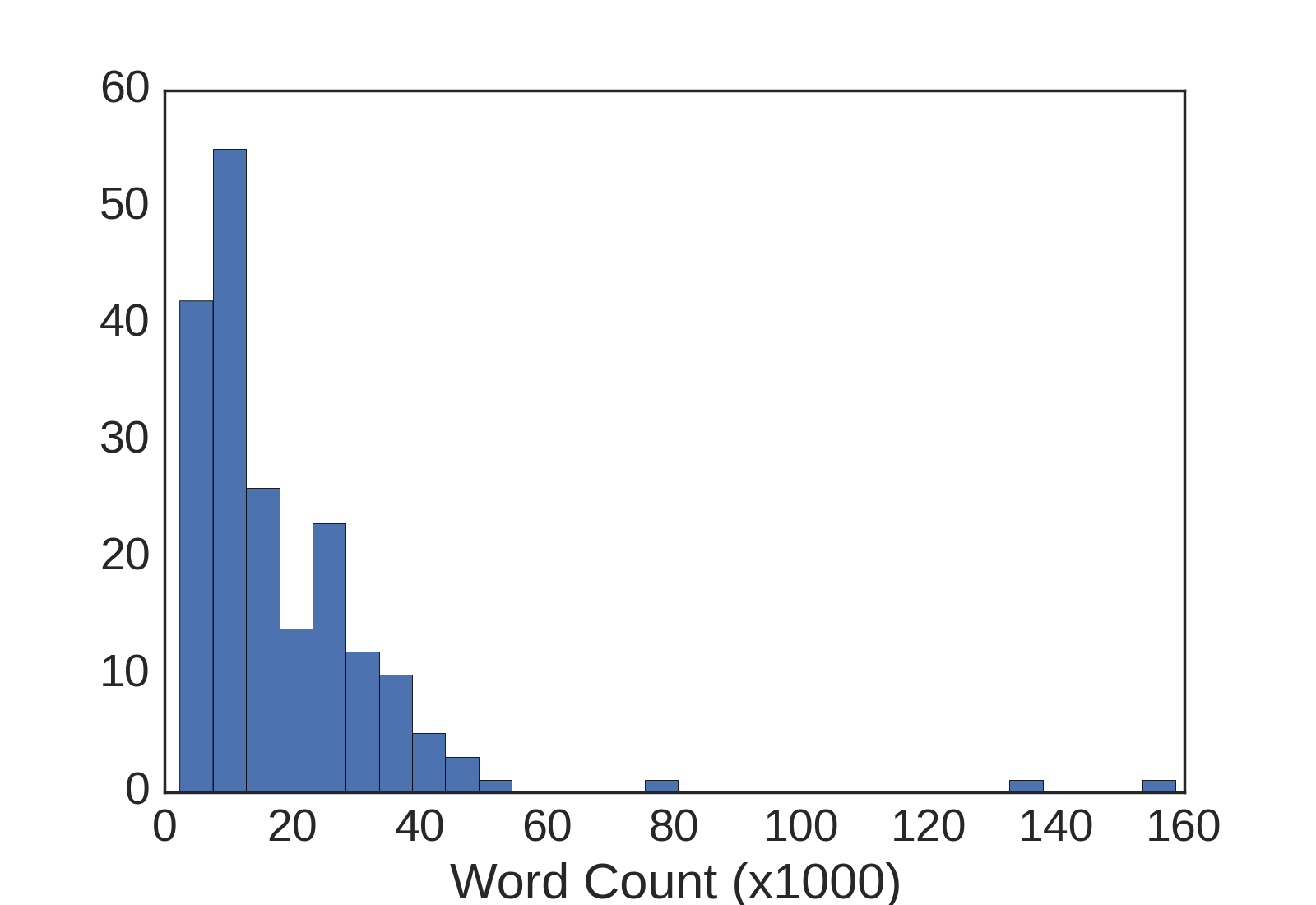}}
\caption{Word Count Histogram}\label{word_count}
\end{figure}

\subsection{Time of Writing Constitutions}
\begin{figure}
\label{yearofwriting}
\centerline{\includegraphics[width=0.5\linewidth]{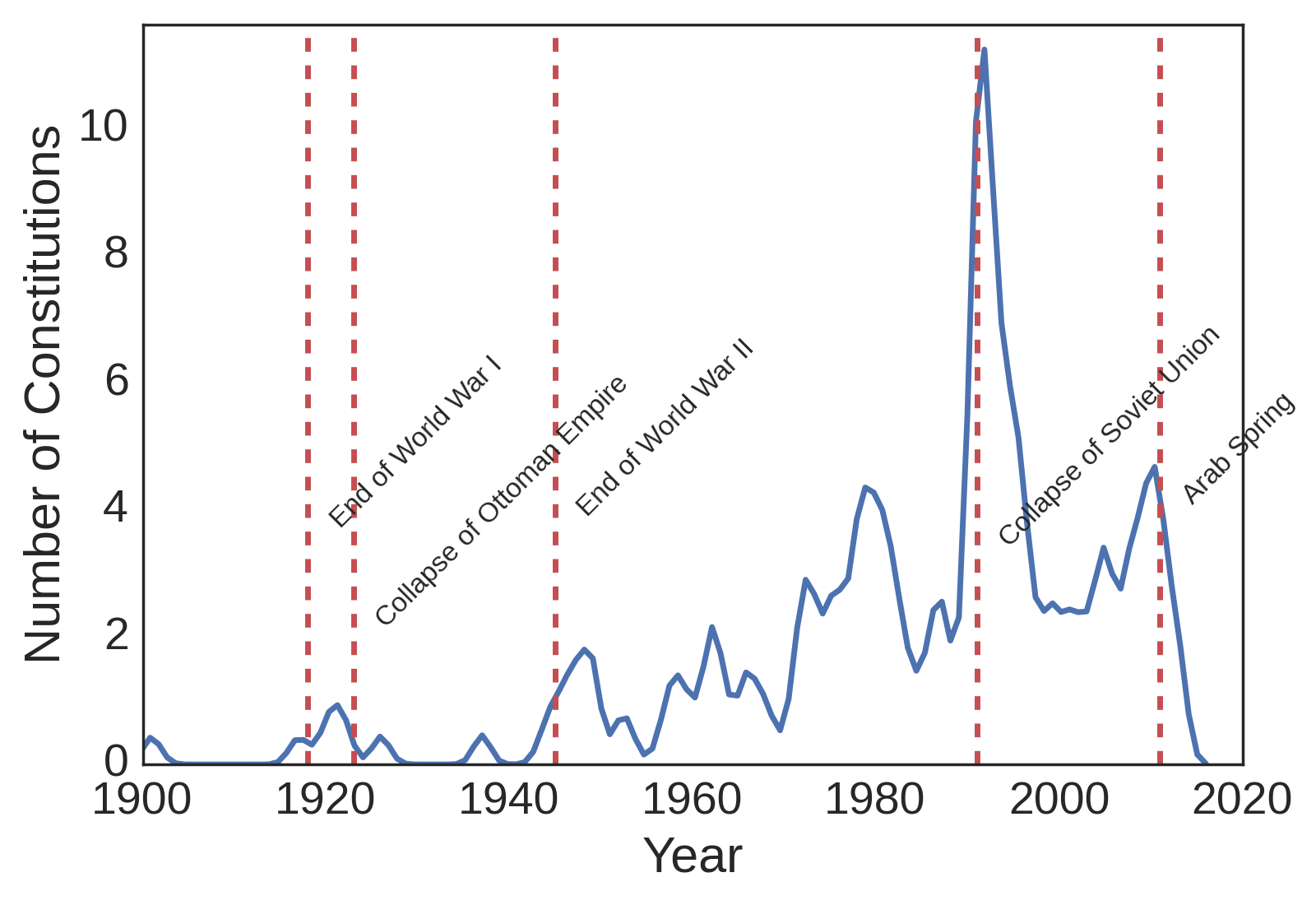}}
\caption{Kernel Density Plot of the Year of Writing of Constitutions Currently in Effect (truncated at 1900 to highlight 20th century political events)}\label{written_dist}
\end{figure}

The oldest constitution in dataset 1 is the United Kingdom’s Magna Carta (1215) followed by the United States (1789).
The year of writing of current constitutions are seen to follow geo-political events (fig \ref{written_dist})\\

\clearpage

\subsection{Temporal Activity}

We consider constitution writing activity, both new constitutions and amendments, over time from dataset III (fig \ref{amendments})

\begin{figure}[ht]
\label{amendments}
\centerline{\includegraphics[width=0.5\linewidth]{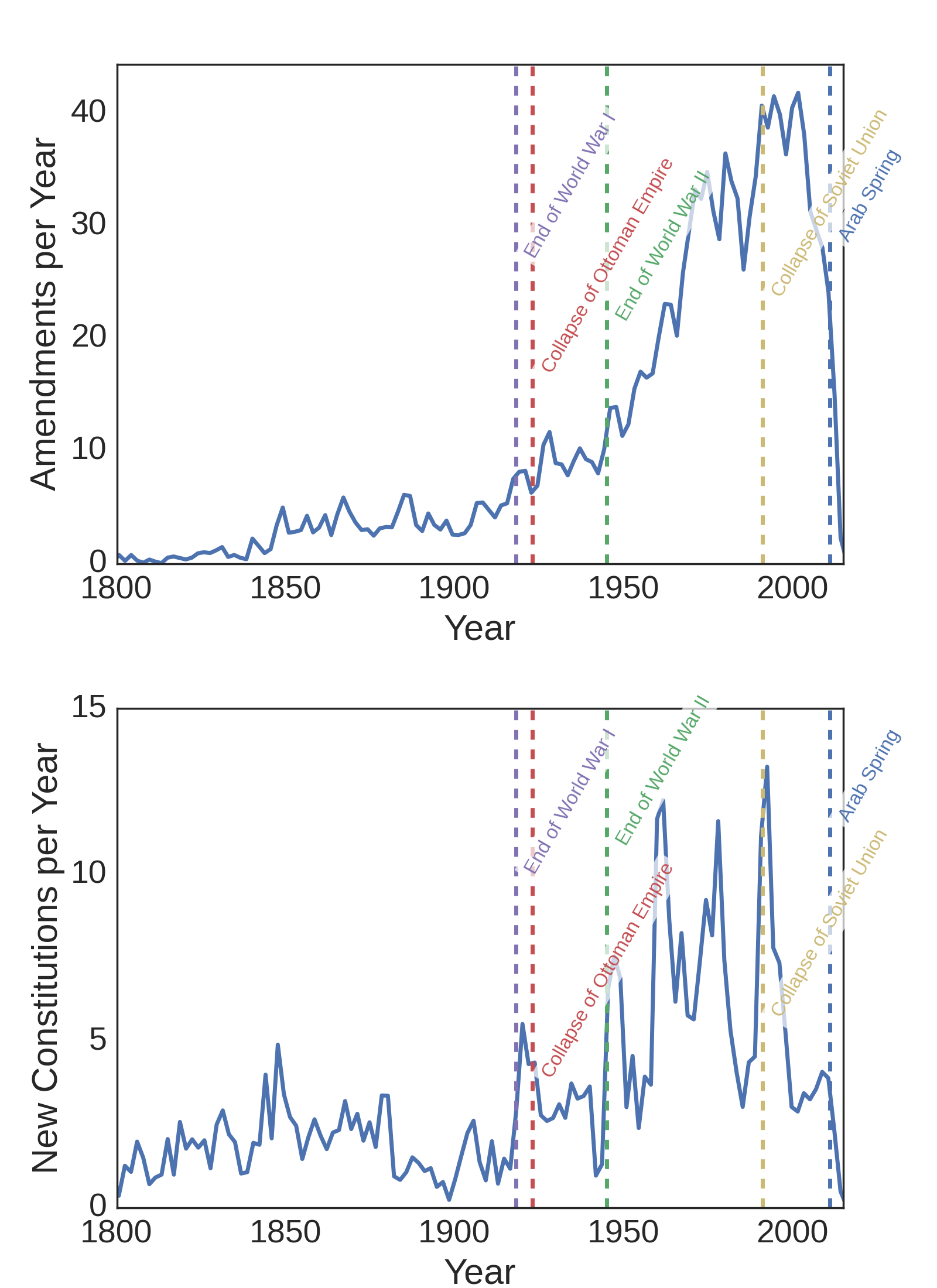}}
\caption{Historical Constitution Amendment Events}
\end{figure}

\begin{table}[]
\label{cluster_corr}
\centering
\caption{Table of pairwise correlations between cluster amendment time series}
\begin{tabular}{|l|l|l|l|}
\hline
Cluster 1    & Cluster 2    & r     & p                  \\\hline
ex-Soviet    & Iberian      & 0.683 & \textless$10^{-5}$ \\
ex-Soviet    & Commonwealth & 0.456 & 0.001              \\
ex-Soviet    & French       & 0.477 & 0.001              \\
Iberian      & Commonwealth & 0.683 & \textless$10^{-5}$ \\
Iberian      & French       & 0.578 & \textless$10^{-5}$ \\
Commonwealth & French       & 0.606 &  \textless$10^{-5}$\\
\hline
\end{tabular}
\label{cluster_corr}
\end{table}

We consider the time series of constitutional amendment activity with 2 year resolution presented in the main paper. The pairwise Pearson correlation coefficient and corresponding p value between the time series describing each cluster is listed in table \ref{cluster_corr}\\

Here we plot the number of constitutions present in the longitudinal dataset, we see distinct increases during periods of decolonisation or independence.

\begin{figure}[h]
\centerline{\includegraphics[width=0.6\linewidth]{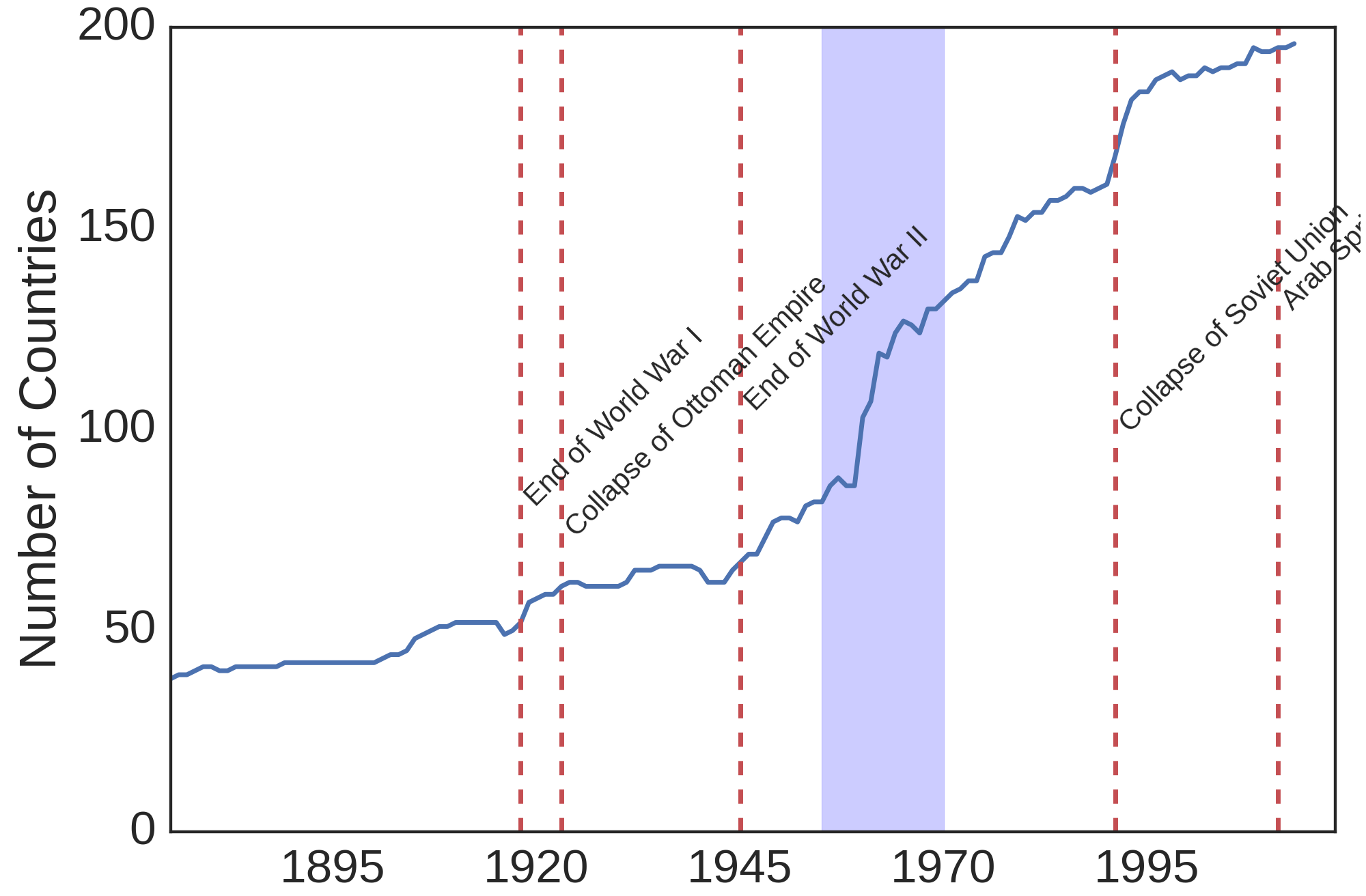}}
\label{countries_time}
\caption{Yearly time series of the number of constitutions in our dataset. The blue shaded area corresponds to the period of 1960-1970 during which many countries achieved independece also known as `decolonisation`}
\end{figure}

\clearpage

\subsection{Comparison of Clusters}

\begin{figure}[ht]
\centerline{\includegraphics[width=0.5\linewidth]{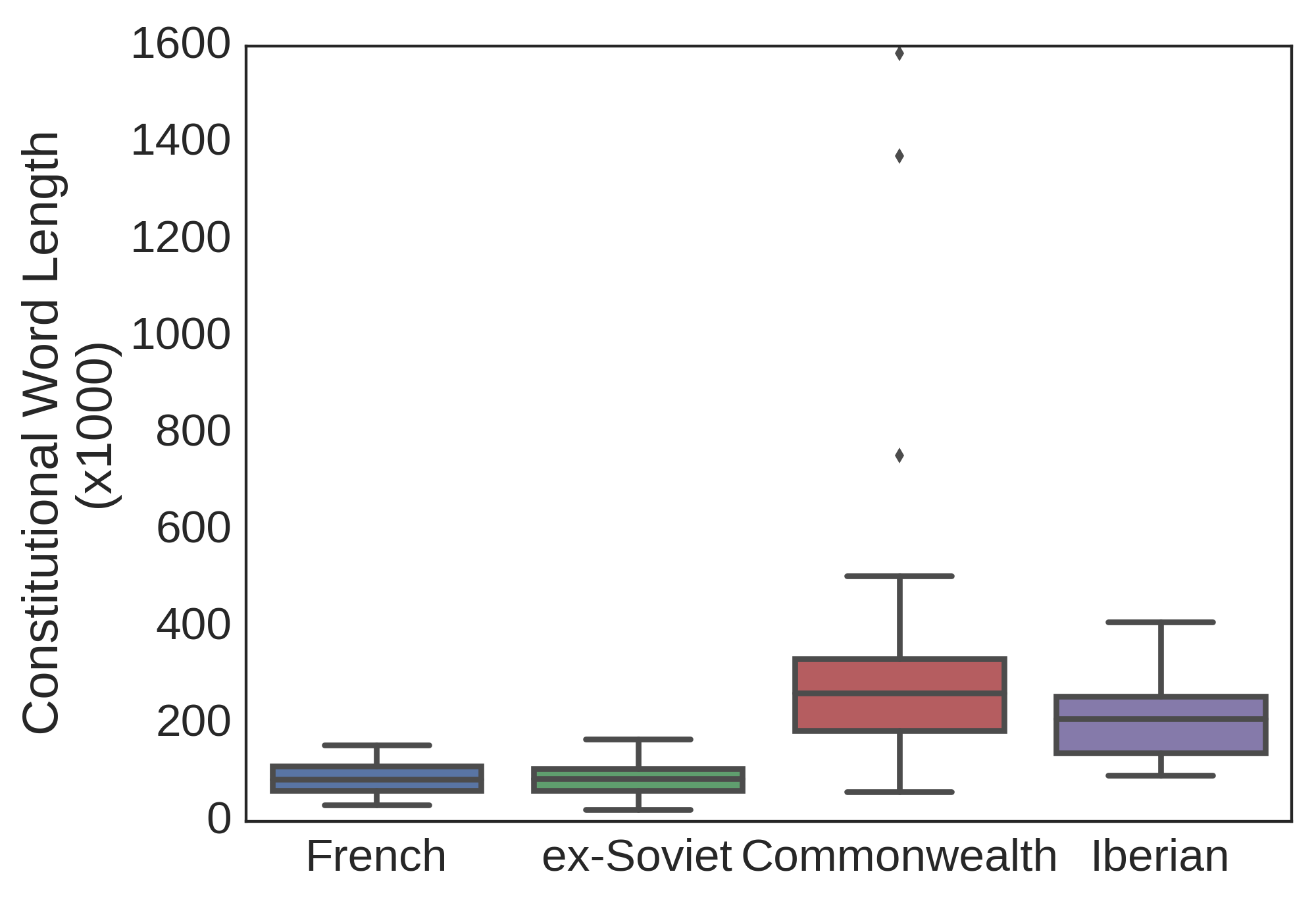}}
\caption{Word Count by Cluster}\label{word_count_cluster}
\end{figure}

We now compare summary statistics between the country clusters derived from the network of constitutional document similarity.
Fig (\ref{word_count_cluster}) demonstrates that the Commonwealth cluster contains vastly longer constitutional documents than the other clusters.

\begin{figure}[h]
\centerline{\includegraphics[width=0.5\linewidth]{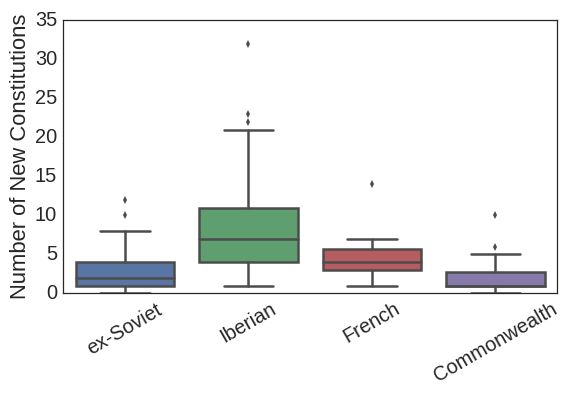}}
\caption{Distribution of number of new constitutions by cluster}\label{amendments_box}
\end{figure}

Next we ask how the number of amendments differ between clusters (figure \ref{amendments_box}).


\begin{figure}[ht]
\centerline{\includegraphics[width=0.5\linewidth]{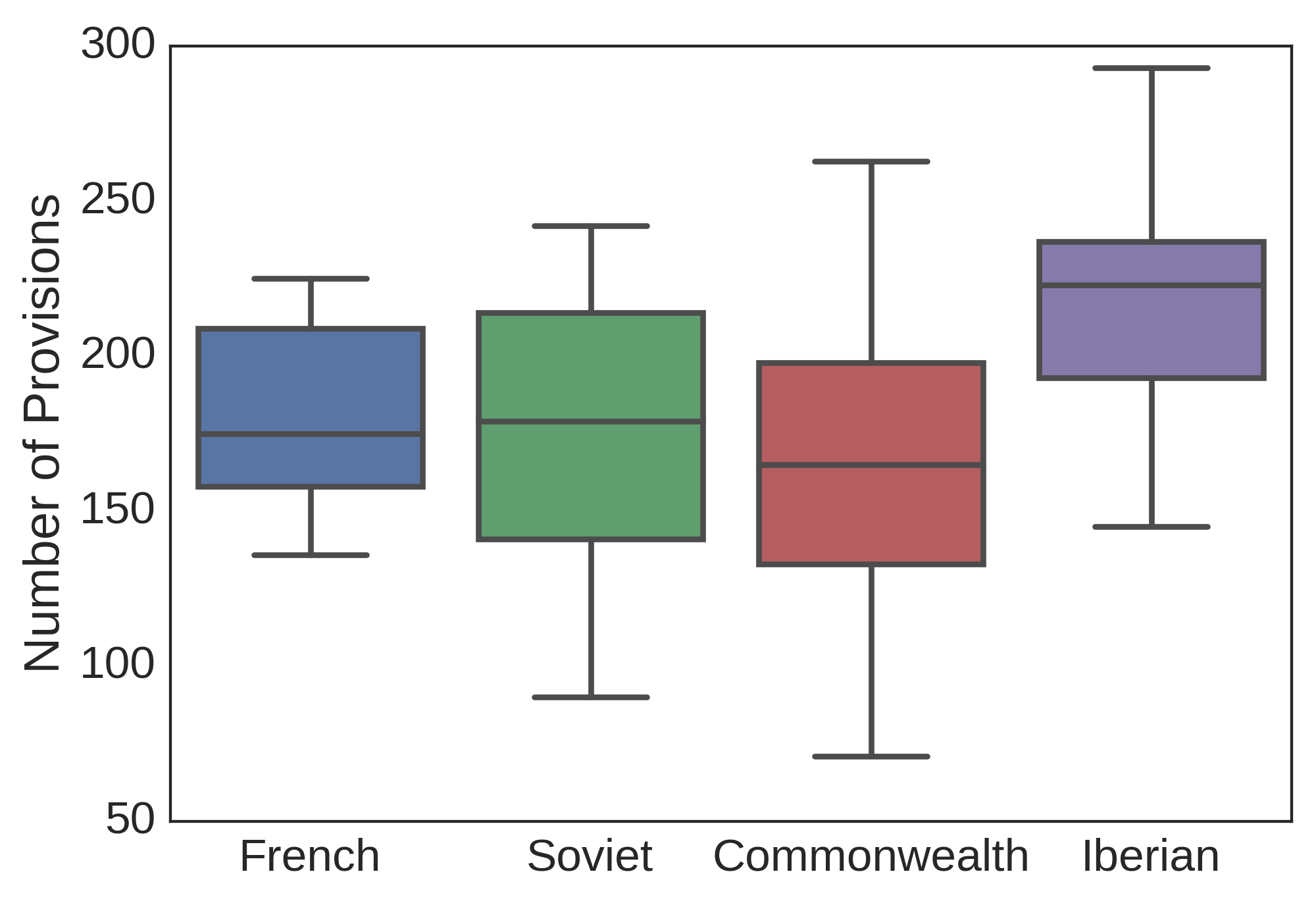}}
\caption{Number of Provisions by Cluster}\label{provisions_boxplot}
\end{figure}

The Iberian cluster not only dominates in the number of new constitutions, but also the number of provisions contained within those constitutions (see Fig. (\ref{provisions_boxplot}))\\

\begin{figure}
    \centering
    \begin{subfigure}[b]{0.6\textwidth}
        \includegraphics[width=\textwidth]{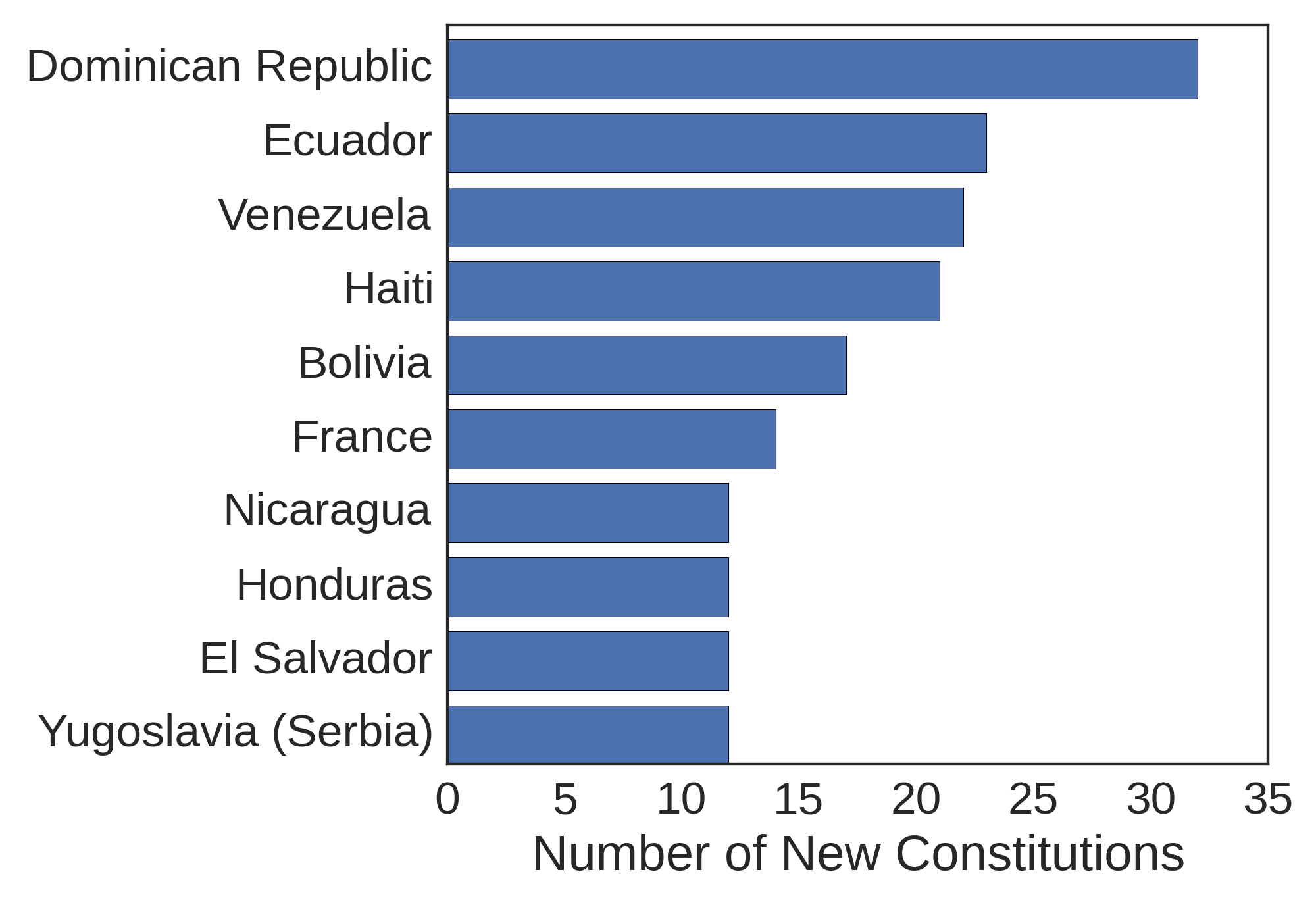}
    \caption{Most New Constitutions}\label{most_new}
    \end{subfigure}

    ~ 

      \begin{subfigure}[b]{0.6\textwidth}
          \includegraphics[width=\textwidth]{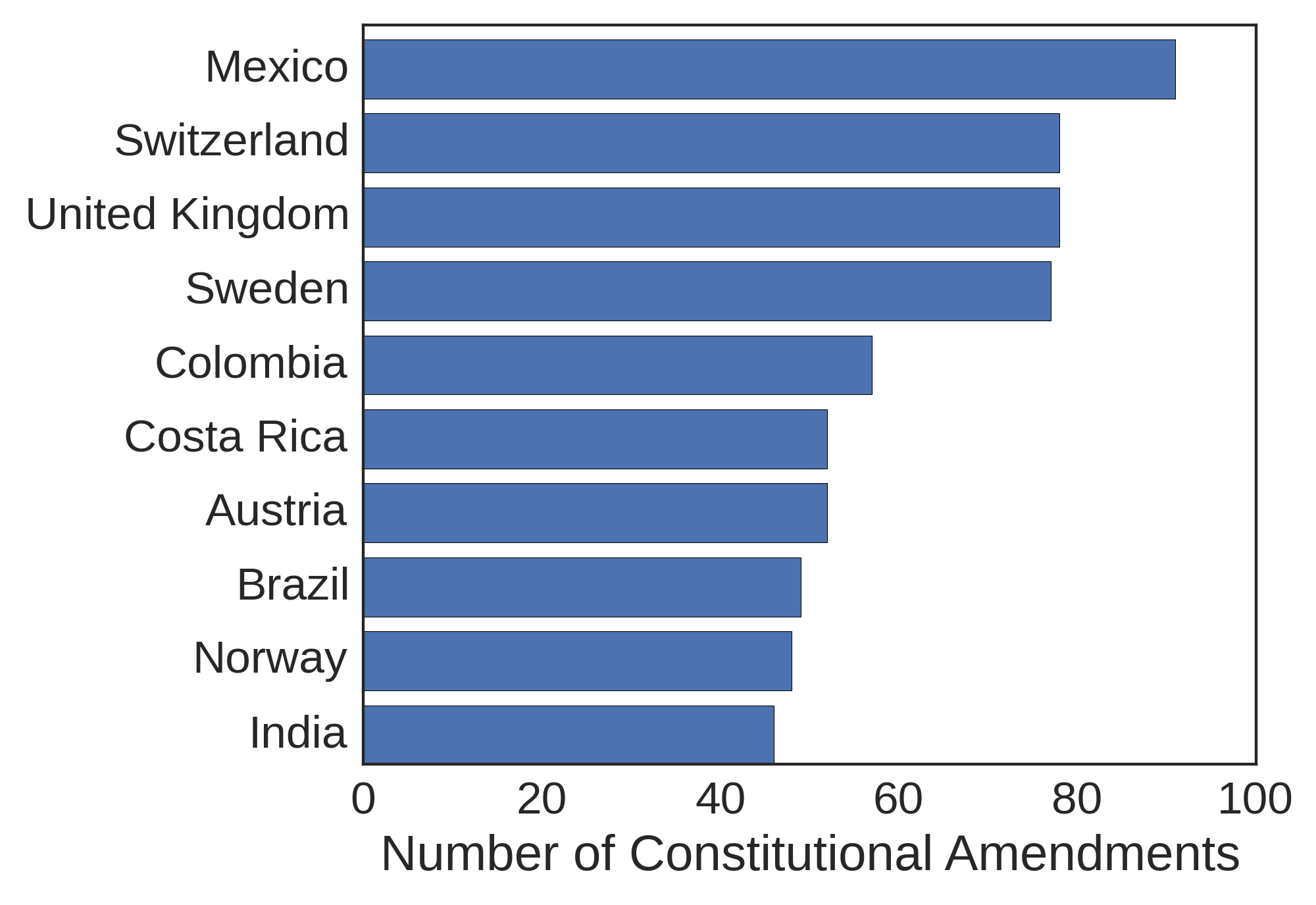}
      \caption{Most Constitutional Amendments}\label{most_amendments}
    \end{subfigure}
    ~ 
    \label{fig:animals}
\end{figure}

We consider the individual countries that have had the greatest number of constitutions and amendments (\ref{most_new} and \ref{most_amendments}).

In order to further examine the differences between clusters, we consider the chronology of new constitutions among members of each cluster. We compare the yearly distribution of new constitution events within these clusters. We observe distinct historical patterns within each. The Iberian cluster also shows an interesting pattern of activity prompted by relatively early colonial independence. However, these early declarations of independence were followed by indigenous led independence movements in the 20th century~\cite{openveins} as more constitutional events were recorded.\\

\begin{figure}[h]
\centerline{\includegraphics[width=1.0\linewidth]{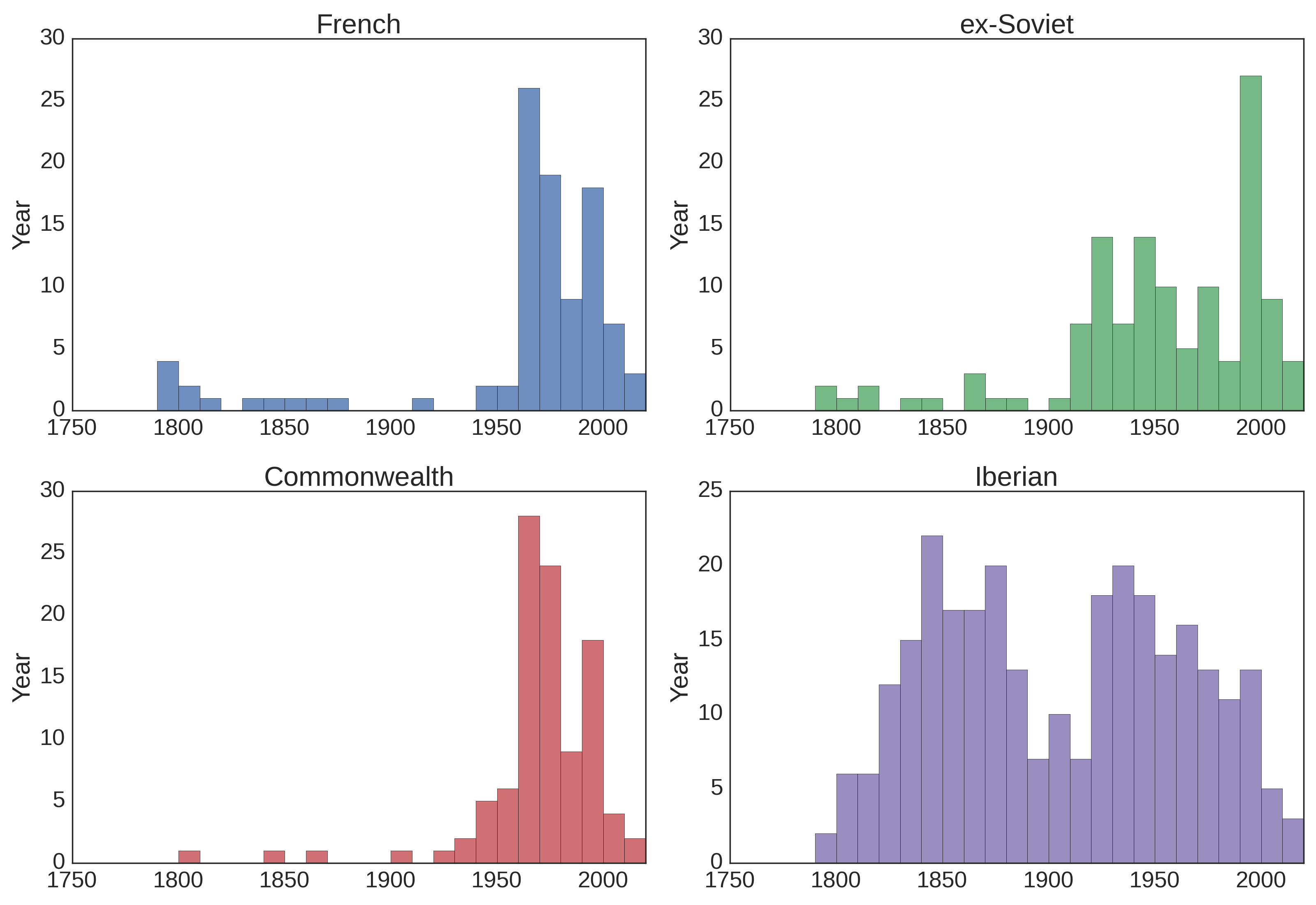}}
\caption{Time distribution of new constitution writing events}\label{amendments}
\end{figure}
Out of the set of labeled provisions, we consider the most commonly and least commonly adopted

\begin{tabular}{|c|c|}
      \hline
      \addheight{\includegraphics[width=0.35\linewidth]{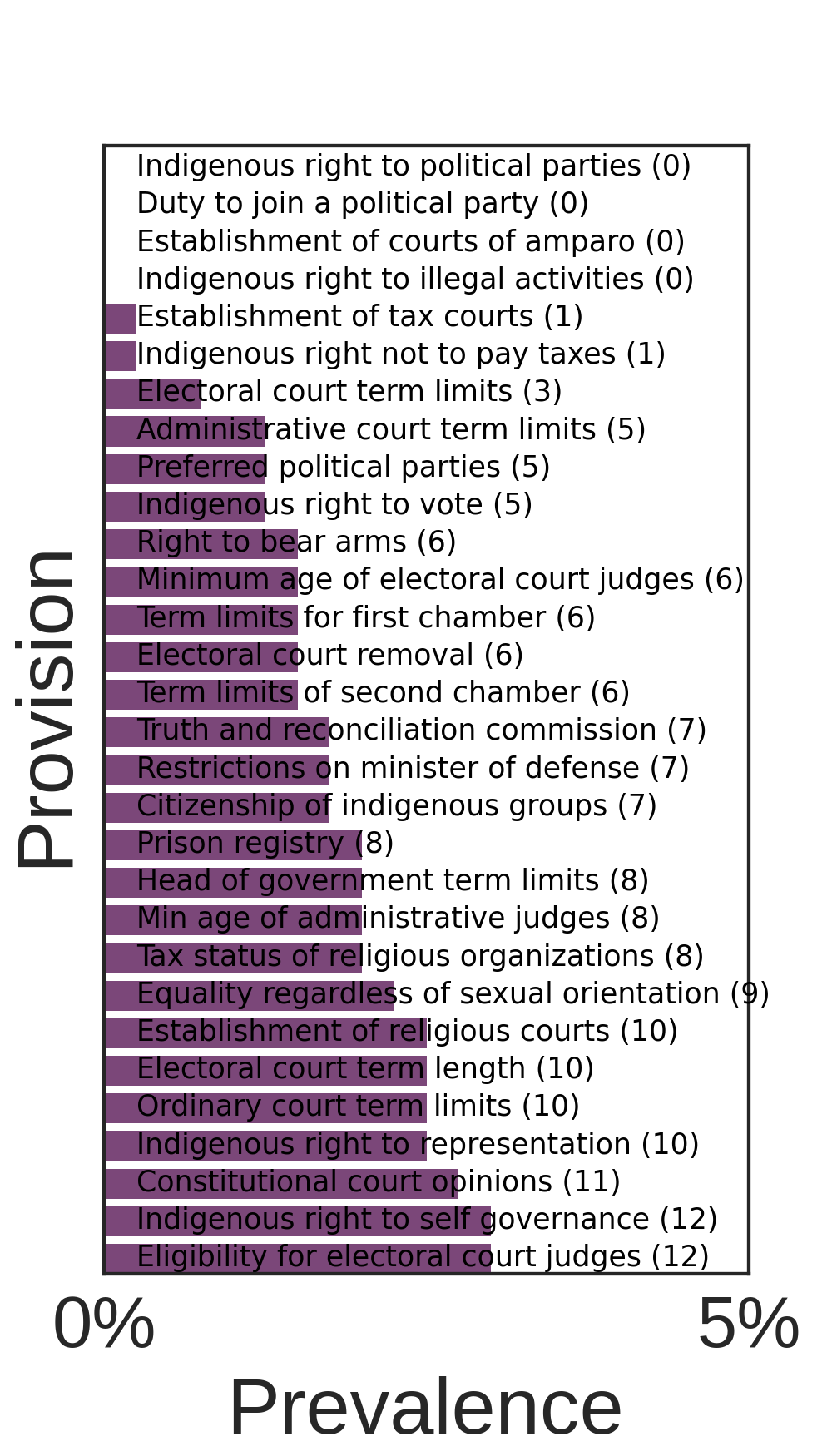}} &
      \addheight{\includegraphics[width=0.35\linewidth]{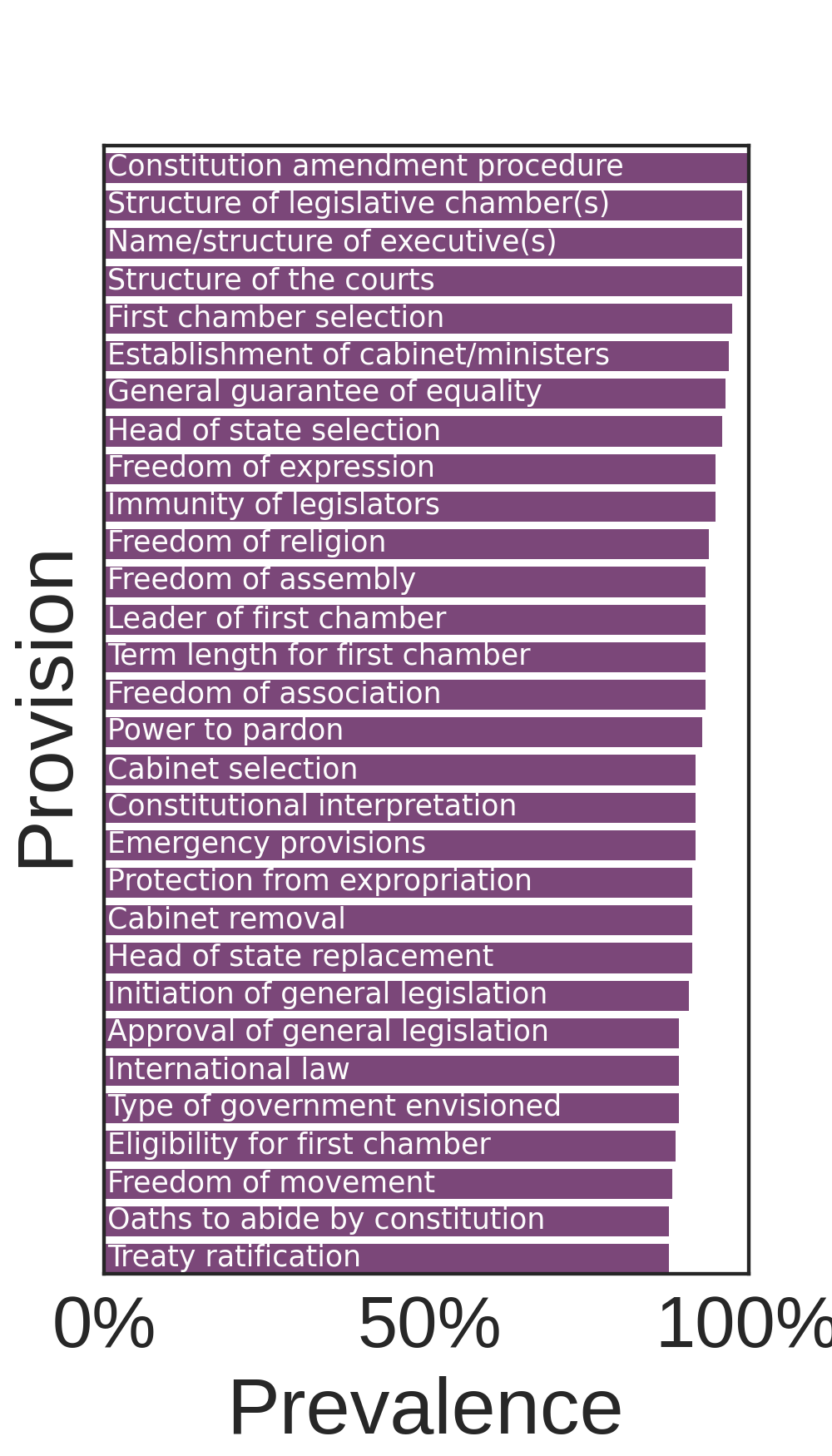}} \\
      \small Least common provisions &  Most common provisions \\
      \hline
\end{tabular}


\clearpage

\section{Constitutional Similarity}

In the following section we consider the Jaccard similarity after removal of stopwords.  The pairwise Jaccard index is calculated between each constitutional document after tokenisation and removal of stopwords give in the Python NLTK library~\cite{nltk} augmented by the following domain specific terms ('shall','article','may','must','page') as well as country names to avoid, for example, trivial clustering based on a long list of other Commonwealth countries. We construct a semantic network where constitutions represent nodes and edges are directed links determined by Jaccard similarity. Clusters of similar documents, representing national constitutions with similar wordings are formed by taking network communities using the spinglass algorithm~\cite{spinglass}.

\subsection{Provisional Text Similarity}

We consider the similarity of each individual provision across all constitutions including that provision. The distribution of similarities is shown below

\begin{figure}[ht]
\centerline{\includegraphics[width=0.5\linewidth]{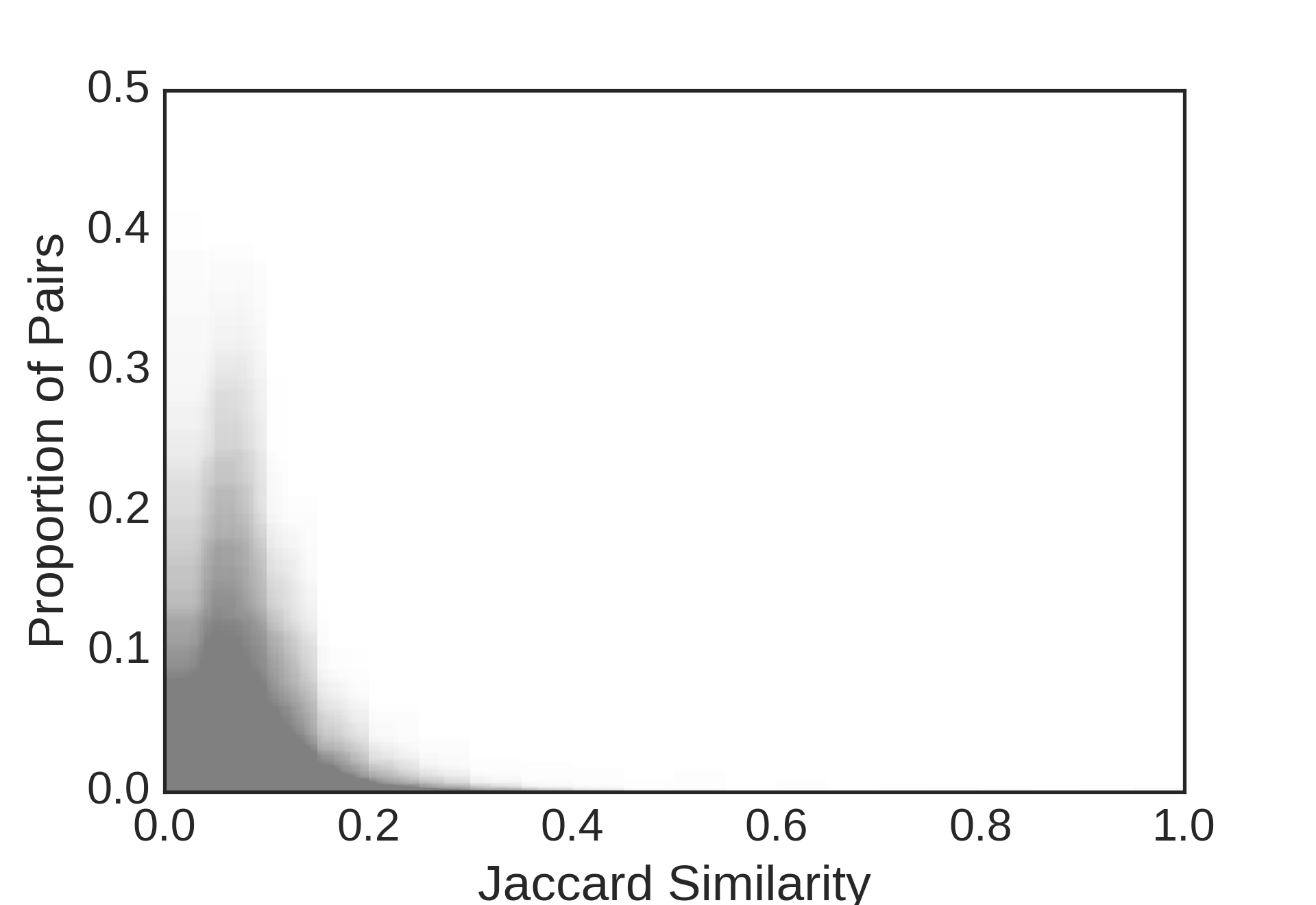}}
\caption{Distribution of Jaccard similarities for all provisions}\label{prov_all}
\end{figure}

The highest similarity is observed with `Examination of witnesses` with a mean similarity of 0.265. The distribution of pairwise similarities for this provision is shown below

\begin{figure}[ht]
\centerline{\includegraphics[width=0.5\linewidth]{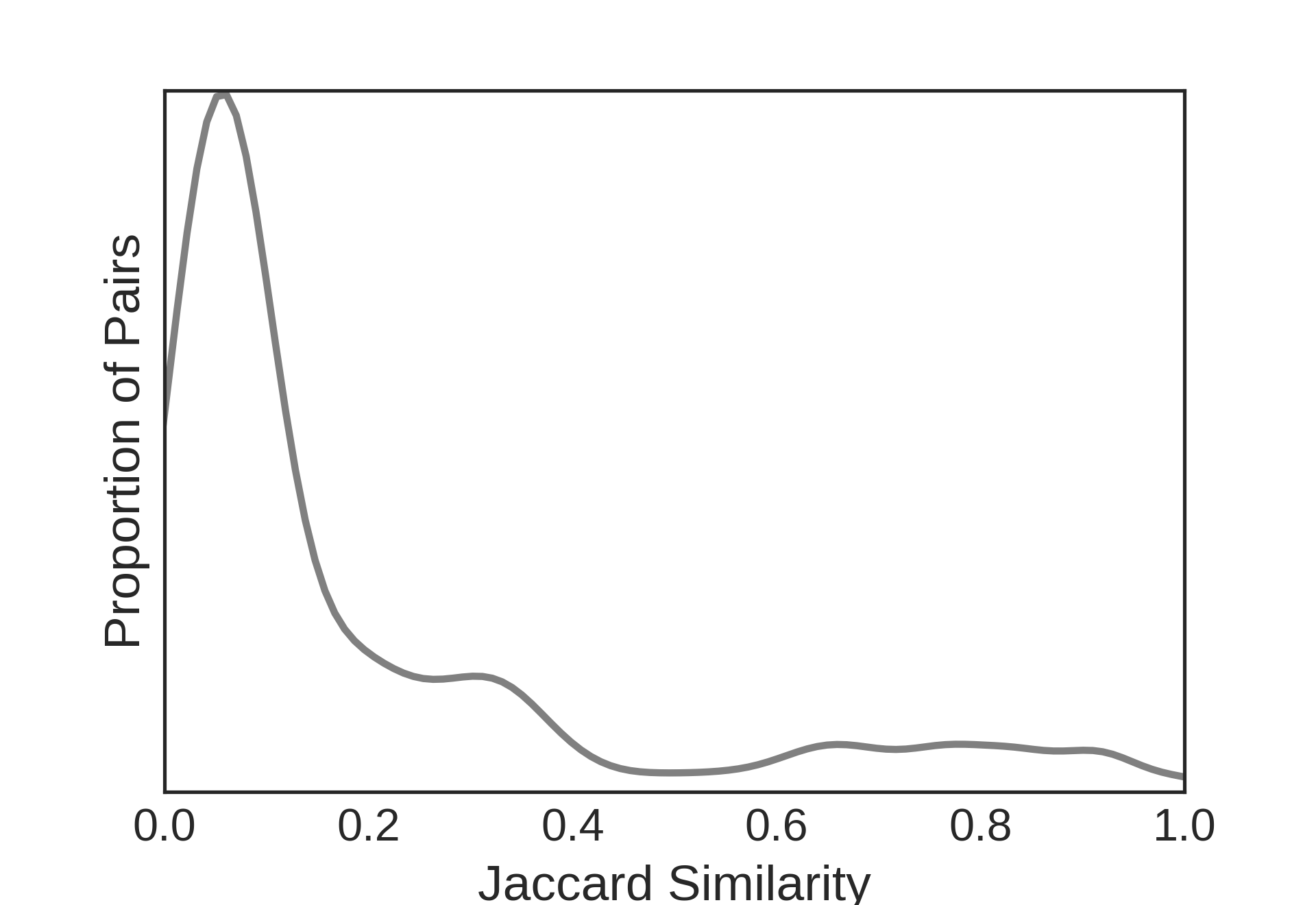}}
\caption{Distribution of Jaccard similarities among all pairs for the provision `Examination of Witnesses`}\label{prov_sim_dist}
\end{figure}

\subsection{Full Constitutional Text Similarity}

We may explore the intrinsic structure of the full text correlation matrix through a hierarchically clustered heatmap. This reveals a clear structure with clusters emerging along several historical and geographical themes moving down the diagonal from top-left to bottom-right;
Commonwealth countries, Francophone countries and former Iberian colonies.\\

\begin{figure}[h]
\centering
\includegraphics[width=0.65\textwidth]{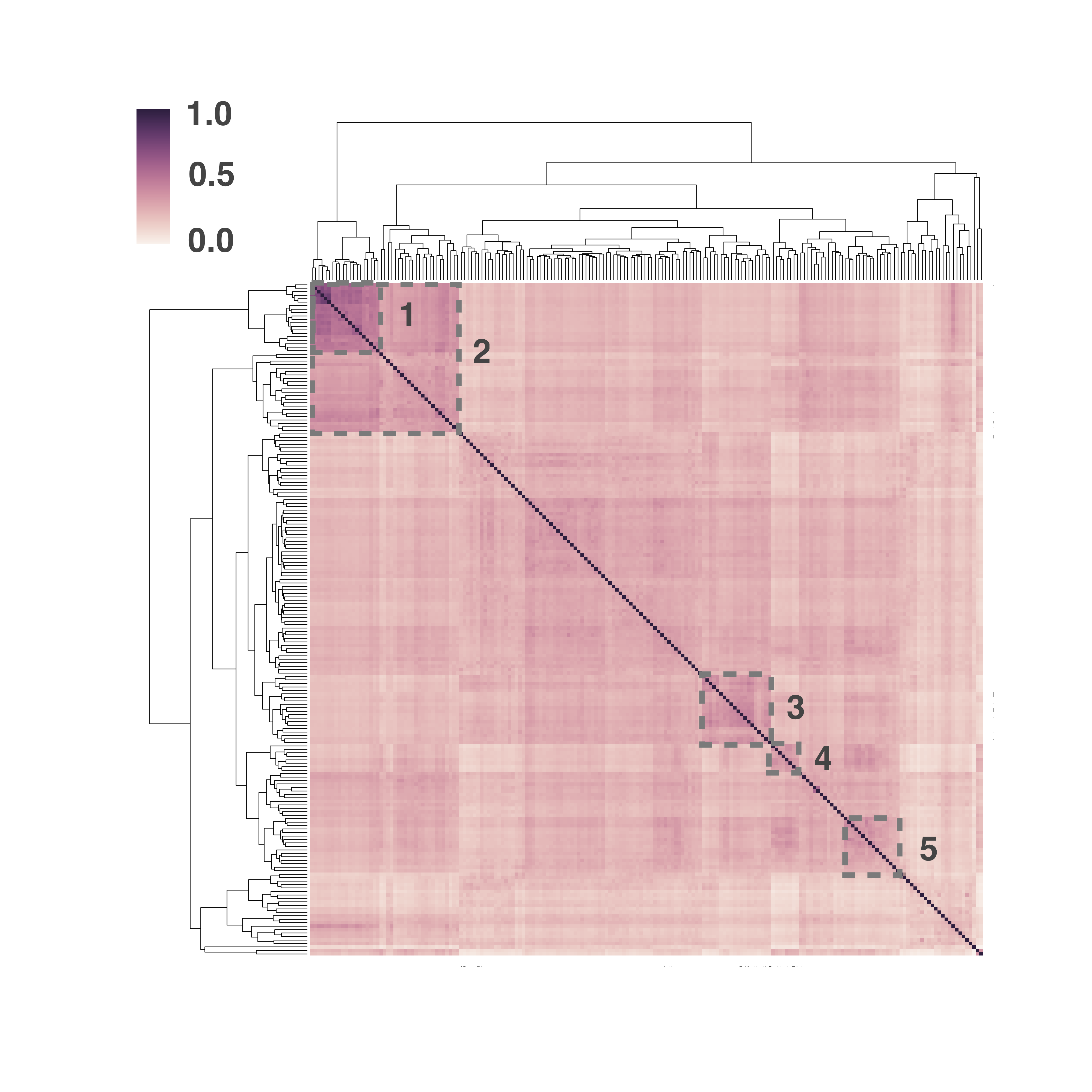}
\caption{\label{fig:frog} Dendrogram of constitutional similarities.}
\end{figure}

The labeled clusters along the diagonal include the following\\

\textbf{Cluster 1:}\\
Sierra Leone
Guyana
Zambia
Trinidad \& Tobago
Malta
Jamaica
Barbados
Bahamas
Belize
Kiribati
Botswana
Solomon Islands
Mauritius
Lesotho
St Lucia
Dominica
St Vincent \& Grenadines
Grenada
St Kitts
Antigua \& Barbuda
\newline
\textbf{Cluster 2:}\\
Singapore
Seychelles
Malaysia
Tuvalu
Swaziland
Gambia
Ghana
Fiji
Bangladesh
Tanzania
Malawi
Namibia
Cyprus
Sri Lanka
Pakistan
Papua New Guinea
Zimbabwe
 Kenya
 Uganda
Samoa
Marshall Islands
Nigeria
\newline
\textbf{Cluster 3:}\\
Djibouti
Mauritania
Gabon
Benin
Algeria
Sao Tome and Principe
Central African Republic
Togo
Burkina Faso
Congo
Chad
Niger Madagascar
Guinea
DRC
Burundi
Morocco
Senegal
Mali
Cote d'Ivoire
\newline
\textbf{Cluster 4:}\\
Ecuador
Bolivia
Guatemala
Paraguay
Colombia
Venezuela
Brazil
Mexico
\newline
\textbf{Cluster 5} \\
Cape Verde
Angola
Mozambique
Dominican Republic
Honduras
El Salvador
Portgual
Panama
Uruguay
Chile
Costa Rica
Haiti
Egyt
Switzerland
Cuba
Iran
\newline

When creating a network from an $N\times N$ matrix of similarities, inverse distances or other weights, there are many ways to define an edge. The most obvious is to consider \textit{weighted} edges with a continuous value given by the degree of similarity defining the strength between i and j. Commonly, a threshold is applied to this set of edges, such that links with a weight below a cut-off value are considered sufficiently insignificant to be excluded and those above are retained. The threshold quantifies a trade-off between a resultant network that is too dense for distinct clusters to emerge and one which is so sparse that the network becomes disconnected.\\

When considering the set of Jacard similarities between constitutional documents as described in the main paper, a difficulty arises since the set of values of pairwise similarities is extremely heterogeneous. This behaviour may be due to a bias based on the highly variable length of the constitutional documents under consideration as noted previously. In addition, applying a minimum threshold will quickly lead to disconnected clusters in the network since the row wise minimum value occurs at the 26th percentile of the entire distribution. In other words, applying a threshold at the largest possible value to retain a connected graph, necessarily retains 74\% of all edges. This density in turn precludes fine grained partition of the network into consistent clusters using most community detection algorithms.\\

\begin{figure}[h]
\centering
\includegraphics[width=0.65\textwidth]{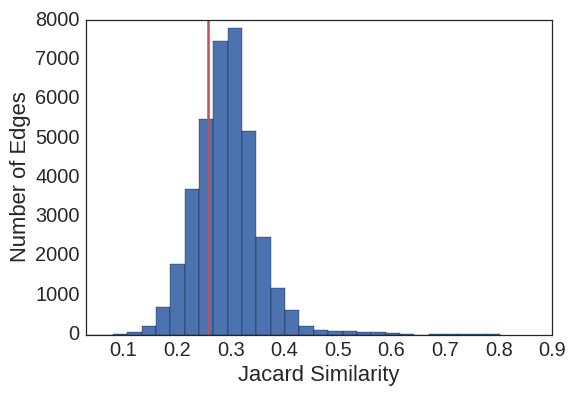}
\caption{\label{fig:frog} Histogram of all edge weights in $N \times N$.}
\end{figure}

Applying the spin-glass algorithm with a threshold of 0.2 yields two clusters\\

\textbf{[0]} Angola, Antigua and Barbuda, Argentina, Austria, Bahamas, Bangladesh,
    Barbados, Belgium, Belize, Bolivia, Botswana, Brazil, Burundi, Canada,
    Cape Verde, Chile, China, Colombia, Democratic Republic of the Congo,
    Congo, Costa Rica, Cuba, Cyprus, Dominica, Dominican Republic, Ecuador,
    Egypt, El Salvador, Fiji, Gambia, German Federal Republic, Ghana, Greece,Grenada, Guatemala, Guinea, Guyana, Haiti, Honduras, India, Iran, Jamaica,Kenya, Kiribati, Lesotho, Malawi, Malaysia, Malta, Marshall Islands,
    Mauritius, Mexico, Morocco, Mozambique, Myanmar, Namibia, Nepal, New
    Zealand, Nicaragua, Niger, Nigeria, Pakistan, Panama, Papua New Guinea,
    Paraguay, Peru, Philippines, Portugal, Romania, St Kitts and Nevis, St
    Lucia, St Vincent and the Grenadines, Samoa, Serbia, Seychelles, Sierra
    Leone, Singapore, Solomon Islands, Somalia, South Africa, South Sudan,
    Spain, Sri Lanka, Sudan, Swaziland, Sweden, Switzerland, East Timor,
    Trinidad and Tobago, Turkey, Tuvalu, Uganda, United Kingdom, Tanzania,
    Uruguay, Venezuela, Socialist Republic of Vietnam, Zambia, Zimbabwe\\

\textbf{[1]} Afghanistan, Albania, Algeria, Andorra, Armenia, Australia, Azerbaijan,
    Bahrain, Belarus, Benin, Bhutan, Bosnia Herzegovina, Brunei, Bulgaria,
    Burkina Faso, Cambodia, Cameroon, Central African Republic, Chad, Comoros,
    Cote D`Ivoire, Croatia, Czech Republic, Denmark, Djibouti, Equatorial
    Guinea, Eritrea, Estonia, Ethiopia, Finland, France, Gabon, Georgia,
    Guinea Bissau, Hungary, Iceland, Indonesia, Iraq, Ireland, Israel, Italy,
    Japan, Jordan, Kazakhstan, Peoples Republic of Korea, Republic of Korea,
    Kosovo, Kuwait, Kyrgyz Republic, Laos, Latvia, Lebanon, Liberia, Libya,
    Liechtenstein, Lithuania, Luxembourg, Macedonia, Madagascar, Maldives,
    Mali, Mauritania, Micronesia, Moldova, Monaco, Mongolia, Montenegro,
    Nauru, Netherlands, Norway, Oman, Palau, Poland, Qatar, Russia, Rwanda,
    Sao Tome and Principe, Saudi Arabia, Senegal, Slovakia, Slovenia, Surinam,
    Syria, Taiwan, Tajikistan, Thailand, Togo, Tonga, Tunisia, Turkmenistan,
    Ukraine, United Arab Emirates, United States of America, Uzbekistan,
    Vanuatu, Yemen\\

These broadly describe the former British, Spanish and Portuguese Empires on one hand and French, Soviet, Middle Eastern, Scandinavian and Asian spheres of influence on the other.\\

Using the same threshold and the leading Eigenvector method yields a similar decomposition\\

\textbf{[0]} Afghanistan, Albania, Algeria, Andorra, Armenia, Australia, Azerbaijan,
    Bahrain, Belarus, Benin, Bhutan, Bosnia Herzegovina, Brunei, Bulgaria,
    Burkina Faso, Cambodia, Cameroon, Central African Republic, Chad, Comoros,
    Cote DIvoire, Croatia, Czech Republic, Denmark, Djibouti, Equatorial
    Guinea, Eritrea, Estonia, Ethiopia, Finland, France, Gabon, Georgia,
    Guinea Bissau, Hungary, Iceland, Indonesia, Iraq, Ireland, Israel, Italy,
    Japan, Jordan, Kazakhstan, Republic of Korea, Kosovo, Kuwait, Kyrgyz
    Republic, Laos, Latvia, Lebanon, Liberia, Libya, Liechtenstein, Lithuania,
    Luxembourg, Macedonia, Madagascar, Maldives, Mali, Mauritania, Micronesia,
    Moldova, Monaco, Mongolia, Montenegro, Nauru, Netherlands, Norway, Oman,
    Palau, Poland, Qatar, Romania, Russia, Rwanda, Sao Tome and Principe,
    Saudi Arabia, Senegal, Slovakia, Slovenia, Surinam, Syria, Taiwan,
    Tajikistan, Thailand, Togo, Tonga, Tunisia, Turkmenistan, Ukraine, United
    Arab Emirates, United States of America, Uzbekistan, Vanuatu, Yemen\\

\textbf{[1]} Angola, Antigua and Barbuda, Argentina, Austria, Bahamas, Bangladesh,
    Barbados, Belgium, Belize, Bolivia, Botswana, Brazil, Burundi, Canada,
    Cape Verde, Chile, China, Colombia, Democratic Republic of the Congo,
    Congo, Costa Rica, Cuba, Cyprus, Dominica, Dominican Republic, Ecuador,
    Egypt, El Salvador, Fiji, Gambia, German Federal Republic, Ghana, Greece,
    Grenada, Guatemala, Guinea, Guyana, Haiti, Honduras, India, Iran, Jamaica,
    Kenya, Kiribati, Peoples Republic of Korea, Lesotho, Malawi, Malaysia,
    Malta, Marshall Islands, Mauritius, Mexico, Morocco, Mozambique, Myanmar,
    Namibia, Nepal, New Zealand, Nicaragua, Niger, Nigeria, Pakistan, Panama,
    Papua New Guinea, Paraguay, Peru, Philippines, Portugal, St Kitts and
    Nevis, St Lucia, St Vincent and the Grenadines, Samoa, Serbia, Seychelles,
    Sierra Leone, Singapore, Solomon Islands, Somalia, South Africa, South
    Sudan, Spain, Sri Lanka, Sudan, Swaziland, Sweden, Switzerland, East
    Timor, Trinidad and Tobago, Turkey, Tuvalu, Uganda, United Kingdom,
    Tanzania, Uruguay, Venezuela, Socialist Republic of Vietnam, Zambia,
    Zimbabwe\\

Likewise the multilevel community detection algorithm returns two clusters

\textbf{[0]} Afghanistan, Albania, Algeria, Andorra, Armenia, Australia, Azerbaijan,
    Bahrain, Belarus, Benin, Bhutan, Bosnia Herzegovina, Brunei, Bulgaria,
    Burkina Faso, Cambodia, Cameroon, Central African Republic, Chad, Comoros,
    Cote DIvoire, Croatia, Czech Republic, Denmark, Djibouti, Equatorial
    Guinea, Eritrea, Estonia, Ethiopia, Finland, France, Gabon, Georgia,
    Guinea Bissau, Hungary, Iceland, Indonesia, Iraq, Ireland, Israel, Italy,
    Japan, Jordan, Kazakhstan, Peoples Republic of Korea, Republic of Korea,
    Kosovo, Kuwait, Kyrgyz Republic, Laos, Latvia, Lebanon, Liberia, Libya,
    Liechtenstein, Lithuania, Luxembourg, Macedonia, Madagascar, Maldives,
    Mali, Mauritania, Micronesia, Moldova, Monaco, Mongolia, Montenegro,
    Nauru, Netherlands, Norway, Oman, Palau, Poland, Qatar, Romania, Russia,
    Rwanda, Sao Tome and Principe, Saudi Arabia, Senegal, Slovakia, Slovenia,
    Surinam, Syria, Taiwan, Tajikistan, Thailand, Togo, Tonga, Tunisia,
    Turkmenistan, Ukraine, United Arab Emirates, United States of America,
    Uzbekistan, Vanuatu, Yemen\\
\textbf{[1]} Angola, Antigua and Barbuda, Argentina, Austria, Bahamas, Bangladesh,
    Barbados, Belgium, Belize, Bolivia, Botswana, Brazil, Burundi, Canada,
    Cape Verde, Chile, China, Colombia, Democratic Republic of the Congo,
    Congo, Costa Rica, Cuba, Cyprus, Dominica, Dominican Republic, Ecuador,
    Egypt, El Salvador, Fiji, Gambia, German Federal Republic, Ghana, Greece,
    Grenada, Guatemala, Guinea, Guyana, Haiti, Honduras, India, Iran, Jamaica,
    Kenya, Kiribati, Lesotho, Malawi, Malaysia, Malta, Marshall Islands,
    Mauritius, Mexico, Morocco, Mozambique, Myanmar, Namibia, Nepal, New
    Zealand, Nicaragua, Niger, Nigeria, Pakistan, Panama, Papua New Guinea,
    Paraguay, Peru, Philippines, Portugal, St Kitts and Nevis, St Lucia, St
    Vincent and the Grenadines, Samoa, Serbia, Seychelles, Sierra Leone,
    Singapore, Solomon Islands, Somalia, South Africa, South Sudan, Spain, Sri
    Lanka, Sudan, Swaziland, Sweden, Switzerland, East Timor, Trinidad and
    Tobago, Turkey, Tuvalu, Uganda, United Kingdom, Tanzania, Uruguay,
    Venezuela, Socialist Republic of Vietnam, Zambia, Zimbabwe\\

However the infomap algorithm returns 96 clusters with little consistency. This may be due to the nature of the network which is necessarily built around a set of pairwise relationships rather than networks upon which dynamical processes such as information flow occur.\\

Since it is apparent that the distribution of similarities with all N-1 other countries differs significantly between all countries, we consider the $N_{closest}$ other countries to each individual country. This effectively controls for the skew present in each countries similarity to all other countries. Ideally we expect ties between node i and j to be most significant when the edge $e_{i,j}$ is both among i's $N_{closest}$ strongest links and j's $N_{closest}$ links. However we require $N_{closest}=67$ before we arrive at a fully connected graph.\\

In addition, when we consider the sum of the N-1(=193) Jacard similarities for each country (out degree using weighted edges), we would expect to see an indication of the most significant or important country constitutions.

\begin{tabular}{ |c| c | }
\hline
Country& Sum of Jaccard similarities\\
\hline
Zambia &	65.075643\\
Sudan	&64.642549\\
Dominica &	64.236225\\
Republic of Korea	&64.131567\\
Russia	&63.698398\\
Serbia&	63.641952\\
Albania	&63.619232\\
Singapore&	62.970728\\
Antigua and Barbuda	&62.844347\\
Guatemala	& 62.759447\\
\hline
\end{tabular}\\

And the lowest\\

\begin{tabular}{ |c| c | }
\hline
Country& Sum of Jaccard similarities\\
\hline
Liberia &	38.125831\\
United States of America	&39.831019\\
Denmark &	39.902812\\
Tonga	&40.771953\\
Mexico	&42.124157\\
Tunisia&	42.352826\\
Bosnia Herzegovina	&43.116865\\
Iraq&	43.998425\\
Uruguay	&44.071001\\
Netherlands	& 44.189905\\
\hline
\end{tabular}
\\

Even normalising the vector of Jaccard indices for all N-1 neighbours i.e. each row in the similarity matrix, we require a threshold no greater than 0.025 to recover a connected graph. At this point the network is sufficiently dense that only one cluster may be recovered.

Considering all of these facts, we therefore relax the condition for preserving an edge and for all of i's $N_{closest}$ neighbours create a directed edge of weight 1. This allows a more fine grained disaggregation of the intuitive clusters found when weighting edges by similarity as described above. Results from this network are reported in the main paper.\\

\textbf{Former Socialist (Cluster 1):} Afghanistan Albania Andorra Armenia Azerbaijan Bahrain Belarus Belgium Bosnia Herzegovina Bulgaria Cambodia China Croatia Czech Republic Denmark Eritrea Estonia Ethiopia Finland Georgia Guinea Bissau Hungary Iceland Indonesia Iraq Japan Jordan Kazakhstan Peoples Republic of Korea Republic of Korea Kosovo Kuwait Kyrgyz Republic Laos Latvia Lebanon Libya Liechtenstein Lithuania Macedonia Micronesia Moldova Monaco Mongolia Montenegro Netherlands Norway Oman Qatar Romania Russia Saudi Arabia Serbia Slovakia Slovenia Syria Taiwan Tajikistan Tunisia Turkmenistan Ukraine United Arab Emirates Uzbekistan Socialist Republic of Vietnam Yemen          \\ \newline
\textbf{Commonwealth (Cluster 2):} Antigua and Barbuda Australia Bahamas Bangladesh Barbados Belize Bhutan Botswana Brunei Canada Cyprus Dominica Fiji Gambia Ghana Grenada Guyana India Ireland Israel Jamaica Kenya Kiribati Lesotho Liberia Malawi Malaysia Maldives Malta Marshall Islands Mauritius Myanmar Namibia Nauru Nepal New Zealand Nigeria Pakistan Palau Papua New Guinea Philippines St Kitts and Nevis St Lucia St Vincent and the Grenadines Samoa Seychelles Sierra Leone Singapore Solomon Islands South Africa South Sudan Sri Lanka Sudan Swaziland Sweden Thailand Tonga Trinidad and Tobago Tuvalu Uganda United Kingdom Tanzania United States of America Vanuatu Zambia Zimbabwe \\ \newline
\textbf{Iberian(Cluster 3):} Angola Argentina Austria Bolivia Brazil Cape Verde Chile Colombia Costa Rica Cuba Dominican Republic Ecuador Egypt El Salvador German Federal Republic Greece Guatemala Haiti Honduras Iran Italy Mexico Mozambique Nicaragua Panama Paraguay Peru Poland Portugal Somalia Spain Surinam Switzerland East Timor Turkey Uruguay Venezuela\\\newline
\textbf{French (Cluster 4):} Algeria Benin Burkina Faso Burundi Cameroon Central African Republic Chad Comoros Democratic Republic of the Congo Congo Cote DIvoire Djibouti Equatorial Guinea France Gabon Guinea Luxembourg Madagascar Mali Mauritania Morocco Niger Rwanda Sao Tome and Principe Senegal Togo\\
\clearpage

Decreasing $N_{closest}=10$ lowers the density of edges allowing an additional cluster to emerge.\\ \newline

\textbf{Ex-Soviet (Cluster 1):}Afghanistan Albania Andorra Argentina Armenia Azerbaijan Belarus Belgium Bosnia Herzegovina Bulgaria Cambodia China Croatia Cuba Czech Republic Denmark Eritrea Estonia Ethiopia Finland Georgia Guinea Bissau Hungary Iceland Indonesia Italy Japan Kazakhstan Peoples Republic of Korea Republic of Korea Kosovo Kyrgyz Republic Laos Latvia Libya Liechtenstein Lithuania Macedonia Moldova Mongolia Montenegro Netherlands Norway Romania Russia Serbia Slovakia Slovenia Somalia Surinam Taiwan Tajikistan Thailand Turkmenistan Ukraine Uzbekistan Socialist Republic of Vietnam\\ \newline
\textbf{Middle Eastern (Cluster 2):}Bahrain Iraq Jordan Kuwait Lebanon Oman Qatar Saudi Arabia Syria Tunisia United Arab Emirates Yemen\\ \newline
\textbf{Ex-French (Cluster 3):}Algeria Benin Burkina Faso Burundi Cameroon Central African Republic Chad Comoros Democratic Republic of the Congo Congo Cote DIvoire Djibouti Equatorial Guinea France Gabon Guinea Luxembourg Madagascar Mali Mauritania Monaco Morocco Niger Sao Tome and Principe Senegal Togo\\ \newline
\textbf{Ex-Iberian (Cluster 4):} Angola Austria Bolivia Brazil Cape Verde Chile Colombia Costa Rica Dominican Republic Ecuador Egypt El Salvador German Federal Republic Greece Guatemala Haiti Honduras Iran Mexico Mozambique Nicaragua Panama Paraguay Peru Poland Portugal Rwanda Spain Switzerland East Timor Turkey Uruguay Venezuela\\ \newline
\textbf{Commonwealth (Cluster 5):}Antigua and Barbuda Australia Bahamas Bangladesh Barbados Belize Bhutan Botswana Brunei Canada Cyprus Dominica Fiji Gambia Ghana Grenada Guyana India Ireland Israel Jamaica Kenya Kiribati Lesotho Liberia Malawi Malaysia Maldives Malta Marshall Islands Mauritius Micronesia Myanmar Namibia Nauru Nepal New Zealand Nigeria Pakistan Palau Papua New Guinea Philippines St Kitts and Nevis St Lucia St Vincent and the Grenadines Samoa Seychelles Sierra Leone Singapore Solomon Islands South Africa South Sudan Sri Lanka Sudan Swaziland Sweden Tonga Trinidad and Tobago Tuvalu Uganda United Kingdom Tanzania United States of America Vanuatu Zambia Zimbabwe\\

\clearpage

\section{Provisional Topics}

\subsection{Provision Decomposition}

We now examine the high level structure of the set of current constitutions. Using dimensionality reduction, we recover sets of provisions that commonly co-occur and best represent the set of constitutions. Henceforth we refer to these sets of co-ocurring provisions as `provisional topics` in analogy with topics (sets of co-ocurring words) found in the text of documents. These provisional topics are illustrated below (fig \ref{prop1}).

\begin{figure}[h!]
\centerline{\includegraphics[width=0.5\linewidth]{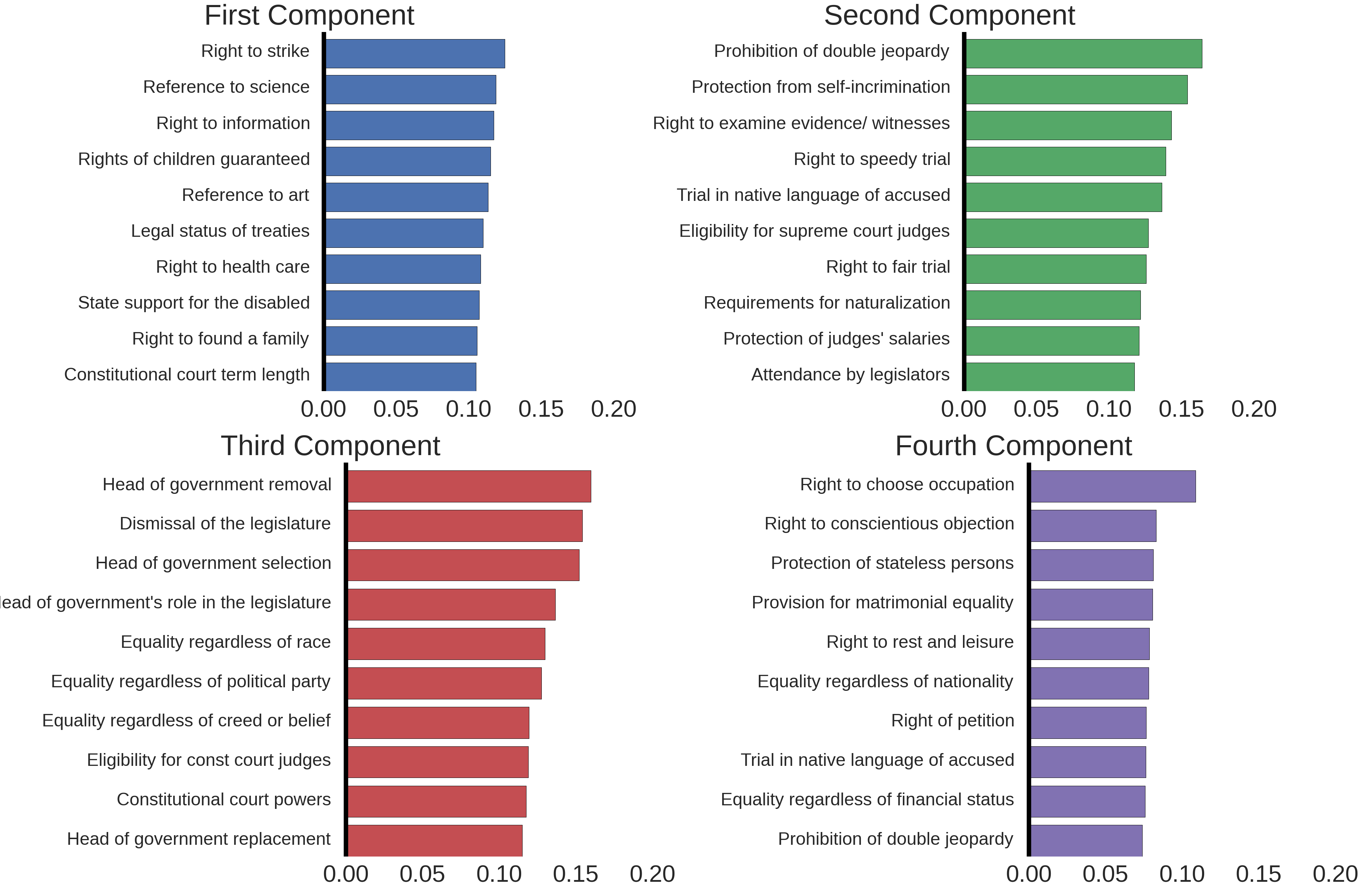}}
\caption{Graphical representation of provisions with largest projections on first 4 components}\label{all_comp}
\end{figure}

We find that the first two singular vectors are characterised by provisions concerning social rights and duties, and legislative process respectively. The third and fourth components are characterised by provisions related to oversight and limits to positions of powers and the structure of the second chamber respectively. In addition we find the first and second vectors allow for clear separation of the Commonwealth and Iberian clusters. However the third and fourth singular vectors do not allow for such a clear separation (see fig (\ref{all_comp})).

\begin{figure}[h]
\centering
\includegraphics[width=0.5\textwidth]{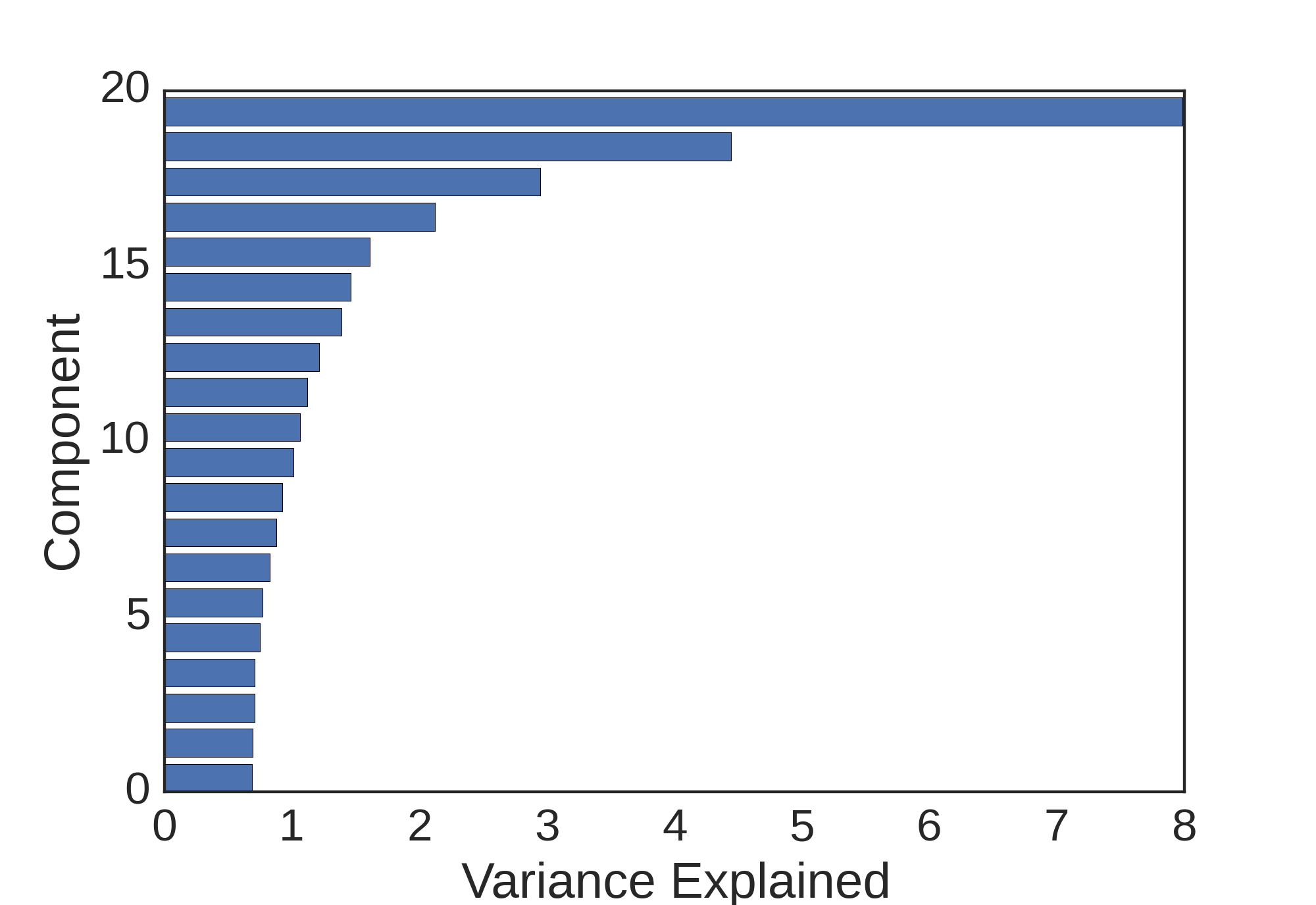}
\caption{\label{fig:scree} Scree plot of decomposition of country-provision matrix }
\end{figure}

In order to resolve the apparent disparity between similarities based on word choice and on provisional fingerprint, we investigate the relationship between the provisional topics and the network clusters discovered based on word choice in the text. We return to the multinomial logit models from the main paper using the network cluster IDs as dependent variables.  From the PCA analysis, we derive a score for each country on each of the 4 components.  A model that includes only these scores has an ePCP of 75.13\%, meaning that it performs better than models that include only colonial history (65.01\%) or legal system type (58.48\%) and almost as well as a model that includes both of the latter (76.24\%).  This indicates that the choice to include or exclude provisions from a constitution can statistically explain most of the similarity in word choice across constitutions.  Finally, a model that includes all three of these variables has an ePCP of 93.14\%, which means that legal system type, colonial history, and provision inclusion/exclusion choices of constitutions perform very well in predicting word choice similarity. Thus, while a given constitution is likely to adopt some new provisions not present within it`s peer constitutions, structure exists within provisions that co-occur within constitutions and this high level structure corresponds to similarities based on word choice. Figure (\ref{fig:scree}) plots the variance explained by each component of the country-provision matrix decomposition. The shape of this plot demonstrates that significant structure exists within this matrix. This is followed by visualisation of the projection onto these components by each country as well as tabulation of the composition of these components.

\begin{figure}[h!]
\centerline{\includegraphics[width=0.6\linewidth]{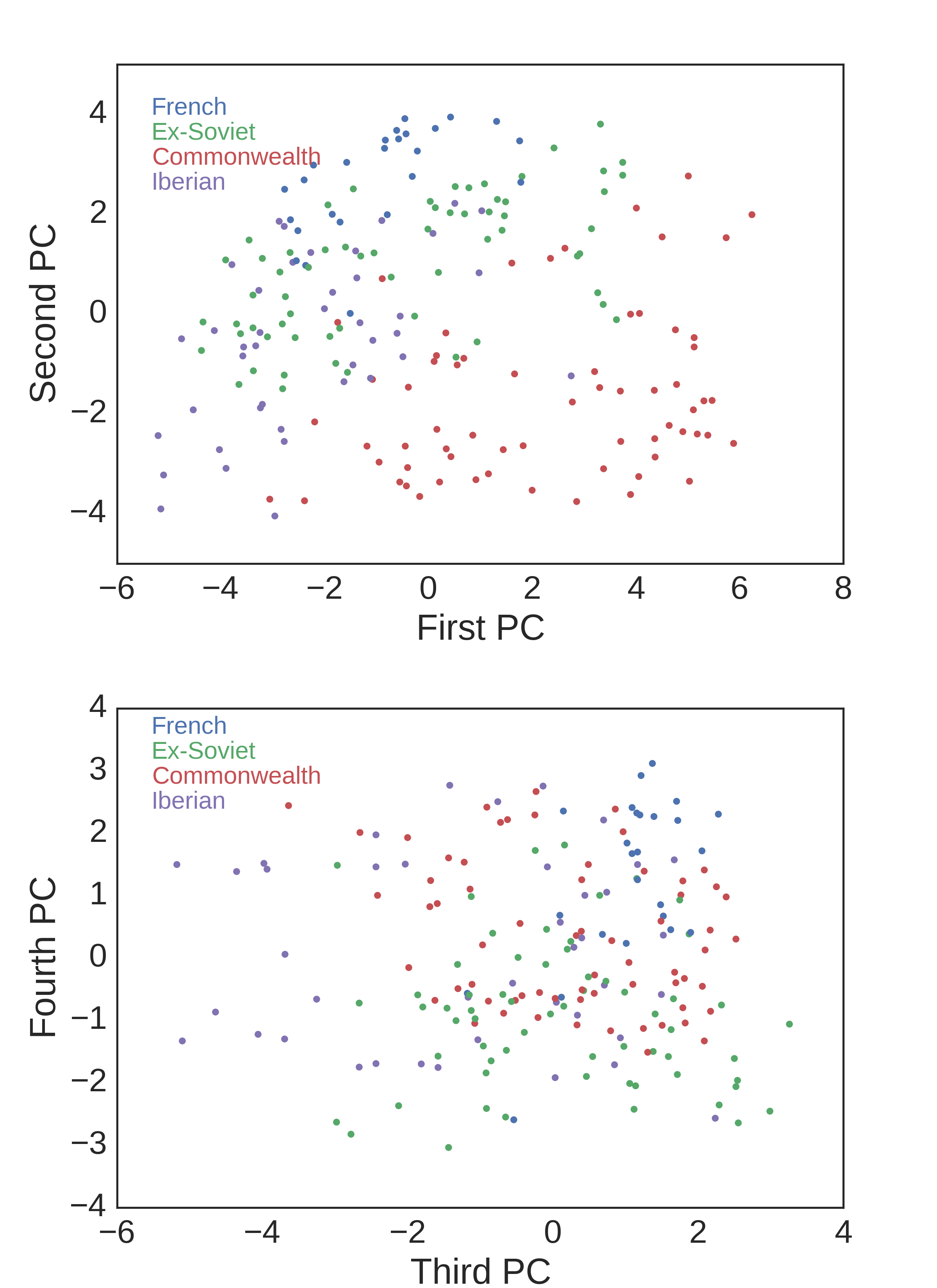}}
\caption{Projection of constitutional traces onto first and second and third and fourth principal components}\label{pca}
\end{figure}

\begin{table}
\label{decomposition}
\begin{tabular}{|l|l|}
\hline
Provision & Weight\\
\hline
Spending Bills               & 0.0458  \\
Tax bills & 0.0396\\
Right to examine evidence/ witnesses&0.0379\\
Electoral districts & 0.0376\\
Finance Bills & 0.0361 \\
\hline
Reference to art & -0.1129\\
Right to information & -0.1181\\
Reference to science & -0.1182\\
Rights of children guaranteed &-0.1182 \\
Right to strike & -0.1250\\
\hline
\end{tabular}
\caption{Top Components of First Singular Vector}

\begin{tabular}{|l|l|}
\hline
Provision & Weight\\
\hline
Establishment of constitutional court&0.0975 \\
Head of government powers&0.0968\\
Constitutional court selection&0.0925\\
Constitutional court powers&0.0889\\
Head of government selection&0.0843 \\
\hline
Trial in native language of accused&-0.1352\\
Right to speedy trial&-0.1383\\
Right to examine evidence/ witnesses&-0.1428\\
Protection from self-incrimination&-0.1532\\
Prohibition of double jeopardy&-0.1636\\
\hline
\end{tabular}
\caption{Top Components of Second Singular Vector}
\begin{tabular}{|l|l|}
\hline
Provision & Weight\\
\hline
Head of government removal&0.1576\\
Dismissal of the legislature&0.1540\\
Head of government selection&0.1501\\
Head of government's role in the legislature&0.1348\\
Equality regardless of race&0.1309\\
\hline
Electoral court selection&-0.0808\\
Right to amparo&-0.0818\\
Electoral court powers&-0.0865\\
State support for the elderly&-0.0870\\
Minimum age of supreme court judges&-0.0901\\
\hline
\end{tabular}
\caption{Top Components of Third Singular Vector}
\begin{tabular}{|l|l|}
\hline
Provision & Weight\\
\hline
Eligibility for second chamber&0.2536\\
Division of labor between chambers&0.2507\\
Leader of second chamber&0.2499\\
Second chamber selection&0.2471\\
Minimum age for second chamber&0.2354\\
\hline
Right to rest and leisure&-0.0793\\
Provision for matrimonial equality&-0.0815\\
Protection of stateless persons&-0.0818\\
Right to conscientious objection&-0.0826\\
Right to choose occupation&-0.1095\\
\hline
\end{tabular}
\caption{Top Components of Fourth Singular Vector}

\end{table}

\begin{table}
\label{decomposition}
\begin{tabular}{|l|l|}
\hline
Provision & Weight\\
\hline
Right to strike(-)&0.1250\\
Rights of children guaranteed(-)&0.1182\\
Reference to science(-)&0.1182\\
Right to information(-)&0.1181\\
Reference to art(-)&0.1129\\
Legal status of treaties(-)&0.1094\\
Right to health care(-)&0.1075\\
State support for the disabled(-)&0.1065\\
Right to found a family(-)&0.1060\\
Constitutional court term length(-)&0.1050\\
Compulsory education(-)&0.1039\\
Free education(-)&0.1036\\
Constitutional court selection(-)&0.1026\\
National anthem(-)&0.1022\\
Right to culture(-)&0.1008\\

\hline
\end{tabular}
\caption{Top Absolute Components of First Singular Vector}

\begin{tabular}{|l|l|}
\hline
Provision & Weight\\
\hline
Prohibition of double jeopardy(-)&0.1636\\
Protection from self-incrimination(-)&0.1532\\
Right to examine evidence/ witnesses(-)&0.1428\\
Right to speedy trial(-)&0.1383\\
Trial in native language of accused(-)&0.1352\\
Eligibility for supreme court judges(-)&0.1266\\
Right to fair trial(-)&0.1250\\
Requirements for naturalization(-)&0.1215\\
Protection of judges' salaries(-)&0.1208\\
Electoral commission(-)&0.1201\\
Attendance by legislators(-)&0.1177\\
Equality regardless of skin color(-)&0.1106\\
Right to renounce citizenship(-)&0.1085\\
Right to appeal judicial decisions(-)&0.1085\\
Tax bills(-)&0.1061\\
\hline
\end{tabular}
\caption{Top Absolute Components of Second Singular Vector}
\begin{tabular}{|l|l|}
\hline
Provision & Weight\\
\hline
Head of government removal(+)&0.1576\\
Dismissal of the legislature(+)&0.1540\\
Head of government selection(+)&0.1501\\
Head of government's role in the legislature(+)&0.1348\\
Equality regardless of race(+)&0.1309\\
Equality regardless of political party(+)&0.1277\\
Equality regardless of creed or belief(+)&0.1235\\
Eligibility for const court judges(+)&0.1165\\
Head of government replacement(+)&0.1155\\
Constitutional court powers(+)&0.1150\\
Equality regardless of origin(+)&0.1120\\
Establishment of constitutional court(+)&0.1097\\
Establishment of judicial council(+)&0.1082\\
Constitutional court term length(+)&0.1075\\
Head of government powers(+)&0.1064\\

\hline
\end{tabular}
\caption{Top Absolute Components of Third Singular Vector}
\begin{tabular}{|l|l|}
\hline
Provision & Weight\\
\hline
Eligibility for second chamber(+)&0.2536\\
Division of labor between chambers(+)&0.2507\\
Leader of second chamber(+)&0.2499\\
Second chamber selection(+)&0.2471\\
Minimum age for second chamber(+)&0.2354\\
Term length of second chamber(+)&0.2236\\
Joint meetings of legislative chambers(+)&0.1982\\
Replacement of legislators(+)&0.1679\\
Size of second chamber(+)&0.1623\\
First chamber reserved policy areas(+)&0.1382\\
Compensation of legislators(+)&0.1161\\
Second chamber reserved policy areas(+)&0.1097\\
Right to choose occupation(-)&0.1095\\
Finance bills(+)&0.1067\\
Establishment of judicial council(+)&0.0997\\
\hline
\end{tabular}
\caption{Top Absolute Components of Third Singular Vector}
\end{table}

\clearpage
\subsection{Convergence of Constitutional Traces}

\begin{figure}[h]
\centerline{
\includegraphics[width=0.7\textwidth]{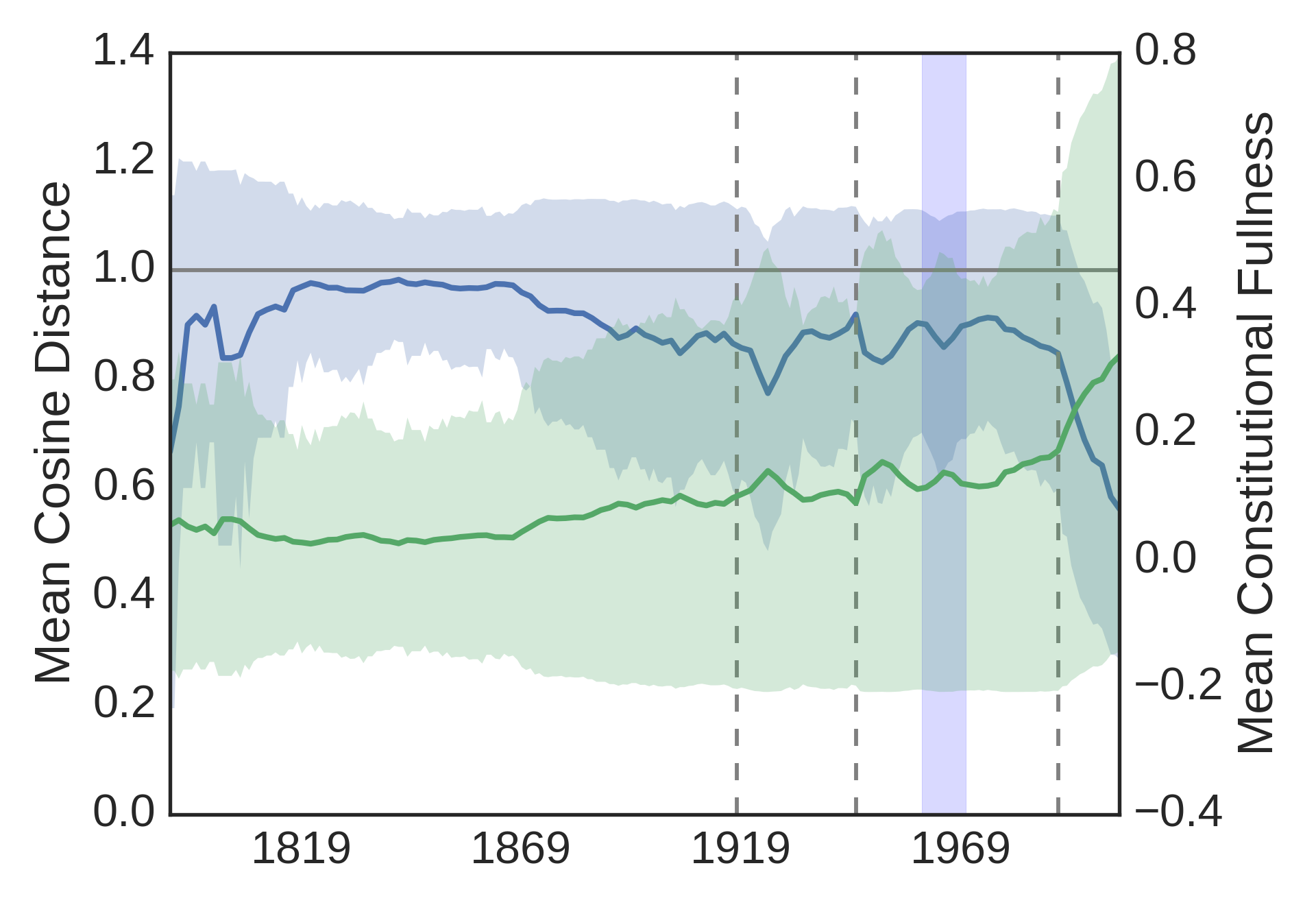}}
\caption{Convergence of constitutional provisional similarity over time. Blue line shows mean cosine distance between pairs of countries with one standard deviation marked on. The green line shows mean `density` of constitutions i.e. proportion of provisions adopted. The years 1919, 1945 and 1991 are marked.}
\label{convergence}
\end{figure}

Figure \ref{convergence} shows both (i) the mean proportion of provisions that are adopted (in green) and (ii) the provisional similarity (measured by the cosine similarity, shown in green) between countries increases over time.

\clearpage

\section{Correlations of Text and Provisional Similarities}

\begin{figure*}[h]
\centerline{\includegraphics[width=0.5\linewidth]{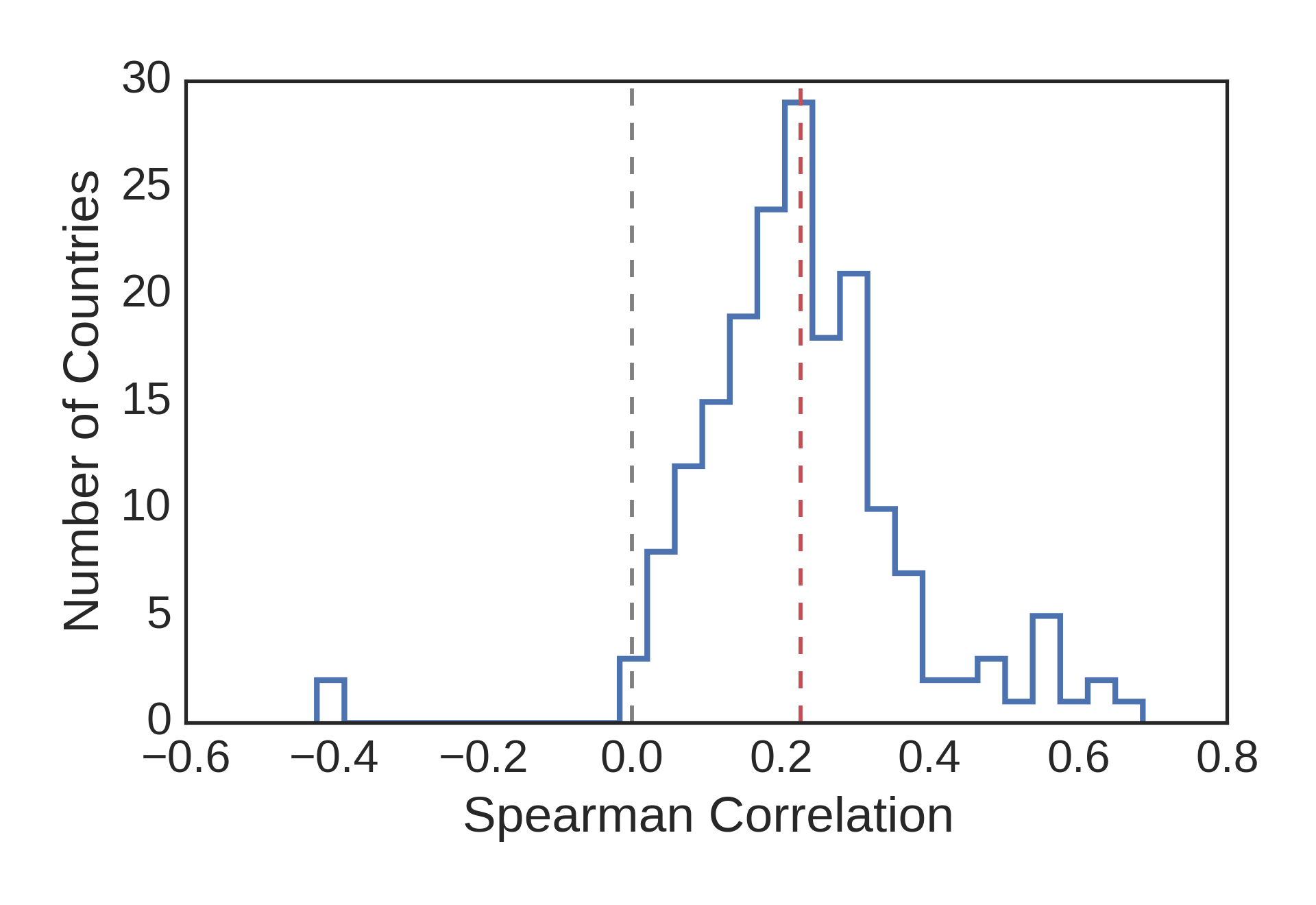}}
\caption{Distribution over countries of Spearman rank correlation coefficients based on text and provisions similarity}\label{spearman_correlations}
\end{figure*}

\begin{figure*}[h]
\centerline{\includegraphics[width=0.5\linewidth]{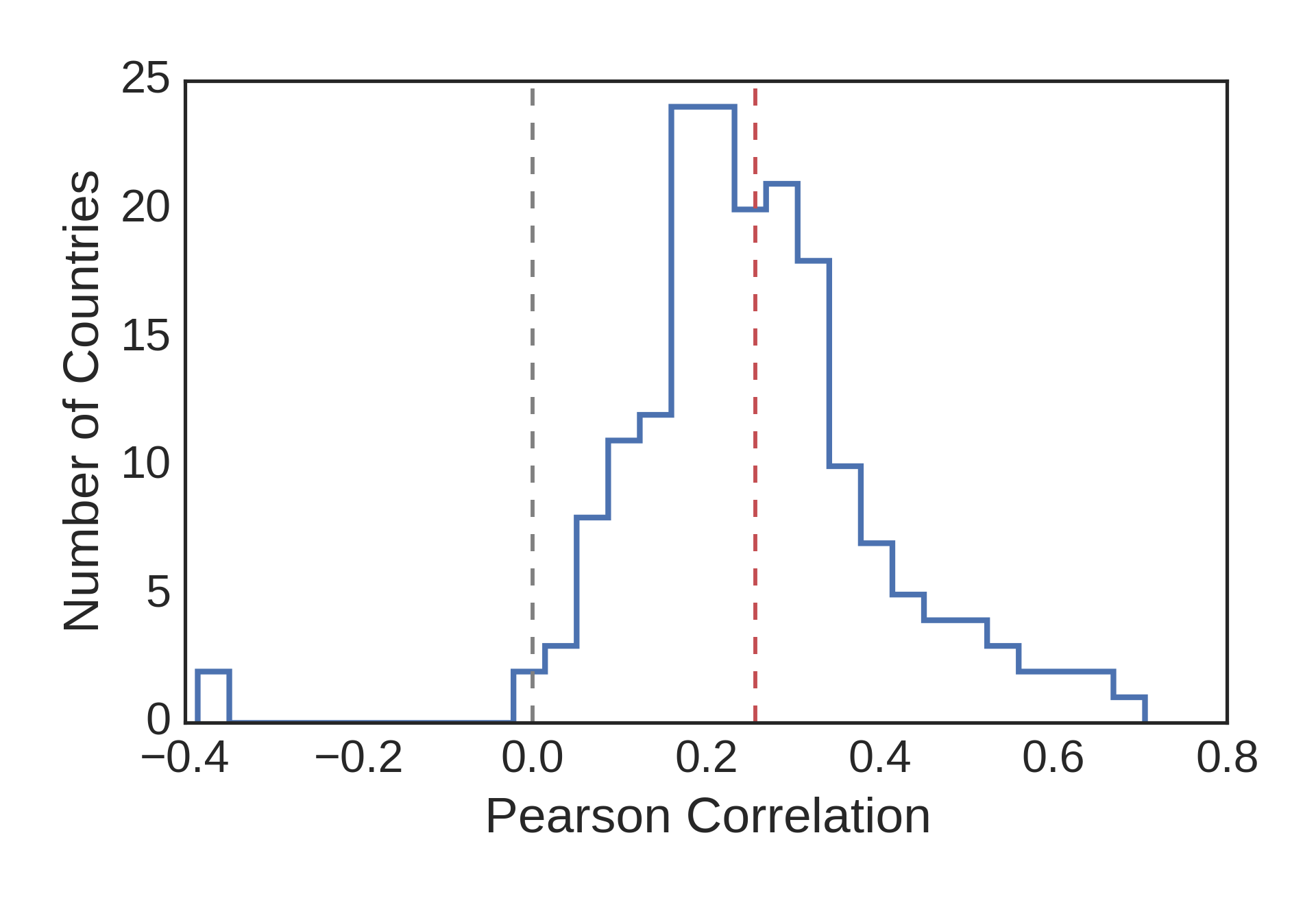}}
\caption{Distribution over countries of Pearson rank correlation coefficients based on text and provisions similarity}\label{pearson_correlations}
\end{figure*}

\clearpage

\section{Insertion/Removal of Provisions upon Independence}

Here we consider the number of provisions present in countries when independence is first achieved i.e. in the first constitution that appears in the CCPNC datsaset\cite{ccpnc}. We compare this number to the number of provisions present in the former Imperial power in that cluster (e.g. Russia, UK, Spain or France) in that same year. Due to shortcomings in this dataset, often no information is available on the presence of these provisions, therefore we consider the first year after independence for which this information is available. Where no future year has this information we look to the first available \textit{past} year for comparison. The boxplot of differences in provisions is plotted below with the Iberian cluster (Spain) omitted due to the sparsity of information available on Spanish provisions before the year 1932.

\begin{figure*}[h]
\centerline{\includegraphics[width=0.6\linewidth]{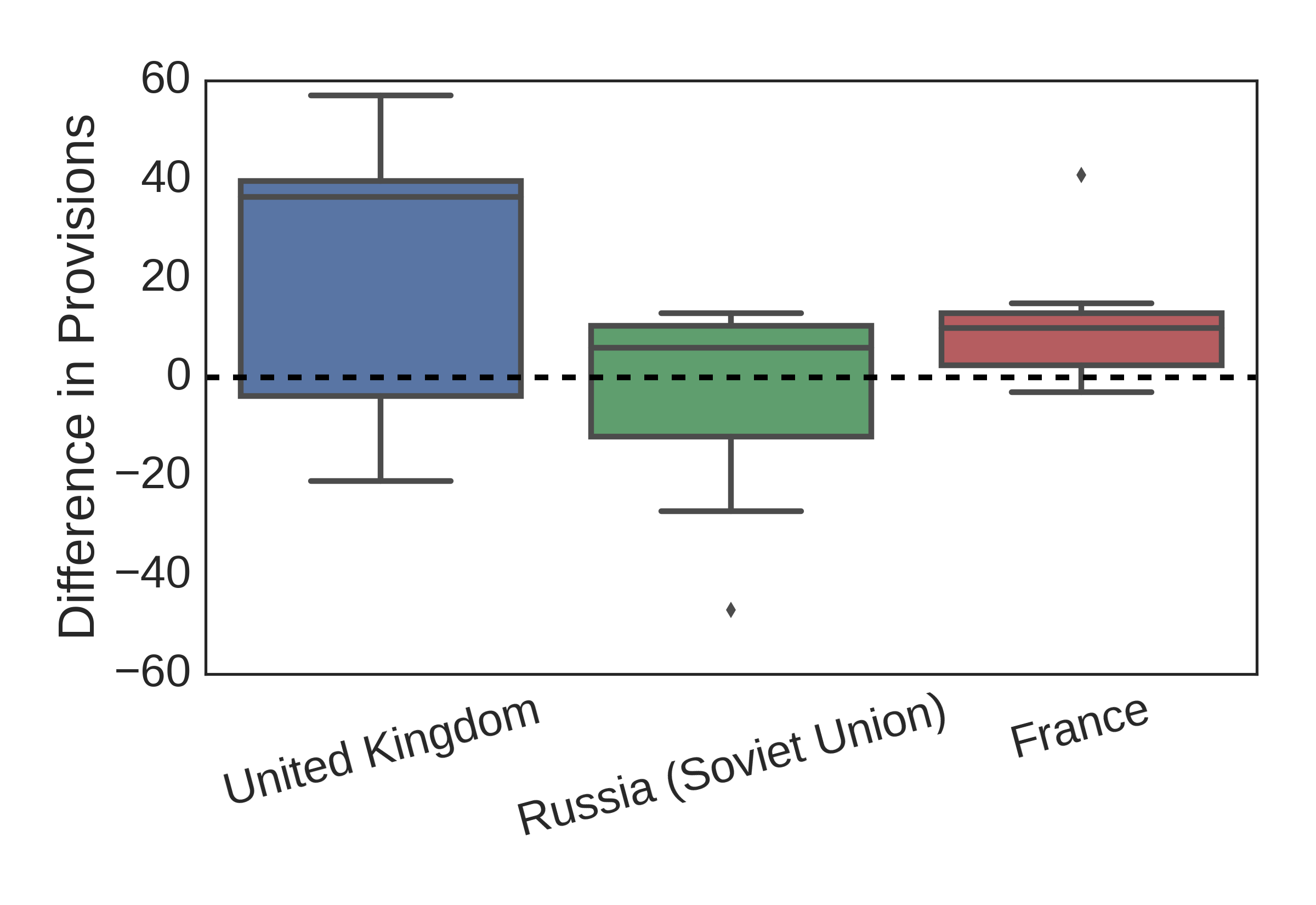}}
\caption{Boxplot of number of provisions added to each constitution relative to parent constitution by cluster.}\label{cluster_provision_differences}
\end{figure*}

\clearpage

\section{Provisional Time Series}

\subsection{Time Series Dimensionality Reduction}

In order to find structure within the historical adoption of each provision, we consider the time series of the proportion of constitutions adopting each provision. On account of the large peaks around 1990 which occur for all provisions, a simple linear measure of similarity such as Pearson correlation coefficient returns artificially large correlations which precludes meaningful clustering. Therefore we perform a dimensionality reduction using a probabilistic PCA~\cite{pca} as implemented in the Python scikit-learn package~\cite{scikit}. This allows for the provisions to be reduced to a linear combination of a smaller number of representative time series. For this purpose we consider the provisional activity over the 100 year period from 1913 to 2013 in order to mitigate the effect of noise arising from a relatively small number of countries having national constitutions at this time.\\

The dimensionality reduction therefore reduces the matrix of observations from $(\textrm{nYears} \times \textrm{nProvisions})=(100 \times 234)$ to $(\textrm{nPrincipal} \times \textrm{nProvisions})$ where $\textrm{nPrinicipal}$ is chosen arbitrarily. However we can see from the scree plot~\cite{clustering} below that the vast majority of the variance of the provisions is captured by the first 4 principal components. The evolution of the principal components are visualised below, along with their respective proportion of the of explained variance.

\begin{figure}[h]
\centerline{\includegraphics[width=0.5\linewidth]{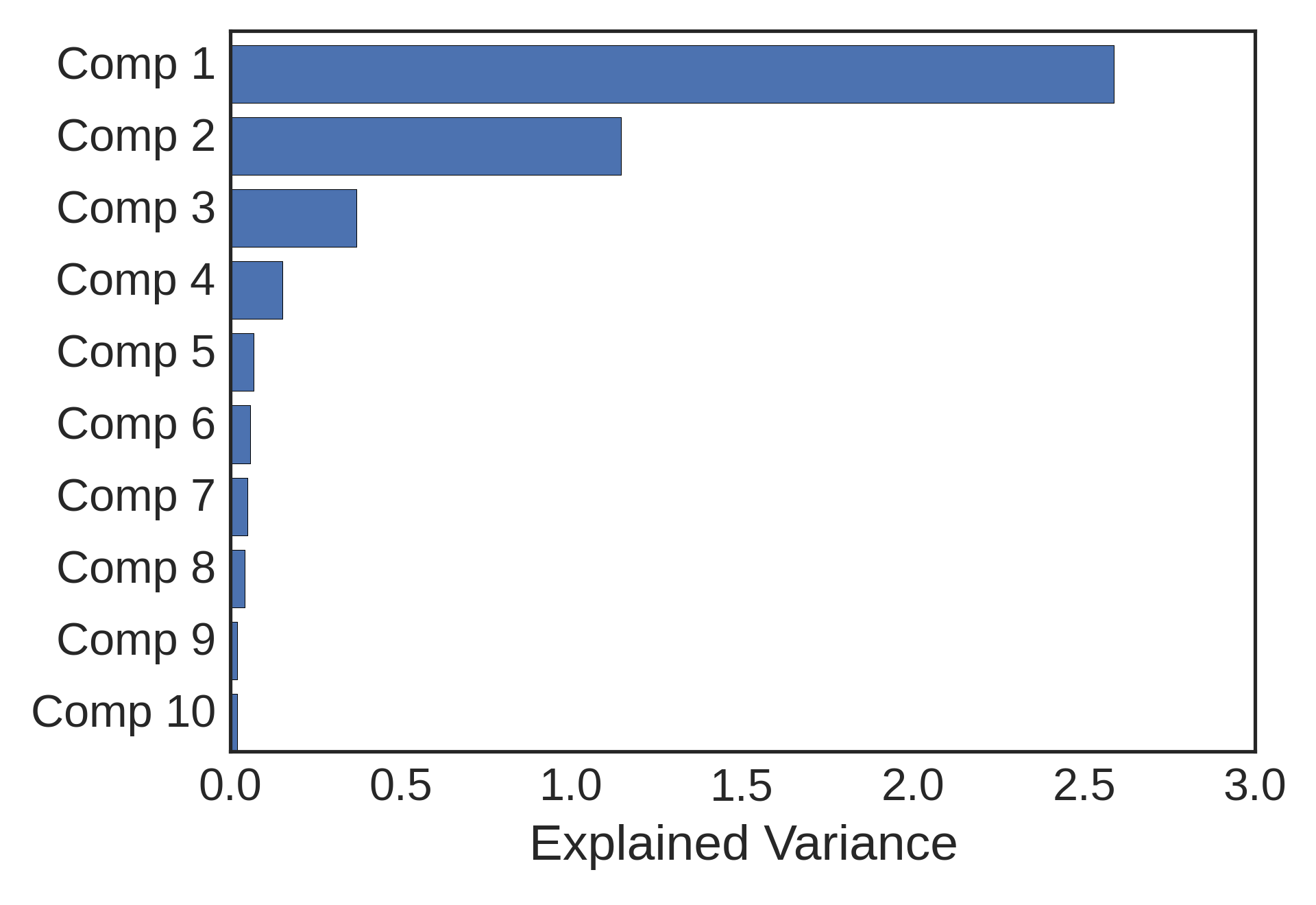}}
\caption{The variance explained by each principal component.}
\end{figure}

The evolution of the principal components are visualised below, along with their respective proportion of explained variance

\begin{figure}[h]
\centerline{\includegraphics[width=0.5\linewidth]{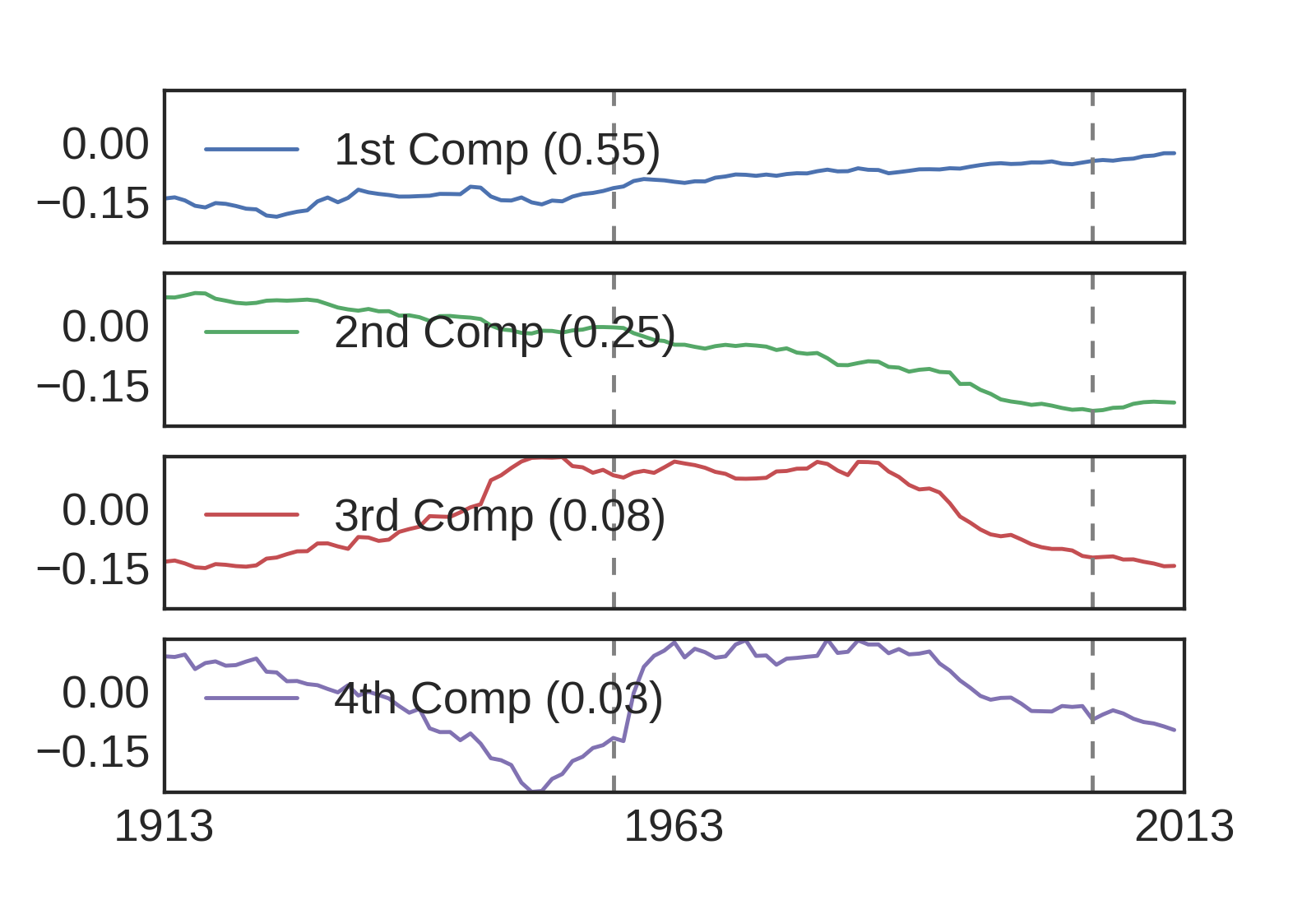}}
\caption{First 4 principal componentsof provisional time series annotated with their respective explained variance}
\end{figure}

\subsection{Clustering in Reduced Space}

Next we consider the clustering of these time series when projected into the reduced 10-dimensional space. We make use of k-means clustering, with k=3,4,5 displayed below. The `elbow' plot below shows the reduction in the total intra-cluster distance to justify our use of k=4\\

\begin{figure}[h]
\centerline{\includegraphics[width=0.7\linewidth]{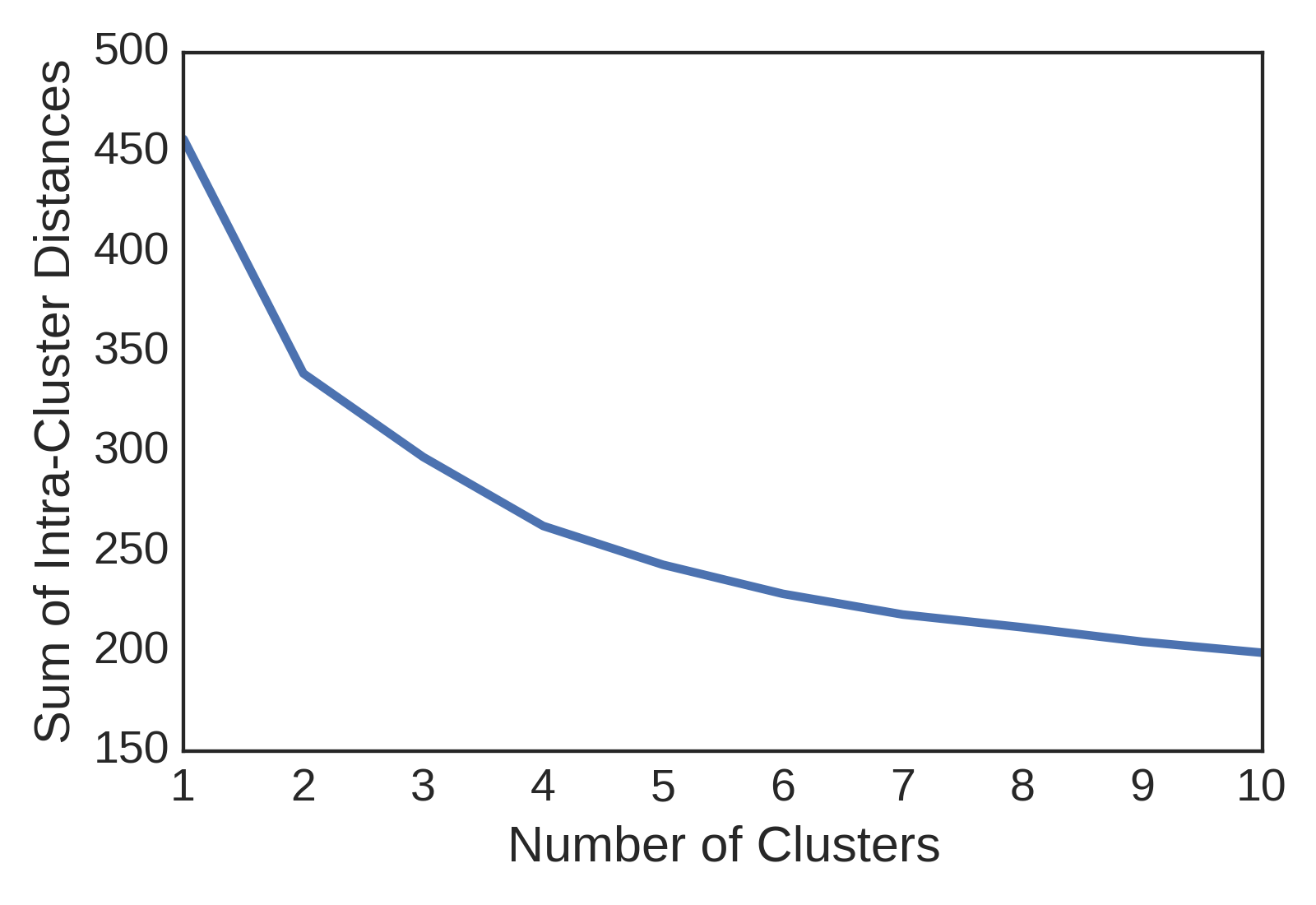}}
\caption{Elbow plot of k-means clustering.}
\end{figure}

Below the full list of provisions in each cluster is enumerated.

\begin{table}[]
\tiny
\centering
\label{my-label}
\begin{tabular}{l}
\textbf{Cluster 1} \\ \hline
\multicolumn{1}{|l|}{Does the constitution provide for matrimonial equality?} \\ \hline
\multicolumn{1}{|l|}{Does the constitution refer to political parties?} \\ \hline
\multicolumn{1}{|l|}{Does the constitution contain provisions concerning the relationship between the constitution and international law?} \\ \hline
\multicolumn{1}{|l|}{Does the constitution provide the right to just remuneration, fair or equal payment for work?} \\ \hline
\multicolumn{1}{|l|}{Does the constitution refer to "democracy" or "democratic"?} \\ \hline
\multicolumn{1}{|l|}{Does the constitution guarantee the rights of children?} \\ \hline
\multicolumn{1}{|l|}{Does the constitution prohibit cruel, inhuman, or degrading treatment?} \\ \hline
\multicolumn{1}{|l|}{Does the constitution provide for a right of rest and leisure?} \\ \hline
\multicolumn{1}{|l|}{Does the constitution place limits on child employment?} \\ \hline
\multicolumn{1}{|l|}{Does the constitution provide for the right to choose ones occupation?} \\ \hline
\multicolumn{1}{|l|}{Does the constitution mention the right to safe/healthy working conditions?} \\ \hline
\multicolumn{1}{|l|}{Does the constitution contain provisions with regard to any additional central independent regulatory agencies ?} \\ \hline
\multicolumn{1}{|l|}{Does the constitution prohibit punishment by laws enacted ex post facto ?} \\ \hline
\multicolumn{1}{|l|}{Does the constitution stipulate that some public office holders take an oath to support or abide by the constitution?} \\ \hline
\multicolumn{1}{|l|}{Is a supermajority needed for passing any legislation?} \\ \hline
\multicolumn{1}{|l|}{Is there a mandatory retirement age for judges?} \\ \hline
\multicolumn{1}{|l|}{Does the constitution refer to a duty of the people to take part in building society or to work for the development of the country?} \\ \hline
\multicolumn{1}{|l|}{Does the constitution provide for the right of some redress in the case of false imprisonment, arrest, or judicial error?} \\ \hline
\multicolumn{1}{|l|}{Does the constitution provide for freedom of movement?} \\ \hline
\multicolumn{1}{|l|}{Does the constitution provide for freedom of opinion, thought, and/or conscience?} \\ \hline
\multicolumn{1}{|l|}{How does the constitution address the state operation of print or electronic media?} \\ \hline
\multicolumn{1}{|l|}{Does the constitution refer to the "free market," "capitalism," or an analogous term?} \\ \hline
\multicolumn{1}{|l|}{Does the constitution include provisions for the meritocratic recruitment of civil servants (e.g. exams or credential requirements)?} \\ \hline
\multicolumn{1}{|l|}{Are any parts of the constitution unamendable?} \\ \hline
\multicolumn{1}{|l|}{Does the constitution contain an explicit declaration regarding the independence of the central judicial organ(s)?} \\ \hline
\multicolumn{1}{|l|}{Does the constitution refer to radio?} \\ \hline
\multicolumn{1}{|l|}{Does the constitution provide the right to counsel if one is indicted or arrested?} \\ \hline
\multicolumn{1}{|l|}{Does the constitution make a claim to universal adult suffrage?} \\ \hline
\multicolumn{1}{|l|}{Does the constitution contain provisions concerning the national flag?} \\ \hline
\multicolumn{1}{|l|}{Does the constitution provide the right to a free and/or competitive market?} \\ \hline
\multicolumn{1}{|l|}{Does the constitution guarantee equal access to higher education?} \\ \hline
\multicolumn{1}{|l|}{Does the constitution provide for a right to an adequate or reasonable standard of living?} \\ \hline
\multicolumn{1}{|l|}{Do defendants have the right to appeal judicial decisions?} \\ \hline
\multicolumn{1}{|l|}{Does the constitution contain a national motto?} \\ \hline
\multicolumn{1}{|l|}{Does the constitution provide for the right to form or to join trade unions?} \\ \hline
\multicolumn{1}{|l|}{Does the constitution provide for the right of protection of one's reputation from libelous actions?} \\ \hline
\multicolumn{1}{|l|}{Does the constitution provide for the extradition of suspected or convicted criminals to other countries?} \\ \hline
\multicolumn{1}{|l|}{Does the constitution refer to "customary" international law or the "law of nations"?} \\ \hline
\multicolumn{1}{|l|}{Does the constitution provide for the right/possibility of pre-trial release?} \\ \hline
\multicolumn{1}{|l|}{Does the constitution contain provisions concerning national integration of ethnic communities?} \\ \hline
\multicolumn{1}{|l|}{In what language is the source document written (not the original, but the one used for coding)?} \\ \hline
\multicolumn{1}{|l|}{Does the constitution mention a state duty to provide work/employment?} \\ \hline
\multicolumn{1}{|l|}{Does the constitution refer to a state duty to protect or promote culture or cultural rights?} \\ \hline
\multicolumn{1}{|l|}{Does the constitution refer to the social security of the society or nation?} \\ \hline
\multicolumn{1}{|l|}{Does the constitution provide for positive obligations to transfer wealth to, or provide opportunity for, particular groups?} \\ \hline
\multicolumn{1}{|l|}{Does the constitution refer to social, political, or economic conditions in the time before the birth of the state or in the time of a former constitution?} \\ \hline
\multicolumn{1}{|l|}{Does the constitution refer to "fraternity" or "solidarity"?} \\ \hline
\multicolumn{1}{|l|}{Does the constitution provide for inheritance rights?} \\ \hline
\multicolumn{1}{|l|}{Does the constitution restrict entry or exit of the states borders?} \\ \hline
\multicolumn{1}{|l|}{Does the constitution provide for the prohibition of double jeopardy (i.e., being tried for the same crime twice)?} \\ \hline
\multicolumn{1}{|l|}{Does the constitution refer to "socialism" or "socialist"?} \\ \hline
\multicolumn{1}{|l|}{Does the constitution contain provisions for the protection of stateless individuals, refugees from other states, or the right to asylum?} \\ \hline
\multicolumn{1}{|l|}{Does the constitution stipulate that courts have to take into account decisions of higher courts?} \\ \hline
\multicolumn{1}{|l|}{Does the constitution mention a state duty to provide health care?} \\ \hline
\multicolumn{1}{|l|}{Does the constitution prohibit torture?} \\ \hline
\multicolumn{1}{|l|}{Does the constitution generally require public trials?} \\ \hline
\multicolumn{1}{|l|}{Does the constitution mention the adoption of national economic plans?} \\ \hline
\multicolumn{1}{|l|}{Does the constitution stipulate a quota for representation of certain groups in the Second Chamber?} \\ \hline
\multicolumn{1}{|l|}{Does the constitution prescribe that electoral ballots be secret?} \\ \hline
\multicolumn{1}{|l|}{Are there provisions for dismissing the Head of Government?} \\ \hline
\multicolumn{1}{|l|}{Does the constitution specify a deputy executive of any kind (e.g., deputy prime minister, vice president)?} \\ \hline
\multicolumn{1}{|l|}{Does the constitution provide for either general or financial support by the government for the elderly?} \\ \hline
\multicolumn{1}{|l|}{Does the constitution provide for either general or financial support by the government for the unemployed?} \\ \hline
\multicolumn{1}{|l|}{Does the constitution provide for either general or financial support by the government for the disabled?} \\ \hline
\multicolumn{1}{|l|}{Does the constitution provide for either general or financial support by the government for children, orphans?} \\ \hline
\multicolumn{1}{|l|}{Does the constitution protect from discrimination/provide for equality for gender?} \\ \hline
\multicolumn{1}{|l|}{Does the constitution protect from discrimination/provide for equality for nationality?} \\ \hline
\multicolumn{1}{|l|}{Does the constitution protect from discrimination/provide for equality for race?} \\ \hline
\multicolumn{1}{|l|}{Does the constitution protect from discrimination/provide for equality for religion?} \\ \hline
\multicolumn{1}{|l|}{Does the constitution protect from discrimination/provide for equality for creed/beliefs?} \\ \hline
\multicolumn{1}{|l|}{Does the constitution protect from discrimination/provide for equality for social status?} \\ \hline
\multicolumn{1}{|l|}{Does the constitution protect from discrimination/provide for equality for parentage?\textless/pre\textgreater} \\ \hline
\end{tabular}
\end{table}

\clearpage

\begin{table}[]
\tiny
\centering
\label{my-label}
\begin{tabular}{|l|}
\hline
\textbf{Cluster 2} \\ \hline
Does the constitution specify that the chambers should meet jointly for any reason? \\ \hline
Does the constitution refer to a duty to join a political party? \\ \hline
Does the constitution mention a state duty to provide work/employment? \\ \hline
Does the constitution mention anything about crimes committed by the previous regime? \\ \hline
Does the constitution prohibit one or more political parties? \\ \hline
Does the constitution provide for immunity for the members of the Legislature under some conditions? \\ \hline
Does the constitution require that the names of those imprisoned be entered in a public registry? \\ \hline
Does the constitution provide the right for same sex marriages? \\ \hline
Does the constitution provide for the right to bear arms? \\ \hline
Does the constitution explicitly mention due process? \\ \hline
How many chambers or houses does the Legislature contain? \\ \hline
Does the constitution refer to a duty to join trade unions? \\ \hline
Does the constitution specify the electoral system for the Second Chamber? \\ \hline
Does the constitution refer to the French declaration of rights (1789)? \\ \hline
Does the constitution refer to the Helsinki Accords (1966)? \\ \hline
Is there citizen involvement in the indicting process (such as a grand jury)? \\ \hline
Is there a constitutional provision for civil marriage?\textless/pre\textgreater \\ \hline
\end{tabular}
\end{table}

\clearpage

\begin{table}[]
\tiny
\centering
\label{my-label}
\begin{tabular}{|l|}
\hline
\textbf{Cluster 3} \\ \hline
Does the constitution provide for an individual right to view government files or documents under at least some conditions? \\ \hline
Does the constitution provide for the right to examine evidence or confront all witnesses? \\ \hline
Does the constitution specify that healthcare should be provided by government free of charge? \\ \hline
Does the constitution require that legislators give up any other profession (i.e. work exclusively as legislators)? \\ \hline
Does the constitution contain provisions for a Judicial Council/Commission? \\ \hline
Is one of the executives explicitly referred to as the "Head of Government"? \\ \hline
Does the constitution mention judicial opinions of the Constitutional Court? \\ \hline
Does the constitution provide for a right to form political parties? \\ \hline
Is there special mention of terrorism and public security provisions regarding terrorism? \\ \hline
Is there a right to exemption from military service for conscientous objectors to war or other groups? \\ \hline
Does the constitution use the words (socio-) economic rights or similar? \\ \hline
Does the constitution require that legislators disclose their earnings and/or assets? \\ \hline
Does the constitution provide for an Ombudsman? \\ \hline
Does the constitution suggest that citizens should have the right to overthrow their government under certain circumstances? \\ \hline
Does the constitution provide for an individual's right to self determination or the right to free development of personality? \\ \hline
Does the constitution provide for the ability of individuals to propose legislative initiatives (referenda from below)? \\ \hline
Do citizens have the right to renounce their citizenship? \\ \hline
Does the constitution provide for a commission for truth and reconciliation? \\ \hline
Does the constitution mention a special regulatory body/institution to oversee the media market? \\ \hline
Does the constitution express a preference for one or more political parties? \\ \hline
Does the constitution provide for the right to a speedy trial? \\ \hline
Does the constitution stipulate a quota for representation of certain groups in the first (or only) chamber? \\ \hline
Does the constitution refer to protection or preservation of the environment? \\ \hline
Does the constitution provide for a people's right of self-determination? \\ \hline
Is one of the executives explicitly referred to as the "Head of State"? \\ \hline
Does the constitution provide the right to found a family? \\ \hline
Does the constitution provide for a right to strike? \\ \hline
Does the constitution specify the trial has to be in a language the accused understands or the right to an interpreter if the accused cannot understand the language? \\ \hline
Does the constitution mention any special procedures for removing members of the constitutional court? \\ \hline
Does the constitution contain a general statement regarding rule of law, legality, or Rechtsstaat (the German equivalent)? \\ \hline
Is there a presumption of innocence in trials? \\ \hline
Does the constitution contain provisions concerning the national anthem? \\ \hline
Does the constitution refer to the "dignity of man" or human "dignity"? \\ \hline
Does the constitution contain provisions concerning international organizations? \\ \hline
Does the constitution provide for the right to shelter or housing? \\ \hline
Does the constitution provide for a right to enjoy the benefits of scientific progress? \\ \hline
Does the constitution provide for an electoral commission or electoral court to oversee the election process? \\ \hline
Does the constitution mention consumer rights or consumer protection? \\ \hline
Does the constitution give juveniles special rights/status in the criminal justice process? \\ \hline
Does the constitution suggest that citizens should have the right to overthrow their government under certain circumstances? \\ \hline
Does the constitution refer to television? \\ \hline
Are rights provisions binding on private parties as well as the state? \\ \hline
Does the constitution mention "foreign investment" or "foreign capital"? \\ \hline
Does the constitution contain provisions for a central bank? \\ \hline
Does the constitution provide the right to a fair trial? \\ \hline
Does the constitution refer to the protection of different languages? \\ \hline
Does the constitution refer to the UN universal Declaration on Human Rights (1948)? \\ \hline
Does the constitution refer to the UN charter Article 45 (1945)? \\ \hline
Does the constitution refer to the European Convention for the Protection of Human Rights and Fundamental Freedoms (1950)? \\ \hline
Does the constitution refer to the International Covenant on Civil and Political Rights (1966)? \\ \hline
Does the constitution refer to the International Covenant on Economic and Social Rights (1966)? \\ \hline
Does the constitution refer to the American Convention on Human Rights (1969)? \\ \hline
Does the constitution refer to the African Charter on Human People's Rights (1981)? \\ \hline
Does the constitution contain provisions for a counter corruption commission? \\ \hline
Is there a special mention of victims rights in the constitution? \\ \hline
If counsel is provided, is it provided at the state's expense? \\ \hline
Does the constitution protect from discrimination/provide for equality for country of origin? \\ \hline
Does the constitution protect from discrimination/provide for equality for language? \\ \hline
Does the constitution protect from discrimination/provide for equality for sexual orientation? \\ \hline
Does the constitution protect from discrimination/provide for equality for age? \\ \hline
Does the constitution protect from discrimination/provide for equality for mental or physical disability? \\ \hline
Does the constitution protect from discrimination/provide for equality for color? \\ \hline
Does the constitution protect from discrimination/provide for equality for financial/propety ownership? \\ \hline
Does the constitution protect from discrimination/provide for equality for tribe/clan? \\ \hline
Does the constitution protect from discrimination/provide for equality for political party?\textless/pre\textgreater \\ \hline
\end{tabular}
\end{table}

\clearpage

\begin{table}[]
\tiny
\centering
\label{my-label}
\begin{tabular}{|l|}
\hline
\textbf{Cluster 4} \\ \hline
Does the constitution contain an explicit declaration regarding the INDEPENDENCE of the central executive organ(s)? \\ \hline
Does the constitution provide for a right to petition for "amparo"? \\ \hline
Does the constitution mention international treaties? \\ \hline
Does the constitution provide for freedom of religion? \\ \hline
Are there provisions for dismissing judges? \\ \hline
Does the constitution prescribe whether or not the meetings of the Legislature are (generally) held in public? \\ \hline
Is the executive identified explicitly as the Head of State or Head of Government? \\ \hline
Does the constitution give the accused a right to silence or protection from self incrimination? \\ \hline
Are there provisions for removing individual legislators? \\ \hline
Does the constitution provide for the right to protection from unjustified restraint (habeas corpus)? \\ \hline
Does the constitution provide for freedom of assembly? \\ \hline
Does the constitution refer to a duty to pay taxes? \\ \hline
Is the military or armed forces mentioned in the constitution? \\ \hline
Does the constitution provide a right to conduct/establish a business? \\ \hline
Does the legislature have the power to investigate the activities of the executive branch? \\ \hline
Does the constitution contain provisions concerning the relationship between the constitution and international law? \\ \hline
Does the constitution provide for freedom of expression or speech? \\ \hline
Does the constitution refer to ownership or possession of natural resources (such as minerals, oil, etc.)? \\ \hline
Does the constitution forbid the detention of debtors? \\ \hline
Does the constitution stipulate that some public office holders take an oath to support or abide by the constitution? \\ \hline
Does the constitution provide for a right of petition? \\ \hline
Does the constitution require a jury or any form of citizen participation in decision making in criminal trials? \\ \hline
Is the state described as either federal, confederal, or unitary? \\ \hline
Does the constitution explicitly state that judicial salaries are protected from governmental intervention? \\ \hline
Does the constitution contain provisions protecting the individual against illegal or ultra-vires administrative actions? \\ \hline
Do constitutional amendments require more than a simple majority by the legislature to be approved? \\ \hline
Does the constitution provide for a central representative body (a legislature)? \\ \hline
Does the constitution prohibit slavery, servitude, or forced labor? \\ \hline
Does the constitution refer to artists or the arts? \\ \hline
Does the constitution provide for freedom of expression or speech? \\ \hline
Does the constitution contain provisions concerning a national or official religion or a national or official church? \\ \hline
Does the constitution refer to "democracy" or "democratic"? \\ \hline
Does the constitution prohibit censorship? \\ \hline
Does the constitution provide for a right to own property? \\ \hline
Does the constitution place any restrictions on the right to vote? \\ \hline
Does the constitution mention bankruptcy law? \\ \hline
Does the legislature have the power to interpellate members of the executive branch?? \\ \hline
Does the constitution provide for freedom of association? \\ \hline
Are there provisions for the secession or withdrawal of parts of the state? \\ \hline
Does the constitution provide for at least one procedure for amending the constitution? \\ \hline
Does the constitution make voting mandatory, at least for some elections? \\ \hline
Are there provisions for dismissing the Head of State? \\ \hline
Does the constitution mention the executive cabinet/ministers? \\ \hline
Does the constitution provide for judicial opinions of the Highest Ordinary Court? \\ \hline
Does the constitution refer to equality before the law, the equal rights of men, or non-discrimination? \\ \hline
Does the constitution specify a census? \\ \hline
Does the constitution provide for naturalized citizens or naturalization? \\ \hline
Does the constitution stipulate that education be free, at least up to some level? \\ \hline
Does the constitution provide for the right to marry? \\ \hline
Does the constitution contain an explicit decree of separation of church and state? \\ \hline
What is the specified level of compensation for expropriation of private property? \\ \hline
Does the constitution stipulate that certain rights are inalienable or inviolable? \\ \hline
Does the constitution contain provisions concerning education? \\ \hline
Does the constitution guarantee academic freedom? \\ \hline
Does the constitution provide for a right of privacy? \\ \hline
Does the constitution regulate the collection of evidence? \\ \hline
Does the constitution provide for a right of testate, or the right to transfer property freely after death? \\ \hline
Does the constitution contain provisions allowing review of the legislation of the constituent units in federations by federal judicial or other central government organs? \\ \hline
Does the constitution mention nulla poena sine lege or the principle that no person should be punished without law? \\ \hline
Does the constitution define the geographic borders/territory of the state? \\ \hline
Does the constitution mention "God" or any other Deities? \\ \hline
Does the constitution refer to artists or the arts? \\ \hline
Can the government expropriate private property under at least some conditions? \\ \hline
Does the constitution refer to a duty of military service? \\ \hline
Are members of the first (or only) chamber elected in the same cohort, or in staggered cohorts? \\ \hline
Does the constitution contain provisions specifying the location of the capital (if so, please specify the location in the comments section)? \\ \hline
Does the constitution mention foreign or international trade? \\ \hline
Does the constitution provide for suspension or restriction of rights during states of emergency? \\ \hline
Is there a mention of telecommunications? \\ \hline
Does the constitution mention the right to transfer property freely? \\ \hline
Are there provisions for the secession or withdrawal of parts of the state? \\ \hline
Does the constitution stipulate that education be compulsory until at least some level? \\ \hline
Does the constitution mention any of the following intellectual property rights: patents? \\ \hline
Does the constitution mention any of the following intellectual property rights: copyrights? \\ \hline
Does the constitution mention any of the following intellectual property rights: trademark? \\ \hline
Does the constitution mention any of the following intellectual property rights: general reference to IP?\textless/pre\textgreater \\ \hline
\end{tabular}
\end{table}

\clearpage

The figure analogous to that in the main paper for k=3 and k=5 are displayed below

\begin{figure}[h]
\centerline{\includegraphics[width=0.75\linewidth]{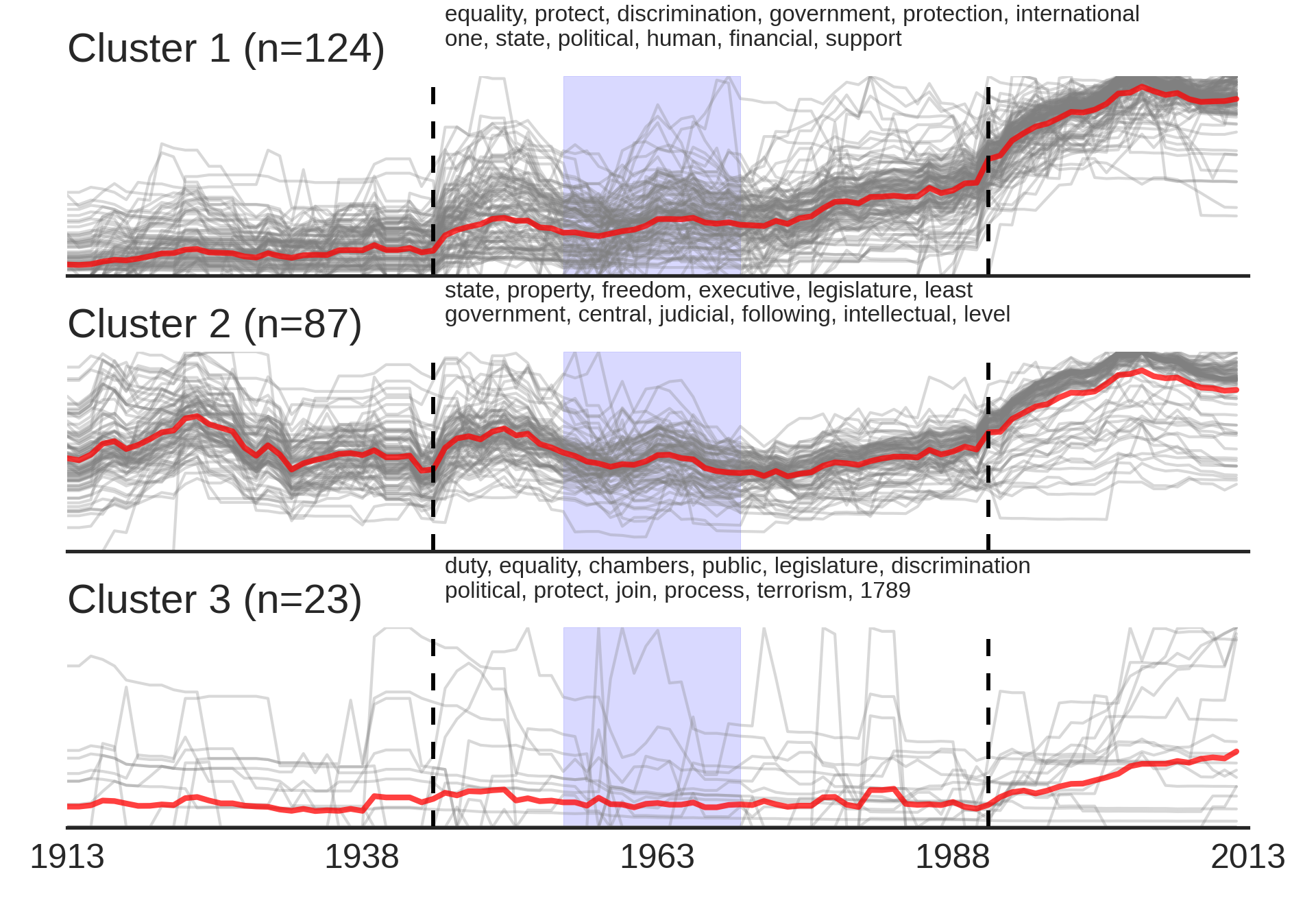}}
\caption{Clusters of provisional time series with child specific provision shown in red. Top terms in descriptions of provisions are included.}
\end{figure}

\begin{figure}[h]
\centerline{\includegraphics[width=0.75\linewidth]{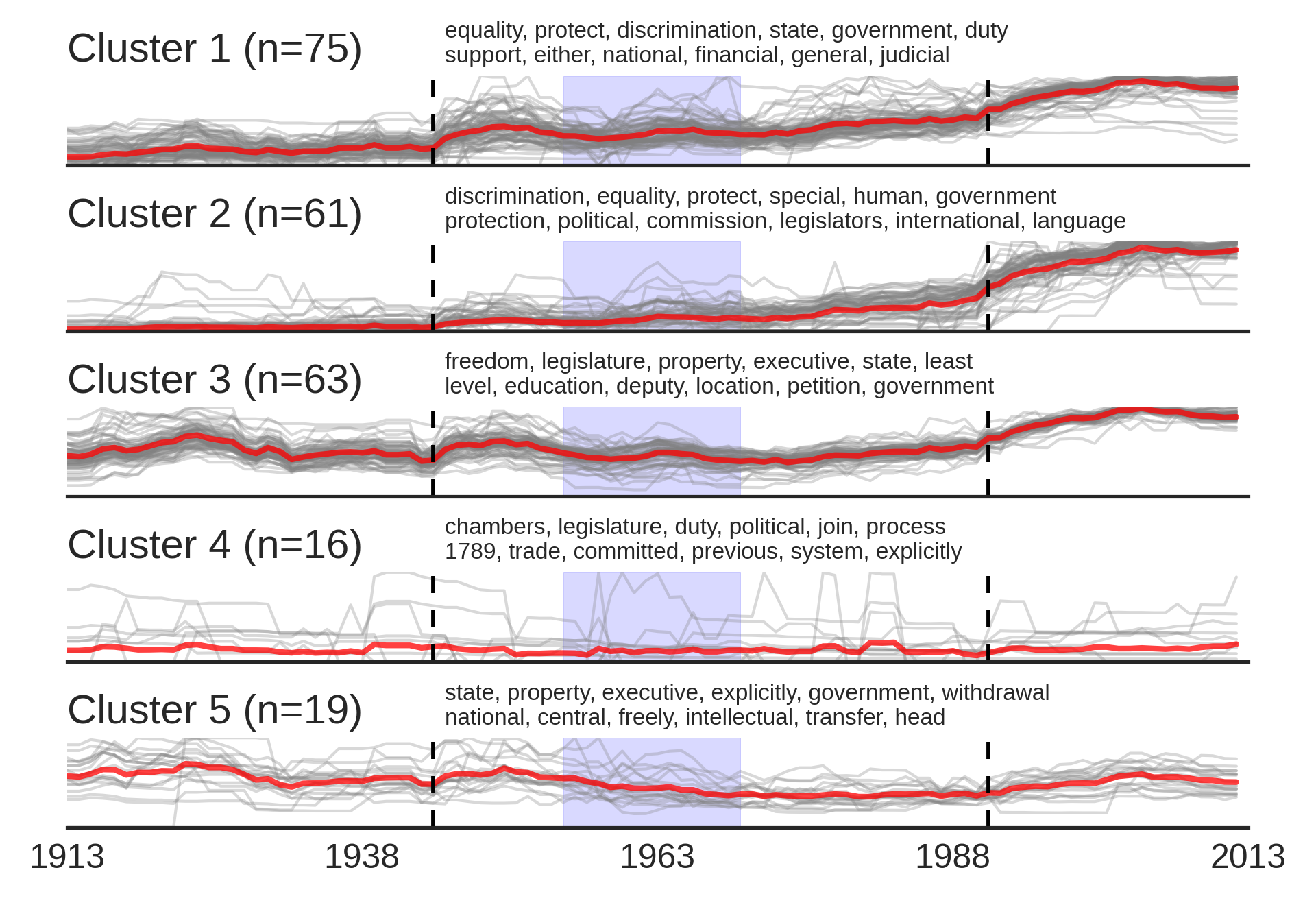}}
\caption{Clusters of provisional time series with child specific provision shown in red. Top terms in descriptions of provisions are included.}
\end{figure}

We present a heatmap of key terms and their appearance in each of the clusters when n=4, this figure accompanies the time series clustering plot in the main paper.

\begin{figure}[h]
\centerline{\includegraphics[width=1.0\linewidth]{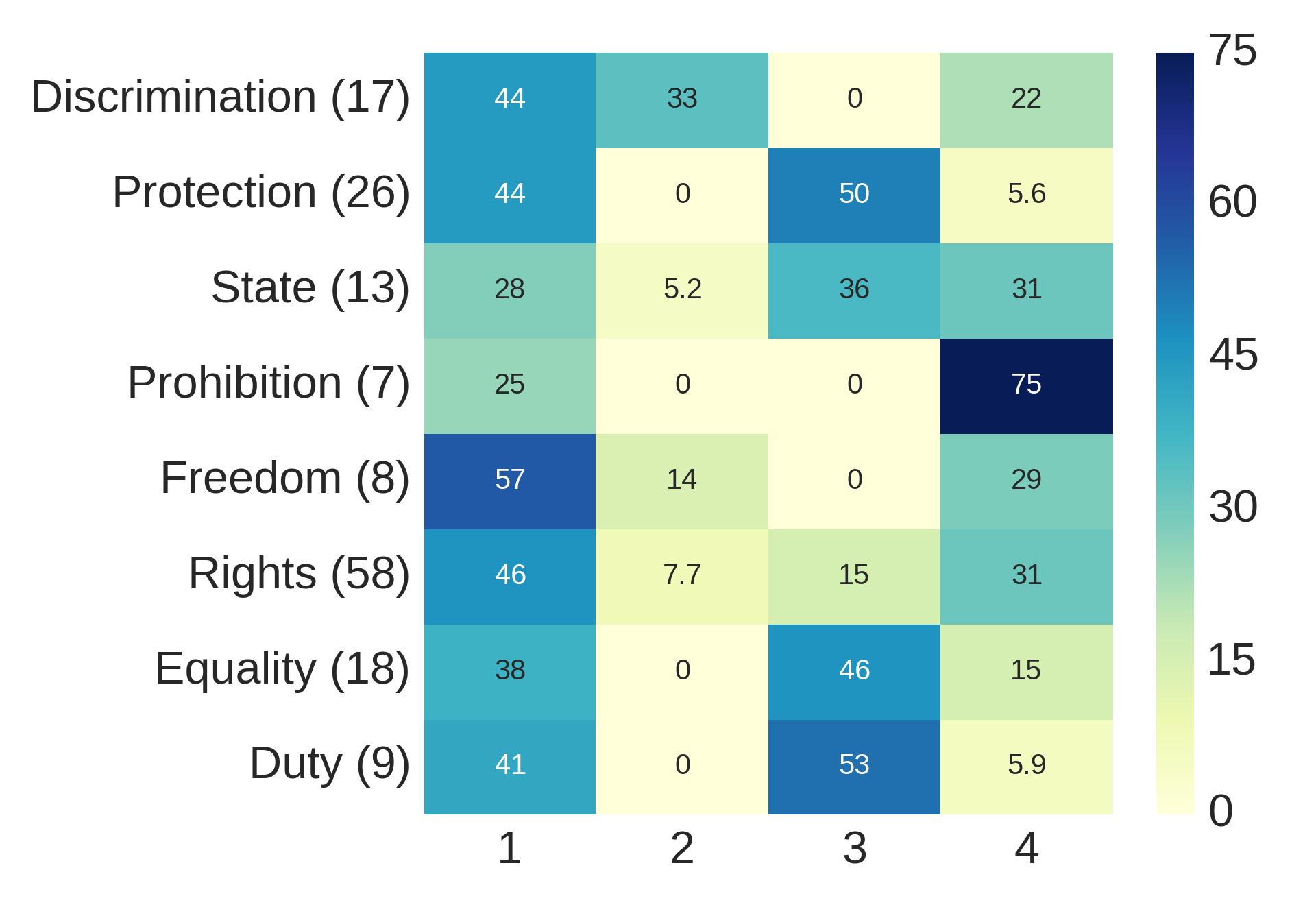}}
\caption{Heatmap showing how frequently key terms such as \textit{discrimination} appear in the provisions in each cluster normalised by how often they appearacross all clusters. \label{time_series_terms}}
\end{figure}

\begin{table}[]
\centering
\caption{Questions and Corresponding codes}
\tiny
\label{my-label}
\begin{tabular}{|l|l|}
\hline
Code & Description \\ \hline
LIFE & Does the constitution provide for a right to life? \\ \hline
HEALTHR & Does the constitution mention the right to health care? \\ \hline
WARAP & Who has the power to approve declarations of war? \\ \hline
INFOACC & Does the constitution provide for an individual right to view government files or documents under at least some conditions? \\ \hline
MATEQUAL & Does the constitution provide for matrimonial equality? \\ \hline
EXAMWIT & Does the constitution provide for the right to examine evidence or confront all witnesses? \\ \hline
CITDEP & Does the constitution grant the government the right to deport citizens or residents? \\ \hline
EXECINDP & Does the constitution contain an explicit declaration regarding the INDEPENDENCE of the central executive organ(s)? \\ \hline
PART & Does the constitution refer to political parties? \\ \hline
AMPARO & "Does the constitution provide for a right to petition for ""amparo""?" \\ \hline
TREAT & Does the constitution mention international treaties? \\ \hline
NAT & Does the constitution contain provisions concerning the relationship between the constitution and international law? \\ \hline
HEALTHF & Does the constitution specify that healthcare should be provided by government free of charge? \\ \hline
REMUNER & Does the constitution provide the right to just remuneration, fair or equal payment for work? \\ \hline
PROFLEG & Does the constitution require that legislators give up any other profession (i.e. work exclusively as legislators)? \\ \hline
JC & Does the constitution contain provisions for a Judicial Council/Commission? \\ \hline
DEMOC & "Does the constitution refer to ""democracy"" or ""democratic""?" \\ \hline
FREEREL & Does the constitution provide for freedom of religion? \\ \hline
JREM & Are there provisions for dismissing judges? \\ \hline
CHILDPRO & Does the constitution guarantee the rights of children? \\ \hline
HOGID & "Is one of the executives explicitly referred to as the ""Head of Government""?" \\ \hline
CRUELTY & Does the constitution prohibit cruel, inhuman, or degrading treatment? \\ \hline
LEISURE & Does the constitution provide for a right of rest and leisure? \\ \hline
PUBMEET & Does the constitution prescribe whether or not the meetings of the Legislature are (generally) held in public? \\ \hline
CHILDWRK & Does the constitution place limits on child employment? \\ \hline
CONCOP & Does the constitution mention judicial opinions of the Constitutional Court? \\ \hline
HOSHOG & Is the executive identified explicitly as the Head of State or Head of Government? \\ \hline
PARTRGHT & Does the constitution provide for a right to form political parties? \\ \hline
MIRANDA & Does the constitution give the accused a right to silence or protection from self incrimination? \\ \hline
TERROR & Is there special mention of terrorism and public security provisions regarding terrorism? \\ \hline
REMLEG & Are there provisions for removing individual legislators? \\ \hline
LEGJOINT & Does the constitution specify that the chambers should meet jointly for any reason? \\ \hline
HABCORP & Does the constitution provide for the right to protection from unjustified restraint (habeas corpus)? \\ \hline
NOMIL & Is there a right to exemption from military service for conscientous objectors to war or other groups? \\ \hline
OCCUPATE & Does the constitution provide for the right to choose ones occupation? \\ \hline
ASSEM & Does the constitution provide for freedom of assembly? \\ \hline
PRTYDUTY & Does the constitution refer to a duty to join a political party? \\ \hline
TAXES & Does the constitution refer to a duty to pay taxes? \\ \hline
SAFEWORK & Does the constitution mention the right to safe/healthy working conditions? \\ \hline
EXINST & Does the constitution contain provisions with regard to any additional central independent regulatory agencies ? \\ \hline
MILITARY & Is the military or armed forces mentioned in the constitution? \\ \hline
BUSINES & Does the constitution provide a right to conduct/establish a business? \\ \hline
EXPOST & Does the constitution prohibit punishment by laws enacted ex post facto ? \\ \hline
SOCECON & Does the constitution use the words (socio-) economic rights or similar? \\ \hline
INVEXE & Does the legislature have the power to investigate the activities of the executive branch? \\ \hline
INTLAW & Does the constitution contain provisions concerning the relationship between the constitution and international law? \\ \hline
ASSETS & Does the constitution require that legislators disclose their earnings and/or assets? \\ \hline
REFEREN & Does the constitution stipulate that some public office holders take an oath to support or abide by the constitution? \\ \hline
EXPRESS & Does the constitution provide for freedom of expression or speech? \\ \hline
LEGSUPR & Is a supermajority needed for passing any legislation? \\ \hline
RESRCE & Does the constitution refer to ownership or possession of natural resources (such as minerals, oil, etc.)? \\ \hline
JUDRETIR & Is there a mandatory retirement age for judges? \\ \hline
DEBTORS & Does the constitution forbid the detention of debtors? \\ \hline
BUILDSOC & Does the constitution refer to a duty of the people to take part in building society or to work for the development of the country? \\ \hline
OATH & Does the constitution stipulate that some public office holders take an oath to support or abide by the constitution? \\ \hline
PETITION & Does the constitution provide for a right of petition? \\ \hline
FALSEIMP & Does the constitution provide for the right of some redress in the case of false imprisonment, arrest, or judicial error? \\ \hline
OMBUDS & Does the constitution provide for an Ombudsman? \\ \hline
\end{tabular}
\end{table}

\begin{table}[]
\centering
\caption{Questions and Corresponding codes cont.}
\tiny
\label{my-label}
\begin{tabular}{|l|l|}
\hline
Code & Description \\ \hline
JURY & Does the constitution require a jury or any form of citizen participation in decision making in criminal trials? \\ \hline
HR & Does the constitution suggest that citizens should have the right to overthrow their government under certain circumstances? \\ \hline
DEVLPERS & Does the constitution provide for an individual's right to self determination or the right to free development of personality? \\ \hline
PROVWORK & Does the constitution mention a state duty to provide work/employment? \\ \hline
FREEMOVE & Does the constitution provide for freedom of movement? \\ \hline
INITIAT & Does the constitution provide for the ability of individuals to propose legislative initiatives (referenda from below)? \\ \hline
FEDUNIT & Is the state described as either federal, confederal, or unitary? \\ \hline
JUDSAL & Does the constitution explicitly state that judicial salaries are protected from governmental intervention? \\ \hline
ILLADMIN & Does the constitution contain provisions protecting the individual against illegal or ultra-vires administrative actions? \\ \hline
OPINION & Does the constitution provide for freedom of opinion, thought, and/or conscience? \\ \hline
AMNDAMAJ & Do constitutional amendments require more than a simple majority by the legislature to be approved? \\ \hline
GOVMED & How does the constitution address the state operation of print or electronic media? \\ \hline
MARKET & "Does the constitution refer to the ""free market,"" ""capitalism,"" or an analogous term?" \\ \hline
LEGISL & Does the constitution provide for a central representative body (a legislature)? \\ \hline
SLAVE & Does the constitution prohibit slavery, servitude, or forced labor? \\ \hline
CITREN & Do citizens have the right to renounce their citizenship? \\ \hline
TRUTHCOM & Does the constitution provide for a commission for truth and reconciliation? \\ \hline
CIVIL & Does the constitution include provisions for the meritocratic recruitment of civil servants (e.g. exams or credential requirements)? \\ \hline
MEDCOM & Does the constitution mention a special regulatory body/institution to oversee the media market? \\ \hline
UNAMEND & Are any parts of the constitution unamendable? \\ \hline
JUDIND & Does the constitution contain an explicit declaration regarding the independence of the central judicial organ(s)? \\ \hline
SCIENCE & Does the constitution refer to artists or the arts? \\ \hline
PRESS & Does the constitution provide for freedom of expression or speech? \\ \hline
RADIO & Does the constitution refer to radio? \\ \hline
OFFREL & Does the constitution contain provisions concerning a national or official religion or a national or official church? \\ \hline
EM & "Does the constitution refer to ""democracy"" or ""democratic""?" \\ \hline
PREVLEAD & Does the constitution mention anything about crimes committed by the previous regime? \\ \hline
PARTPRF & Does the constitution express a preference for one or more political parties? \\ \hline
COUNS & Does the constitution provide the right to counsel if one is indicted or arrested? \\ \hline
PARTPRH & Does the constitution prohibit one or more political parties? \\ \hline
CENSOR & Does the constitution prohibit censorship? \\ \hline
PROPRGHT & Does the constitution provide for a right to own property? \\ \hline
VOTEUN & Does the constitution make a claim to universal adult suffrage? \\ \hline
SPEEDTRI & Does the constitution provide for the right to a speedy trial? \\ \hline
VOTERES & Does the constitution place any restrictions on the right to vote? \\ \hline
FLAG & Does the constitution contain provisions concerning the national flag? \\ \hline
LHQUOTA & Does the constitution stipulate a quota for representation of certain groups in the first (or only) chamber? \\ \hline
ENV & Does the constitution refer to protection or preservation of the environment? \\ \hline
FREECOMP & Does the constitution provide the right to a free and/or competitive market? \\ \hline
SELFDET & Does the constitution provide for a people's right of self-determination? \\ \hline
HOSID & "Is one of the executives explicitly referred to as the ""Head of State""?" \\ \hline
FNDFAM & Does the constitution provide the ribght to found a family? \\ \hline
BANKRUPT & Does the constitution mention bankruptcy law? \\ \hline
ACHIGHED & Does the constitution guarantee equal access to higher education? \\ \hline
INTEXEC & Does the legislature have the power to interpellate members of the executive branch?? \\ \hline
ASSOC & Does the constitution provide for freedom of association? \\ \hline
SECCESS & Are there provisions for the secession or withdrawal of parts of the state? \\ \hline
STANDLIV & Does the constitution provide for a right to an adequate or reasonable standard of living? \\ \hline
STRIKE & Does the constitution provide for a right to strike? \\ \hline
RGHTAPP & Do defendants have the right to appeal judicial decisions? \\ \hline
MOTTO & Does the constitution contain a national motto? \\ \hline
IMMUNITY & Does the constitution provide for immunity for the members of the Legislature under some conditions? \\ \hline
JOINTRDE & Does the constitution provide for the right to form or to join trade unions? \\ \hline
AMEND & Does the constitution provide for at least one procedure for amending the constitution? \\ \hline
TRILANG & Does the constitution specify the trial has to be in a language the accused understands or the right to an interpreter if the accused cannot understand the language? \\ \hline
LIBEL & Does the constitution provide for the right of protection of one's reputation from libelous actions? \\ \hline
PRISONRG & Does the constitution require that the names of those imprisoned be entered in a public registry? \\ \hline
COMPVOTE & Does the constitution make voting mandatory, at least for some elections? \\ \hline
CONREM & Does the constitution mention any special procedures for removing members of the constitutional court? \\ \hline
EXCRIM & Does the constitution provide for the extradition of suspected or convicted criminals to other countries? \\ \hline
HOSDISS & Are there provisions for dismissing the Head of State? \\ \hline
CUSTLAW & "Does the constitution refer to ""customary"" international law or the ""law of nations""?" \\ \hline
RULELAW & Does the constitution contain a general statement regarding rule of law, legality, or Rechtsstaat (the German equivalent)? \\ \hline
PREREL & Does the constitution provide for the right/possibility of pre-trial release? \\ \hline
ETHINCL & Does the constitution contain provisions concerning national integration of ethnic communities? \\ \hline
SAMESEXM & Does the constitution provide the right for same sex marriages? \\ \hline
LANG & In what language is the source document written (not the original, but the one used for coding)? \\ \hline
PRESINOC & Is there a presumption of innocence in trials? \\ \hline
ANTHEM & Does the constitution contain provisions concerning the national anthem? \\ \hline
WORK & Does the constitution mention a state duty to provide work/employment? \\ \hline
ARMS & Does the constitution provide for the right to bear arms? \\ \hline
CABINET & Does the constitution mention the executive cabinet/ministers? \\ \hline
DUEPROC & Does the constitution explicitly mention due process? \\ \hline
CULTRGHT & Does the constitution refer to a state duty to protect or promote culture or cultural rights? \\ \hline
SOCSEC & Does the constitution refer to the social security of the society or nation? \\ \hline
OPGROUP & Does the constitution provide for positive obligations to transfer wealth to, or provide opportunity for, particular groups? \\ \hline
PREVCOND & Does the constitution refer to social, political, or economic conditions in the time before the birth of the state or in the time of a former constitution? \\ \hline
DIGNITY & "Does the constitution refer to the ""dignity of man"" or human ""dignity""?" \\ \hline
HOCOP & Does the constitution provide for judicial opinions of the Highest Ordinary Court? \\ \hline
SOLID & "Does the constitution refer to ""fraternity"" or ""solidarity""?" \\ \hline
INHERIT & Does the constitution provide for inheritance rights? \\ \hline
EQUAL & Does the constitution refer to equality before the law, the equal rights of men, or non-discrimination? \\ \hline
RESENEX & Does the constitution restrict entry or exit of the states borders? \\ \hline
INTORGS & Does the constitution contain provisions concerning international organizations? \\ \hline
CENSUS & Does the constitution specify a census? \\ \hline
HOUSENUM & How many chambers or houses does the Legislature contain? \\ \hline
SHELTER & Does the constitution provide for the right to shelter or housing? \\ \hline
SCIFREE & Does the constitution provide for a right to enjoy the benefits of scientific progress? \\ \hline
NATCIT & Does the constitution provide for naturalized citizens or naturalization? \\ \hline
EDFREE & Does the constitution stipulate that education be free, at least up to some level? \\ \hline
MARRIAGE & Does the constitution provide for the right to marry? \\ \hline
SEPREL & Does the constitution contain an explicit decree of separation of church and state? \\ \hline
OVERSGHT & Does the constitution provide for an electoral commission or electoral court to oversee the election process? \\ \hline
EXPRCOMP & What is the specified level of compensation for expropriation of private property? \\ \hline
INALRGHT & Does the constitution stipulate that certain rights are inalienable or inviolable? \\ \hline
CONRIGHT & Does the constitution mention consumer rights or consumer protection? \\ \hline
DOUBJEP & Does the constitution provide for the prohibition of double jeopardy (i.e., being tried for the same crime twice)? \\ \hline
EDUCATE & Does the constitution contain provisions concerning education? \\ \hline
SOCIALSM & "Does the constitution refer to ""socialism"" or ""socialist""?" \\ \hline
ACFREE & Does the constitution guarantee academic freedom? \\ \hline
PRIVACY & Does the constitution provide for a right of privacy? \\ \hline
TRADEUN & Does the constitution refer to a duty to join trade unions? \\ \hline
JUVENILE & Does the constitution give juveniles special rights/status in the criminal justice process? \\ \hline
EVIDENCE & Does the constitution regulate the collection of evidence? \\ \hline
OVERTHRW & Does the constitution suggest that citizens should have the right to overthrow their government under certain circumstances? \\ \hline
ASYLUM & Does the constitution contain provisions for the protection of stateless individuals, refugees from other states, or the right to asylum? \\ \hline
JUDPREC & Does the constitution stipulate that courts have to take into account decisions of higher courts? \\ \hline
TESTATE & Does the constitution provide for a right of testate, or the right to transfer property freely after death? \\ \hline
PROVHLTH & Does the constitution mention a state duty to provide health care? \\ \hline
TORTURE & Does the constitution prohibit torture? \\ \hline
TV & Does the constitution refer to television? \\ \hline
FEDREV & Does the constitution contain provisions allowing review of the legislation of the constituent units in federations by federal judicial or other central government organs? \\ \hline
WOLAW & Does the constitution mention nulla poena sine lege or the principle that no person should be punished without law? \\ \hline
TERR & Does the constitution define the geographic borders/territory of the state? \\ \hline
UHELSYS & Does the constitution specify the electoral system for the Second Chamber? \\ \hline
GOD & "Does the constitution mention ""God"" or any other Deities?" \\ \hline
ARTISTS & Does the constitution refer to artists or the arts? \\ \hline
EXPROP & Can the government expropriate private property under at least some conditions? \\ \hline
BINDING & Are rights provisions binding on private parties as well as the state? \\ \hline
MILSERV & Does the constitution refer to a duty of military service? \\ \hline
FORINVES & "Does the constitution mention ""foreign investment"" or ""foreign capital""?" \\ \hline
LHCOHORT & Are members of the first (or only) chamber elected in the same cohort, or in staggered cohorts? \\ \hline
PUBTRI & Does the constitution generally require public trials? \\ \hline
ECONPLAN & Does the constitution mention the adoption of national economic plans? \\ \hline
\end{tabular}
\end{table}

\begin{table}[]
\centering
\caption{Questions and Corresponding codes cont.}
\tiny
\label{my-label}
\begin{tabular}{|l|l|}
\hline
Code & Description \\ \hline

CAPITAL & Does the constitution contain provisions specifying the location of the capital (if so, please specify the location in the comments section)? \\ \hline
FORTRAD & Does the constitution mention foreign or international trade? \\ \hline
BANK & Does the constitution contain provisions for a central bank? \\ \hline
FAIRTRI & Does the constitution provide the right to a fair trial? \\ \hline
EMRIGHTS & Does the constitution provide for suspension or restriction of rights during states of emergency? \\ \hline
TELECOM & Is there a mention of telecommunications? \\ \hline
TRANSFER & Does the constitution mention the right to transfer property freely? \\ \hline
LANGPROT & Does the constitution refer to the protection of different languages? \\ \hline
ACCESS & Are there provisions for the secession or withdrawal of parts of the state? \\ \hline
EDCOMP & Does the constitution stipulate that education be compulsory until at least some level? \\ \hline
UHQUOTA & Does the constitution stipulate a quota for representation of certain groups in the Second Chamber? \\ \hline
FREEELEC & Does the constitution prescribe that electoral ballots be secret? \\ \hline
INTRGHT\_1 & Does the constitution refer to the UN universal Declaration on Human Rights (1948)? \\ \hline
INTRGHT\_2 & Does the constitution refer to the French declaration of rights (1789)? \\ \hline
INTRGHT\_3 & Does the constitution refer to the UN charter Article 45 (1945)? \\ \hline
INTRGHT\_4 & Does the constitution refer to the European Convention for the Protection of Human Rights and Fundamental Freedoms (1950)? \\ \hline
INTRGHT\_5 & Does the constitution refer to the International Covenant on Civil and Political Rights (1966)? \\ \hline
INTRGHT\_6 & Does the constitution refer to the International Covenant on Economic and Social Rights (1966)? \\ \hline
INTRGHT\_7 & Does the constitution refer to the American Convention on Human Rights (1969)? \\ \hline
INTRGHT\_8 & Does the constitution refer to the Helsinki Accords (1966)? \\ \hline
INTRGHT\_9 & Does the constitution refer to the African Charter on Human People's Rights (1981)? \\ \hline
CC & Does the constitution contain provisions for a counter corruption commission? \\ \hline
HOGDISS & Are there provisions for dismissing the Head of Government? \\ \hline
DEPEXEC & Does the constitution specify a deputy executive of any kind (e.g., deputy prime minister, vice president)? \\ \hline
FINSUP\_1 & Does the constitution provide for either general or financial support by the government for the elderly? \\ \hline
FINSUP\_2 & Does the constitution provide for either general or financial support by the government for the unemployed? \\ \hline
FINSUP\_3 & Does the constitution provide for either general or financial support by the government for the disabled? \\ \hline
FINSUP\_4 & Does the constitution provide for either general or financial support by the government for children, orphans? \\ \hline
GRJURY & Is there citizen involvement in the indicting process (such as a grand jury)? \\ \hline
CIVMAR & Is there a constitutional provision for civil marriage? \\ \hline
VICRIGHT & Is there a special mention of victims rights in the constitution? \\ \hline
COUNSCOS & If counsel is provided, is it provided at the state's expense? \\ \hline
INTPROP\_1 & Does the constitution mention any of the following intellectual property rights: patents? \\ \hline
INTPROP\_2 & Does the constitution mention any of the following intellectual property rights: copyrights? \\ \hline
INTPROP\_3 & Does the constitution mention any of the following intellectual property rights: trademark? \\ \hline
INTPROP\_4 & Does the constitution mention any of the following intellectual property rights: general reference to IP? \\ \hline
EQUALGR\_1 & Does the constitution protect from discrimination/provide for equality for gender? \\ \hline
EQUALGR\_2 & Does the constitution protect from discrimination/provide for equality for nationality? \\ \hline
EQUALGR\_3 & Does the constitution protect from discrimination/provide for equality for country of origin? \\ \hline
EQUALGR\_4 & Does the constitution protect from discrimination/provide for equality for race? \\ \hline
EQUALGR\_5 & Does the constitution protect from discrimination/provide for equality for language? \\ \hline
EQUALGR\_6 & Does the constitution protect from discrimination/provide for equality for religion? \\ \hline
EQUALGR\_7 & Does the constitution protect from discrimination/provide for equality for sexual orientation? \\ \hline
EQUALGR\_8 & Does the constitution protect from discrimination/provide for equality for age? \\ \hline
EQUALGR\_9 & Does the constitution protect from discrimination/provide for equality for mental or physical disability? \\ \hline
EQUALGR\_10 & Does the constitution protect from discrimination/provide for equality for color? \\ \hline
EQUALGR\_11 & Does the constitution protect from discrimination/provide for equality for creed/beliefs? \\ \hline
EQUALGR\_12 & Does the constitution protect from discrimination/provide for equality for social status? \\ \hline
EQUALGR\_13 & Does the constitution protect from discrimination/provide for equality for financial/propety ownership? \\ \hline
EQUALGR\_14 & Does the constitution protect from discrimination/provide for equality for tribe/clan? \\ \hline
EQUALGR\_15 & Does the constitution protect from discrimination/provide for equality for political party? \\ \hline
EQUALGR\_16 & Does the constitution protect from discrimination/provide for equality for parentage? \\ \hline
\end{tabular}
\end{table}

\clearpage

\section{Provisions for Children \& Young People}

\begin{figure*}[h]
\centerline{\includegraphics[width=1.0\linewidth]{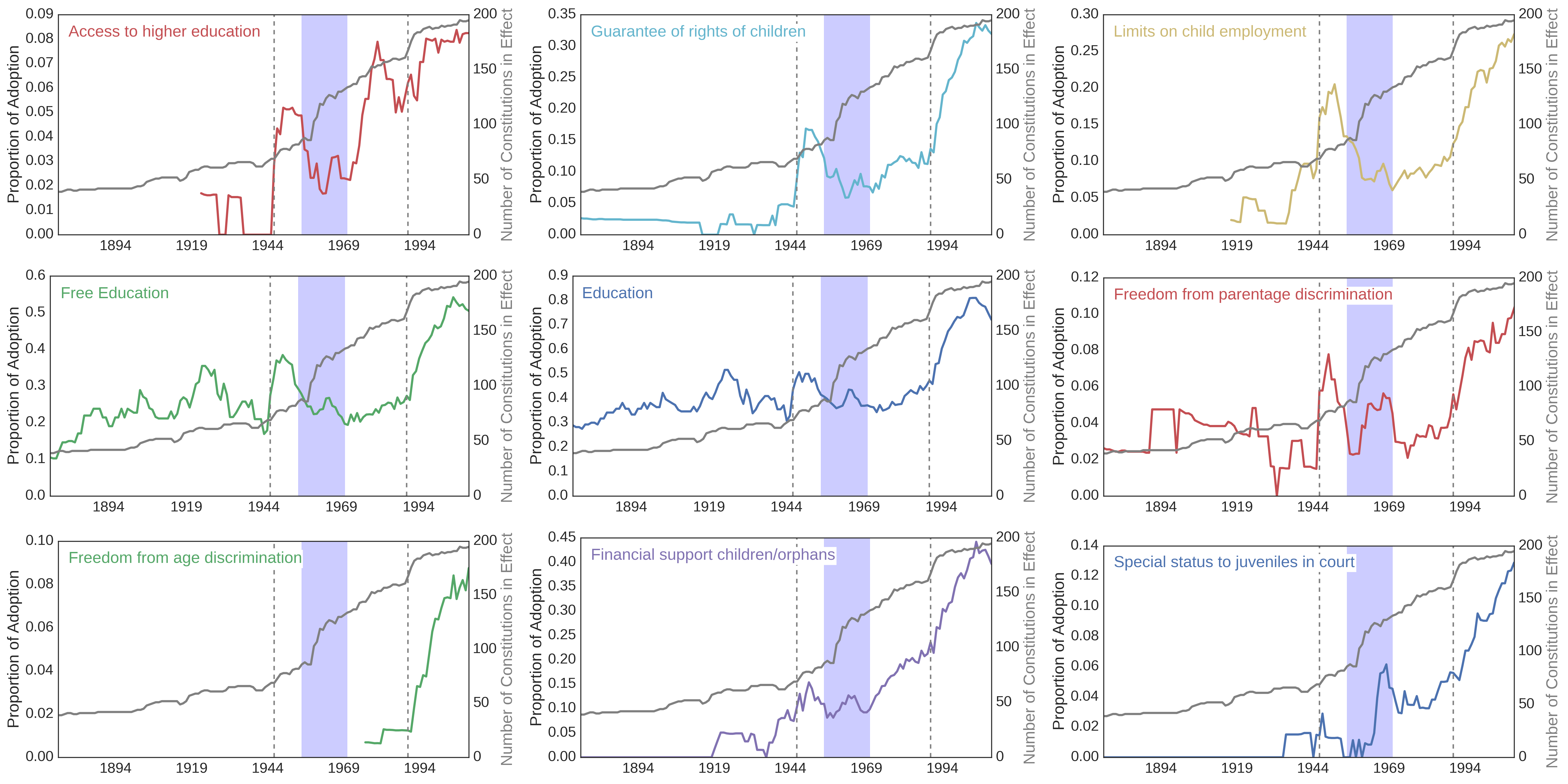}}
\caption{Proportion of adoption, defined as the proportion of independent countries in existence that year including that provision in their constitution, of nine provisions relevant to children over time (coloured lines) each compared to the number of countries (dark line).}\label{child_other}
\end{figure*}

\clearpage

\section{Network-driven and Temporal-driven Provisional Adoption}

\subsection{Network Character of Provisions}

Written explicitly we compute the expected conditional probability of a random pair of countries coadopting a provision given their constitutions were written in the same yer as

\begin{equation}
P(\textrm{coadopt}\mid\textrm{year of writing})=\frac{P(\textrm{coadopt}\cap\textrm{year of writing})}{P(\textrm{year of writing})}
\end{equation}

\begin{equation}
E(P(\textrm{coadopt}\mid\textrm{year}))=\sum_{y}P(\textrm{coadopt}\mid\textrm{year=y})P(\textrm{year=y})
\end{equation}

\subsection{Temporal Character of Provisions}

The expected probability of temporal coadoption of a provision adopted by $n_{prov}$ out of $n$ countries in years $[y_{1}...y_{n_{prov}}]$ is defined as

\begin{equation}
\sum_{i=1,n}\sum_{j=i+1,n} \delta (y_{i},y_{j})/(n\times(n-1))
\end{equation}

\subsection{Network-Temporal Disentanglement}

We further investigate the dynamics of provisions classified as being dominated by temporal or network effects. This classification is strongly confounded by the highly non-uniform number of countries per year. Here, in analogy to forces acting on physical bodies, we consider the rate of change of adoption of a provision with respect to the rate of increase in countries. Written explicitly

\begin{equation}
v_{provision}-v_{countries}=\frac{\Delta n_{provisions}}{\Delta t}-
\frac{\Delta n_{countries}}{\Delta t}
\end{equation}

Since the proportion of countries adopting a provision can never exceed unity, the quantity $n_{provisions}-n_{countries}$ is always negative. In analogy with Newton's first law of motion which states that \textit{a body will continue to remain at rest or to move at a constant velocity unless acted upon by an external force}, we consider the increase in provisional adoption relative to the increase in countries in order to identify periods of time in which a strong exogeneous force acts to increase the rate of adoption.\\

\begin{figure*}[h]
\centerline{\includegraphics[width=0.8\linewidth]{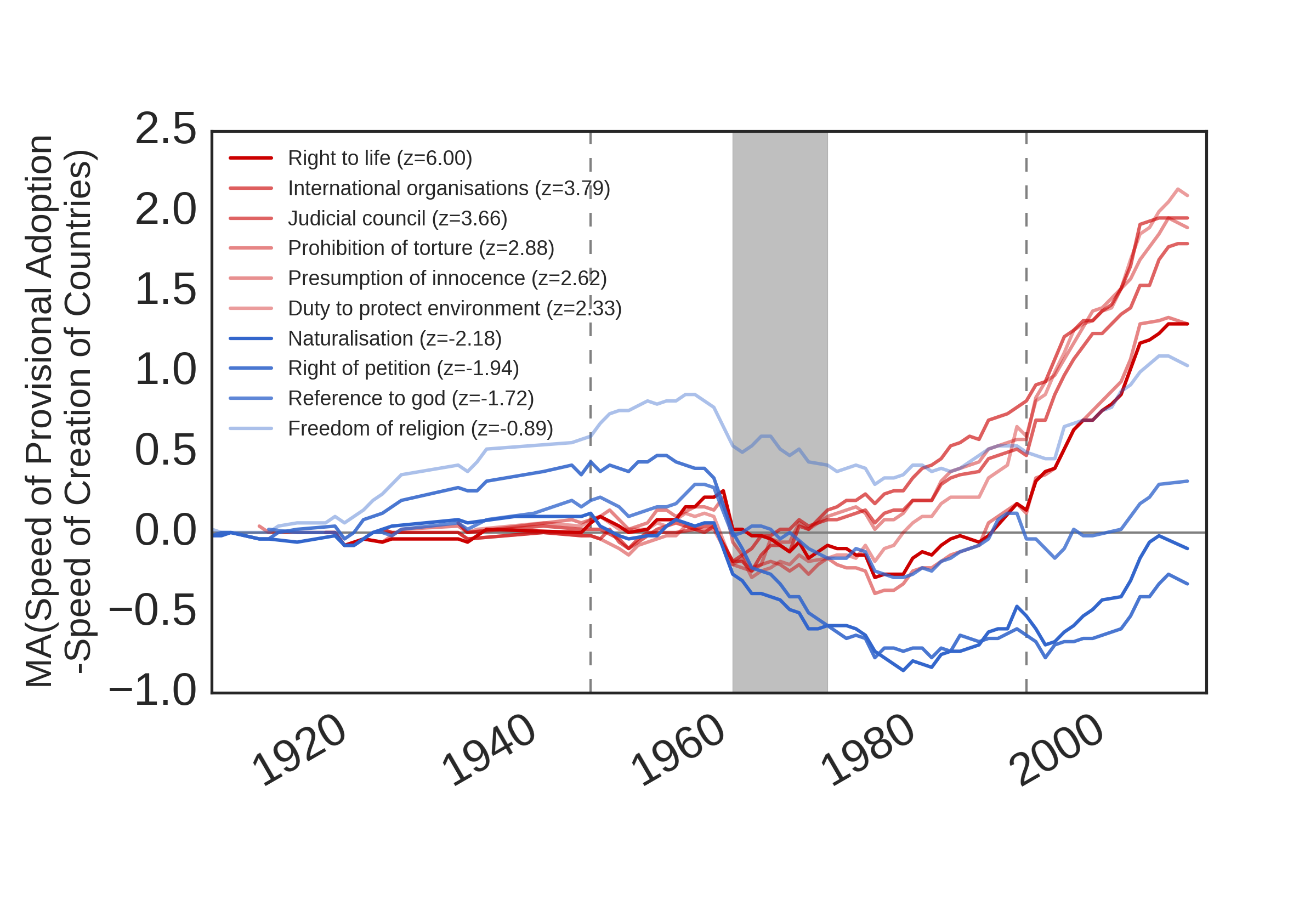}}
\caption{Time series of speed of provisional adoption relative to speed of country creation. Provisions classified as temporal/network driven in the main paper are labelled red/blue. Opacity is relative to magnitude of z-score. Provisional adoption speeds are defined relative to the value on the year of first adoption.}\label{provision_speed}
\end{figure*}

Figure (\ref{provision_speed}) shows the 50 year moving average of the relative speeds. A clear clustering in temporal and network driven provisions is seen. Temporal provisions enjoy a steep rise in recent decades whereas network provisions oscillate less strongly around zero.

\begin{table}[]
  \tiny
  \centering
  \begin{tabularx}{20cm}{@{}>{\bfseries}l@{\hspace{.5em}}X@{}}
    Z-Score & Provision\\\hline
    -2.181 & Does the constitution provide for a right of petition?\\ \hline
-1.944 & Does the constitution provide for naturalized citizens or naturalization?\\ \hline
-1.740 & Are there provisions for dismissing judges?\\ \hline
-1.725 & Does the constitution contain provisions concerning education?\\ \hline
-1.715 & Does the constitution provide for freedom of religion?\\ \hline
-1.594 & Does the constitution specify a deputy executive of any kind (e.g., deputy prime minister, vice president)?\\ \hline
-1.566 & Does the constitution stipulate that some public office holders take an oath to support or abide by the constitution?\\ \hline
-1.528 & Does the constitution provide for freedom of assembly?\\ \hline
-1.391 & Does the constitution provide for freedom of association?\\ \hline
-1.391 & Does the constitution explicitly state that judicial salaries are protected from governmental intervention?\\ \hline
-1.387 & Does the constitution mention international treaties?\\ \hline
-1.369 & Does the constitution provide for a central representative body (a legislature)?\\ \hline
-1.342 & Are there provisions for removing individual legislators?\\ \hline
-1.319 & Does the constitution refer to equality before the law, the equal rights of men, or non-discrimination?\\ \hline
-1.232 & Does the constitution provide for the right/possibility of pre-trial release?\\ \hline
-1.150 & Does the constitution provide for the prohibition of double jeopardy (i.e., being tried for the same crime twice)?\\ \hline
-1.103 & Does the constitution give the accused a right to silence or protection from self incrimination?\\ \hline
-1.099 & Does the constitution provide for freedom of expression or speech?\\ \hline
-1.085 & Does the constitution mention the executive cabinet/ministers?\\ \hline
-1.069 & Does the constitution provide for a right to own property?\\ \hline
-0.987 & Does the constitution contain provisions concerning the relationship between the constitution and international law?\\ \hline
-0.974 & Does the constitution provide for suspension or restriction of rights during states of emergency?\\ \hline
-0.962 & Is the military or armed forces mentioned in the constitution?\\ \hline
-0.951 & Does the constitution specify a census?\\ \hline
-0.947 & Does the legislature have the power to interpellate members of the executive branch??\\ \hline
-0.909 & Does the constitution regulate the collection of evidence?\\ \hline
-0.898 & Does the constitution provide for freedom of expression or speech?\\ \hline
-0.887 & "Does the constitution mention ""God"" or any other Deities?"\\ \hline
-0.849 & Does the constitution generally require public trials?\\ \hline
-0.832 & Does the constitution provide for at least one procedure for amending the constitution?\\ \hline
-0.831 & Can the government expropriate private property under at least some conditions?\\ \hline
-0.823 & Does the constitution mention nulla poena sine lege or the principle that no person should be punished without law?\\ \hline
-0.810 & Are members of the first (or only) chamber elected in the same cohort, or in staggered cohorts?\\ \hline
-0.784 & Does the constitution require a jury or any form of citizen participation in decision making in criminal trials?\\ \hline
-0.758 & Does the constitution refer to a duty to pay taxes?\\ \hline
-0.745 & Does the constitution provide for a right of privacy?\\ \hline
-0.724 & Does the constitution provide for the extradition of suspected or convicted criminals to other countries?\\ \hline
-0.698 & Does the constitution guarantee academic freedom?\\ \hline
-0.697 & Does the constitution refer to ownership or possession of natural resources (such as minerals, oil, etc.)?\\ \hline
-0.691 & Does the constitution mention the right to transfer property freely?\\ \hline
-0.667 & Does the constitution refer to a duty of military service?\\ \hline
-0.659 & Does the legislature have the power to investigate the activities of the executive branch?\\ \hline
-0.651 & Does the constitution place any restrictions on the right to vote?\\ \hline
-0.617 & Does the constitution refer to artists or the arts?\\ \hline
-0.609 & "Does the constitution refer to ""customary"" international law or the ""law of nations""?"\\ \hline
-0.563 & Does the constitution contain provisions protecting the individual against illegal or ultra-vires administrative actions?\\ \hline
-0.550 & Does the constitution provide for immunity for the members of the Legislature under some conditions?\\ \hline
-0.543 & Does the constitution mention bankruptcy law?\\ \hline
-0.541 & Does the constitution provide for the right to protection from unjustified restraint (habeas corpus)?\\ \hline
-0.535 & Does the constitution prohibit slavery, servitude, or forced labor?\\ \hline
-0.534 & Does the constitution include provisions for the meritocratic recruitment of civil servants (e.g. exams or credential requirements)?\\ \hline
-0.507 & Does the constitution provide the right to a fair trial?\\ \hline
-0.506 & Does the constitution provide for the right to a speedy trial?\\ \hline
-0.505 & Does the constitution stipulate that certain rights are inalienable or inviolable?\\ \hline

\end{tabularx}
\caption{Large negative Zscore provisions i.e. Cluster dependent}

\end{table}

\clearpage

\begin{table}[]
  \tiny
  \centering
  \begin{tabularx}{20cm}{@{}>{\bfseries}l@{\hspace{.5em}}X@{}}
    Z-Score & Provision\\\hline

    6.00 & Does the constitution refer to protection or preservation of the environment?\\ \hline
    3.98 & Does the constitution prohibit cruel, inhuman, or degrading treatment?\\ \hline
    3.79 & Does the constitution contain provisions for a Judicial Council/Commission?\\ \hline
    3.66 & Is there a presumption of innocence in trials?\\ \hline
    2.87 & Does the constitution contain provisions concerning international organizations?\\ \hline
    2.61 & Does the constitution prohibit torture?\\ \hline
    2.33 & Does the constitution provide for a right to life?\\ \hline
    2.07 & Does the constitution stipulate that some public office holders take an oath to support or abide by the constitution?\\ \hline
    2.04 & Does the constitution provide for a right to form political parties?\\ \hline
    2.02 & Does the constitution contain a general statement regarding rule of law, legality, or Rechtsstaat (the German equivalent)?\\ \hline
    1.98 & "Does the constitution refer to the ""dignity of man"" or human ""dignity""?"\\ \hline
    1.86 & Does the constitution refer to a state duty to protect or promote culture or cultural rights?\\ \hline
    1.80 & Does the constitution refer to the protection of different languages?\\ \hline
    1.76 & Does the constitution contain provisions concerning the national anthem?\\ \hline
    1.72 & Does the constitution mention the right to health care?\\ \hline
    1.58 & Does the constitution provide the right to counsel if one is indicted or arrested?\\ \hline
    1.52 & Does the constitution provide for a right to strike?\\ \hline
    1.35 & Does the constitution provide the right to found a family?\\ \hline
    1.33 & Does the constitution protect from discrimination/provide for equality for race?\\ \hline
    1.23 & Does the constitution protect from discrimination/provide for equality for social status?\\ \hline
    1.20 & Does the constitution provide for an Ombudsman?\\ \hline
    1.12 & Does the constitution provide for the right to form or to join trade unions?\\ \hline
    1.10 & Does the constitution guarantee the rights of children?\\ \hline
    1.10 & Does the constitution protect from discrimination/provide for equality for gender?\\ \hline
    1.09 & Does the constitution provide for an individual right to view government files or documents under at least some conditions?\\ \hline
    1.06 & Does the constitution provide for either general or financial support by the government for the disabled?\\ \hline
    1.05 & Does the constitution provide for the right of protection of one's reputation from libelous actions?\\ \hline
    1.05 & Does the constitution refer to political parties?\\ \hline
    1.00 & Does the constitution provide for freedom of movement?\\ \hline
    0.98 & Does the constitution protect from discrimination/provide for equality for religion?\\ \hline
    0.89 & Does the constitution protect from discrimination/provide for equality for language?\\ \hline
    0.88 & Does the constitution mention the adoption of national economic plans?\\ \hline
    0.81 & Does the constitution provide the right to just remuneration, fair or equal payment for work?\\ \hline
    0.81 & Does the constitution provide a right to conduct/establish a business?\\ \hline
    0.78 & Are there provisions for dismissing the Head of Government?\\ \hline
    0.76 & Does the constitution contain provisions concerning the national flag?\\ \hline
    0.75 & Does the constitution use the words (socio-) economic rights or similar?\\ \hline
    0.75 & Does the constitution protect from discrimination/provide for equality for country of origin?\\ \hline
    0.69 & "Is one of the executives explicitly referred to as the ""Head of State""?"\\ \hline
    0.69 & Does the constitution provide for either general or financial support by the government for children, orphans?\\ \hline
    0.67 & Does the constitution contain provisions for a central bank?\\ \hline
    0.64 & Does the constitution protect from discrimination/provide for equality for political party?\\ \hline
    0.63 & In what language is the source document written (not the original, but the one used for coding)?\\ \hline
    0.63 & Does the constitution provide for a right of rest and leisure?\\ \hline
    0.62 & Does the constitution mention the right to safe/healthy working conditions?\\ \hline
    0.60 & Does the constitution provide for either general or financial support by the government for the elderly?\\ \hline
    0.54 & Does the constitution stipulate that education be compulsory until at least some level?\\ \hline
    0.52 & "Does the constitution refer to ""fraternity"" or ""solidarity""?"\\ \hline
    0.52 & Is there a right to exemption from military service for conscientous objectors to war or other groups?\\ \hline

  \end{tabularx}
  \caption{Large Z-score provisions i.e. Time dependent}
  \end{table}

  \section{Provision Hierarchy and Ranking}

  We investigate the presence of hierarchy: that is to say that the adoption of X tends to follow the adoption of Y. This is formalised as follows: first a network of provisions is constructed with edges between provision $i$ and $j$ determined by how often $i$ is adopted before $j$ considering all countries adopting both $i$ and $j$.

  \begin{equation}
  w_{i,j}=\sum_{c \in \mathcal{C}_{i}\,\cap\,\mathcal{C}_{j}} \phi^{c}(i,j)
  \end{equation}

  Where $c$ sums over all countries adopting both $i$ and $j$ (i.e. the intersection of $\mathcal{C}_{i}$ and $\mathcal{C}_{j}$) and $\phi^{c}(i,j)$ is a binary valued variable indicating if country $c$ adopts provision $i$ strictly before $j$. Correspondingly for countries

  \begin{equation}
  w_{i,j}=\sum_{p \in (\mathcal{P}_{i}\,\cap\,\mathcal{P}_{j})} \phi^{\,p}(i,j)
  \end{equation}

  Where $p$ sums over all provisions adopted by both $i$ and $j$ (i.e. the intersection of $\mathcal{P}_{i}$ and $\mathcal{P}_{j}$) and $\phi^{\,p}(i,j)$ is a binary valued variable indicating if country $i$ adopts $p$ before $j$ adopts $p$.

  \begin{equation}
  \begin{split}
  \phi^{\,p}(i,j) &\,= 1,\, t^{c,i}_{\textrm{adoption}}<t^{c,j}_{\textrm{adoption}}\\
  &=0\,,\, \textrm{otherwise}
  \end{split}
  \end{equation}

  Where $t^{c,i}_{adoption}$ is the year of first adoption of provision $i$ in country $c$.\\

  From this directed network, we seek to find the \textit{minimum violation ranking}~\cite{mvr}; a ranking of the nodes that minimises the sum of edge weights from a lower ranked node to higher ranked node. This measure has previously been used to uncover hierarchical structure in academic institutions~\cite{hiring}. The optimal MVR is found using Markov Chain Monte-Carlo sampling algorithm using a zero temperature Metropolis-Hastings acceptance function.\\

  Here we list the results of the full ranking of the provisions presented in the hierarchy in the main paper.

    \begin{table}[]
      \tiny
      \centering
      \begin{tabularx}{20cm}{@{}>{\bfseries}l@{\hspace{.5em}}X@{}}
        Ranking & Provision\\\hline

        1.0 & Does the constitution provide for a central representative body (a legislature)? \\ \hline
2.0 & Does the constitution mention the executive cabinet/ministers? \\ \hline
3.0 & Is the military or armed forces mentioned in the constitution? \\ \hline
4.0 & Does the constitution mention international treaties? \\ \hline
5.0 & Does the constitution provide for at least one procedure for amending the constitution? \\ \hline
6.0 & Does the constitution refer to equality before the law, the equal rights of men, or non-discrimination? \\ \hline
7.0 & Does the constitution stipulate that some public office holders take an oath to support or abide by the constitution? \\ \hline
8.0 & Does the constitution contain provisions concerning education? \\ \hline
9.0 & Does the constitution contain provisions concerning the relationship between the constitution and international law? \\ \hline
10.0 & Does the constitution contain provisions concerning a national or official religion or a national or official church? \\ \hline
11.0 & Does the constitution place any restrictions on the right to vote? \\ \hline
12.0 & Does the constitution provide for freedom of religion? \\ \hline
13.0 & "Does the constitution refer to ""democracy"" or ""democratic""?" \\ \hline
14.0 & Can the government expropriate private property under at least some conditions? \\ \hline
15.0 & Does the constitution provide for freedom of expression or speech? \\ \hline
16.0 & Does the constitution provide for freedom of assembly? \\ \hline
17.0 & Does the constitution provide for freedom of association? \\ \hline
18.0 & Does the constitution provide for a right of privacy? \\ \hline
19.0 & Are there provisions for dismissing judges? \\ \hline
20.0 & Does the constitution provide for a right to own property? \\ \hline
21.0 & Does the constitution provide for the right to protection from unjustified restraint (habeas corpus)? \\ \hline
22.0 & Does the constitution mention nulla poena sine lege or the principle that no person should be punished without law? \\ \hline
23.0 & "Does the constitution mention ""God"" or any other Deities?" \\ \hline
24.0 & Does the constitution contain an explicit declaration regarding the independence of the central judicial organ(s)? \\ \hline
25.0 & "Does the constitution refer to ""democracy"" or ""democratic""?" \\ \hline
26.0 & Does the constitution provide for naturalized citizens or naturalization? \\ \hline
27.0 & Do constitutional amendments require more than a simple majority by the legislature to be approved? \\ \hline
28.0 & Does the constitution prescribe whether or not the meetings of the Legislature are (generally) held in public? \\ \hline
29.0 & Does the constitution provide for freedom of opinion, thought, and/or conscience? \\ \hline
30.0 & Does the constitution provide for the extradition of suspected or convicted criminals to other countries? \\ \hline
31.0 & Does the constitution contain a national motto? \\ \hline
32.0 & Are there provisions for dismissing the Head of State? \\ \hline
33.0 & Does the constitution refer to the French declaration of rights (1789)? \\ \hline
34.0 & Does the constitution provide for freedom of expression or speech? \\ \hline
35.0 & Does the constitution provide for a right of petition? \\ \hline
36.0 & Does the constitution prohibit punishment by laws enacted ex post facto ? \\ \hline
37.0 & Is there a mention of telecommunications? \\ \hline
38.0 & Does the constitution regulate the collection of evidence? \\ \hline
39.0 & Does the constitution stipulate that certain rights are inalienable or inviolable? \\ \hline
40.0 & Does the constitution generally require public trials? \\ \hline
41.0 & Does the constitution prescribe that electoral ballots be secret? \\ \hline
42.0 & Does the constitution contain provisions specifying the location of the capital (if so, please specify the location in the comments section)? \\ \hline
43.0 & Are there provisions for dismissing the Head of Government? \\ \hline
44.0 & Does the constitution provide for freedom of movement? \\ \hline
45.0 & Does the constitution provide for suspension or restriction of rights during states of emergency? \\ \hline
46.0 & Are there provisions for removing individual legislators? \\ \hline
47.0 & Does the constitution specify a deputy executive of any kind (e.g., deputy prime minister, vice president)? \\ \hline
48.0 & Does the constitution specify a census? \\ \hline
49.0 & Does the constitution protect from discrimination/provide for equality for tribe/clan? \\ \hline
50.0 & Does the constitution define the geographic borders/territory of the state? \\ \hline
51.0 & Does the constitution contain provisions concerning the national flag? \\ \hline
52.0 & Does the constitution refer to a duty to pay taxes? \\ \hline
53.0 & What is the specified level of compensation for expropriation of private property? \\ \hline
54.0 & Does the constitution refer to ownership or possession of natural resources (such as minerals, oil, etc.)? \\ \hline
55.0 & Is the executive identified explicitly as the Head of State or Head of Government? \\ \hline
56.0 & Are there provisions for the secession or withdrawal of parts of the state? \\ \hline
57.0 & Does the constitution make a claim to universal adult suffrage? \\ \hline
58.0 & Does the constitution protect from discrimination/provide for equality for religion? \\ \hline
59.0 & Does the legislature have the power to interpellate members of the executive branch?? \\ \hline
60.0 & Are any parts of the constitution unamendable? \\ \hline
61.0 & Does the constitution refer to a duty of military service? \\ \hline
62.0 & Does the constitution stipulate that education be free, at least up to some level? \\ \hline
63.0 & Does the constitution refer to artists or the arts? \\ \hline
64.0 & Who has the power to approve declarations of war? \\ \hline
65.0 & Does the constitution refer to artists or the arts? \\ \hline
66.0 & Does the constitution contain an explicit declaration regarding the INDEPENDENCE of the central executive organ(s)? \\ \hline
67.0 & Does the constitution mention foreign or international trade? \\ \hline
68.0 & Does the constitution require a jury or any form of citizen participation in decision making in criminal trials? \\ \hline
69.0 & Are members of the first (or only) chamber elected in the same cohort, or in staggered cohorts? \\ \hline
70.0 & Does the constitution contain provisions allowing review of the legislation of the constituent units in federations by federal judicial or other central government organs? \\ \hline
71.0 & Is the state described as either federal, confederal, or unitary? \\ \hline
72.0 & Does the constitution provide for the right to choose ones occupation? \\ \hline
73.0 & Does the constitution prohibit censorship? \\ \hline
74.0 & Does the constitution give the accused a right to silence or protection from self incrimination? \\ \hline
75.0 & Does the constitution mention any of the following intellectual property rights: patents? \\ \hline
76.0 & In what language is the source document written (not the original, but the one used for coding)? \\ \hline
77.0 & Does the constitution contain provisions with regard to any additional central independent regulatory agencies ? \\ \hline
78.0 & Does the constitution prohibit slavery, servitude, or forced labor? \\ \hline
79.0 & Does the constitution mention the right to transfer property freely? \\ \hline
80.0 & Does the constitution provide for the right to bear arms? \\ \hline
81.0 & Is there citizen involvement in the indicting process (such as a grand jury)? \\ \hline
82.0 & Does the constitution specify the electoral system for the Second Chamber? \\ \hline
83.0 & Does the constitution stipulate that education be compulsory until at least some level? \\ \hline
84.0 & Does the constitution guarantee academic freedom? \\ \hline
85.0 & Does the constitution provide for judicial opinions of the Highest Ordinary Court? \\ \hline
86.0 & Does the constitution protect from discrimination/provide for equality for race? \\ \hline
87.0 & Does the constitution explicitly state that judicial salaries are protected from governmental intervention? \\ \hline
88.0 & Is there a constitutional provision for civil marriage? \\ \hline
89.0 & Does the constitution refer to political parties? \\ \hline
90.0 & Does the constitution require that the names of those imprisoned be entered in a public registry? \\ \hline
91.0 & Does the constitution provide for the right to form or to join trade unions? \\ \hline
92.0 & Does the constitution provide for a right to life? \\ \hline
93.0 & Does the constitution specify that the chambers should meet jointly for any reason? \\ \hline
94.0 & Is there a mandatory retirement age for judges? \\ \hline
95.0 & Does the constitution provide for the right of some redress in the case of false imprisonment, arrest, or judicial error? \\ \hline
96.0 & Does the constitution provide for immunity for the members of the Legislature under some conditions? \\ \hline
97.0 & Does the constitution provide for the right/possibility of pre-trial release? \\ \hline
98.0 & Does the constitution contain an explicit decree of separation of church and state? \\ \hline
99.0 & Does the constitution contain provisions protecting the individual against illegal or ultra-vires administrative actions? \\ \hline
100.0 & Does the constitution mention bankruptcy law? \\ \hline
101.0 & Does the constitution stipulate that some public office holders take an oath to support or abide by the constitution? \\ \hline
102.0 & Does the legislature have the power to investigate the activities of the executive branch? \\ \hline
103.0 & "Does the constitution refer to ""fraternity"" or ""solidarity""?" \\ \hline
104.0 & Does the constitution protect from discrimination/provide for equality for gender? \\ \hline
105.0 & Does the constitution refer to the UN universal Declaration on Human Rights (1948)? \\ \hline
106.0 & Does the constitution protect from discrimination/provide for equality for creed/beliefs? \\ \hline
107.0 & Does the constitution require that legislators give up any other profession (i.e. work exclusively as legislators)? \\ \hline
108.0 & Does the constitution contain provisions concerning the relationship between the constitution and international law? \\ \hline
109.0 & Does the constitution contain provisions for the protection of stateless individuals, refugees from other states, or the right to asylum? \\ \hline
110.0 & Is a supermajority needed for passing any legislation? \\ \hline
111.0 & Does the constitution make voting mandatory, at least for some elections? \\ \hline
112.0 & Does the constitution mention the adoption of national economic plans? \\ \hline
113.0 & Does the constitution mention a state duty to provide work/employment? \\ \hline
114.0 & Does the constitution provide the right to counsel if one is indicted or arrested? \\ \hline
115.0 & Does the constitution specify the trial has to be in a language the accused understands or the right to an interpreter if the accused cannot understand the language? \\ \hline
116.0 & "Is one of the executives explicitly referred to as the ""Head of State""?" \\ \hline
117.0 & "Does the constitution refer to ""socialism"" or ""socialist""?" \\ \hline
118.0 & Are there provisions for the secession or withdrawal of parts of the state? \\ \hline
119.0 & Does the constitution protect from discrimination/provide for equality for country of origin? \\ \hline
120.0 & Does the constitution prohibit torture? \\ \hline
121.0 & Is there a presumption of innocence in trials? \\ \hline
122.0 & Does the constitution refer to social, political, or economic conditions in the time before the birth of the state or in the time of a former constitution? \\ \hline
123.0 & Does the constitution prohibit cruel, inhuman, or degrading treatment? \\ \hline
124.0 & Does the constitution forbid the detention of debtors? \\ \hline
125.0 & Does the constitution protect from discrimination/provide for equality for social status? \\ \hline
126.0 & Does the constitution provide for a right of rest and leisure? \\ \hline
127.0 & Does the constitution protect from discrimination/provide for equality for color? \\ \hline
128.0 & Does the constitution grant the government the right to deport citizens or residents? \\ \hline

\end{tabularx}
\caption{Final ranking of provisions}
\end{table}

\begin{table}[]
  \tiny
  \centering
  \begin{tabularx}{20cm}{@{}>{\bfseries}l@{\hspace{.5em}}X@{}}
    Ranking & Provision\\\hline

129.0 & Does the constitution refer to radio? \\ \hline
130.0 & Does the constitution stipulate a quota for representation of certain groups in the first (or only) chamber? \\ \hline
131.0 & Does the constitution refer to a duty to join a political party? \\ \hline
132.0 & Does the constitution provide a right to conduct/establish a business? \\ \hline
133.0 & Does the constitution provide for a right of testate, or the right to transfer property freely after death? \\ \hline
134.0 & Does the constitution protect from discrimination/provide for equality for nationality? \\ \hline
135.0 & Does the constitution mention any of the following intellectual property rights: copyrights? \\ \hline
136.0 & Does the constitution provide for the right of protection of one's reputation from libelous actions? \\ \hline
137.0 & Does the constitution mention any of the following intellectual property rights: general reference to IP? \\ \hline
138.0 & Does the constitution provide for the prohibition of double jeopardy (i.e., being tried for the same crime twice)? \\ \hline
139.0 & Is there a right to exemption from military service for conscientous objectors to war or other groups? \\ \hline
140.0 & Does the constitution include provisions for the meritocratic recruitment of civil servants (e.g. exams or credential requirements)? \\ \hline
141.0 & Does the constitution restrict entry or exit of the states borders? \\ \hline
142.0 & Does the constitution provide for either general or financial support by the government for the elderly? \\ \hline
143.0 & Does the constitution refer to a state duty to protect or promote culture or cultural rights? \\ \hline
144.0 & Does the constitution provide for either general or financial support by the government for the disabled? \\ \hline
145.0 & Does the constitution mention a state duty to provide work/employment? \\ \hline
146.0 & Does the constitution mention any of the following intellectual property rights: trademark? \\ \hline
147.0 & Does the constitution mention a state duty to provide health care? \\ \hline
148.0 & Does the constitution provide for either general or financial support by the government for the unemployed? \\ \hline
149.0 & Does the constitution contain provisions concerning the national anthem? \\ \hline
150.0 & Does the constitution provide for inheritance rights? \\ \hline
151.0 & Does the constitution contain provisions for a Judicial Council/Commission? \\ \hline
152.0 & Does the constitution provide for the right to a speedy trial? \\ \hline
153.0 & Does the constitution stipulate that courts have to take into account decisions of higher courts? \\ \hline
154.0 & "Does the constitution refer to ""customary"" international law or the ""law of nations""?" \\ \hline
155.0 & Does the constitution contain provisions concerning international organizations? \\ \hline
156.0 & Do defendants have the right to appeal judicial decisions? \\ \hline
157.0 & Does the constitution prohibit one or more political parties? \\ \hline
158.0 & Does the constitution refer to a duty of the people to take part in building society or to work for the development of the country? \\ \hline
159.0 & "Is one of the executives explicitly referred to as the ""Head of Government""?" \\ \hline
160.0 & Does the constitution provide for either general or financial support by the government for children, orphans? \\ \hline
161.0 & Does the constitution provide for positive obligations to transfer wealth to, or provide opportunity for, particular groups? \\ \hline
162.0 & Does the constitution provide for an electoral commission or electoral court to oversee the election process? \\ \hline
163.0 & Does the constitution provide the right to just remuneration, fair or equal payment for work? \\ \hline
164.0 & Does the constitution stipulate a quota for representation of certain groups in the Second Chamber? \\ \hline
165.0 & "Does the constitution refer to the ""dignity of man"" or human ""dignity""?" \\ \hline
166.0 & Does the constitution mention the right to safe/healthy working conditions? \\ \hline
167.0 & Does the constitution protect from discrimination/provide for equality for parentage? \\ \hline
168.0 & "Does the constitution provide for a right to petition for ""amparo""?" \\ \hline
169.0 & Does the constitution express a preference for one or more political parties? \\ \hline
170.0 & Does the constitution provide for a right to form political parties? \\ \hline
171.0 & Does the constitution refer to the social security of the society or nation? \\ \hline
172.0 & Does the constitution provide for a right to strike? \\ \hline
173.0 & Does the constitution explicitly mention due process? \\ \hline
174.0 & Does the constitution refer to the UN charter Article 45 (1945)? \\ \hline
175.0 & Does the constitution refer to protection or preservation of the environment? \\ \hline
176.0 & Does the constitution contain provisions concerning national integration of ethnic communities? \\ \hline
177.0 & Does the constitution place limits on child employment? \\ \hline
178.0 & Does the constitution contain provisions for a central bank? \\ \hline
179.0 & Does the constitution provide the right to a free and/or competitive market? \\ \hline
180.0 & Does the constitution provide for the right to marry? \\ \hline
181.0 & Does the constitution give juveniles special rights/status in the criminal justice process? \\ \hline
182.0 & Does the constitution guarantee the rights of children? \\ \hline
183.0 & Does the constitution provide for a people's right of self-determination? \\ \hline
184.0 & Does the constitution provide for matrimonial equality? \\ \hline
185.0 & "Does the constitution mention ""foreign investment"" or ""foreign capital""?" \\ \hline
186.0 & Does the constitution mention the right to health care? \\ \hline
187.0 & How does the constitution address the state operation of print or electronic media? \\ \hline
188.0 & Does the constitution provide for an individual's right to self determination or the right to free development of personality? \\ \hline
189.0 & Does the constitution refer to television? \\ \hline
190.0 & Does the constitution protect from discrimination/provide for equality for language? \\ \hline
191.0 & Are rights provisions binding on private parties as well as the state? \\ \hline
192.0 & Does the constitution provide for a right to an adequate or reasonable standard of living? \\ \hline
193.0 & Does the constitution guarantee equal access to higher education? \\ \hline
194.0 & Does the constitution protect from discrimination/provide for equality for political party? \\ \hline
195.0 & Does the constitution suggest that citizens should have the right to overthrow their government under certain circumstances? \\ \hline
196.0 & Does the constitution refer to the African Charter on Human People's Rights (1981)? \\ \hline
197.0 & Does the constitution refer to the protection of different languages? \\ \hline
198.0 & "Does the constitution refer to the ""free market,"" ""capitalism,"" or an analogous term?" \\ \hline
199.0 & Do citizens have the right to renounce their citizenship? \\ \hline
200.0 & Does the constitution protect from discrimination/provide for equality for financial/propety ownership? \\ \hline
201.0 & Does the constitution mention any special procedures for removing members of the constitutional court? \\ \hline
202.0 & Does the constitution use the words (socio-) economic rights or similar? \\ \hline
203.0 & Does the constitution provide the right to found a family? \\ \hline
204.0 & Does the constitution refer to the Helsinki Accords (1966)? \\ \hline
205.0 & Does the constitution refer to the International Covenant on Civil and Political Rights (1966)? \\ \hline
206.0 & Does the constitution refer to the International Covenant on Economic and Social Rights (1966)? \\ \hline
207.0 & Does the constitution require that legislators disclose their earnings and/or assets? \\ \hline
208.0 & Does the constitution provide for the ability of individuals to propose legislative initiatives (referenda from below)? \\ \hline
209.0 & Does the constitution mention judicial opinions of the Constitutional Court? \\ \hline
210.0 & Does the constitution contain a general statement regarding rule of law, legality, or Rechtsstaat (the German equivalent)? \\ \hline
211.0 & If counsel is provided, is it provided at the state's expense? \\ \hline
212.0 & Does the constitution refer to the American Convention on Human Rights (1969)? \\ \hline
213.0 & Does the constitution specify that healthcare should be provided by government free of charge? \\ \hline
214.0 & Does the constitution provide the right to a fair trial? \\ \hline
215.0 & Does the constitution provide for an Ombudsman? \\ \hline
216.0 & Does the constitution provide the right for same sex marriages? \\ \hline
217.0 & Does the constitution mention a special regulatory body/institution to oversee the media market? \\ \hline
218.0 & Does the constitution contain provisions for a counter corruption commission? \\ \hline
219.0 & Does the constitution mention anything about crimes committed by the previous regime? \\ \hline
220.0 & Does the constitution provide for a commission for truth and reconciliation? \\ \hline
221.0 & Is there special mention of terrorism and public security provisions regarding terrorism? \\ \hline
222.0 & Does the constitution provide for the right to shelter or housing? \\ \hline
223.0 & Does the constitution provide for an individual right to view government files or documents under at least some conditions? \\ \hline
224.0 & How many chambers or houses does the Legislature contain? \\ \hline
225.0 & Does the constitution refer to a duty to join trade unions? \\ \hline
226.0 & Is there a special mention of victims rights in the constitution? \\ \hline
227.0 & Does the constitution provide for a right to enjoy the benefits of scientific progress? \\ \hline
228.0 & Does the constitution refer to the European Convention for the Protection of Human Rights and Fundamental Freedoms (1950)? \\ \hline
229.0 & Does the constitution mention consumer rights or consumer protection? \\ \hline
230.0 & Does the constitution protect from discrimination/provide for equality for age? \\ \hline
231.0 & Does the constitution protect from discrimination/provide for equality for mental or physical disability? \\ \hline
232.0 & Does the constitution suggest that citizens should have the right to overthrow their government under certain circumstances? \\ \hline
233.0 & Does the constitution protect from discrimination/provide for equality for sexual orientation? \\ \hline
234.0 & Does the constitution provide for the right to examine evidence or confront all witnesses? \\ \hline

\end{tabularx}
\caption{Final ranking of provisions}
\end{table}